\documentclass{article}
\usepackage{sao2}
\usepackage{psfig}
\issue{2003, 55, 90--132  }
\begin{document}
\title{Radio spectra of the WMAP catalog sources }
\author{S.A. Trushkin}
\institute{\saoname}
\date{May 13, 2003}{May 29, 2003}
\maketitle

\begin{abstract}
Compiled radio spectra are presented for 208 extragalactic sources from the
catalog created from the WMAP satellite all-sky survey data in a range of
23--94 GHz taken during the first year of its operation in orbit.
205 out of 208 WMAP sources are reliably identified with radio sources
from other catalogs, including also four out of five sources unidentified
by the WMAP survey authors. We have found 203 WMAP sources to have optical
identification: 141 quasars, 29 galaxies, 19 active galactic nuclei, 19 BL
Lac-type objects and one planetary nebula, IC\,418.
Simultaneous measurements of flux densities  for 26 sources at five
frequencies, 2.3, 3.9, 7.7, 11.2 and 21.7 GHz, were made with the radio
telescope RATAN-600 in 2003 March. 25 sources were detected at all the
frequencies, and only one, WMAP~0517--0546, unidentified in other catalogs
was not detected in our observations and is likely to be spurious. Using
the database CATS we found a large number of identifications in different
radio catalogs and in several long-term monitoring programs (GBI, UMRAO, SEST, IRAM, and
others). The total number of flux density measurements that we collected
in the frequency range from 10 MHz to 245 GHz for the identified WMAP sources
is over 206,000, varying from 10--20 points for some southern sources to a
few thousand points for sources from long-term monitorings. A considerable
number of WMAP sources are variable on time scales from hours to dozens of
years, nevertheless, the obtained integral spectra, including measurements
in different years, may be of particular interest because these bright
     sources have been studied for more than forty years. The general
character of the spectrum (power, flat, with a turnovers at low or high
frequencies) of the WMAP sources remains unchanged. No significant
systematic differences in flux densities measured in the WMAP observations
as compared to earlier measurements at close frequencies for sources with
a small variability index are seen in the spectra.
\end{abstract}

\section{Introduction}

During the first year of observations with the satellite Wilkinson
Microwave Anisotropy Probe (WMAP) with a high sensitivity, the entire sky
was mapped at five frequencies, 23, 33, 41, 61 and 94 GHz
(Bennett et al. 2003, hereafter Paper 1). In the course of data reduction for
removing foreground emission, 208 radio sources were detected on the maps
obtained out of the Galactic plane (Bennett et al. 2003, hereafter Paper 2).
They were found as a result of reduction of the obtained sky maps with
$|b|>10^\circ$, which included the filtration of spatial harmonics,
searches for local peaks with an amplitude $> 5\sigma$, where $\sigma$ is
the local noise level, and Gaussian analysis of these peaks with
preliminary removal of the local foreground.
There are 191 flux measurements at the frequency 23\,GHz, 185 at 33\,GHz,
182 at 41\,GHz, 137 at 61\,GHz and 60 at 94\,GHz. The minimum measured
flux is 0.3\,Jy at 33\,GHz, 0.4\,Jy at 41\,GHz, 0.5\,Jy at 23 and
61\,GHz and 1.1\,Jy at 94\,GHz.

Then, the authors (Paper 2) performed cross identification of the obtained WMAP
catalog with the sources from the catalogs GB6 (Gregory et al. 1996),
PMN (Griffith et al. 1994, 1995; Wright et al. 1996) and from the complete
sample of sources brighter than 1\,Jy at 5 \,GHz (K\"{u}hr et al. 1981).
The sources were thought to be identified if they were separated by less
than $11'$ (that is, the coordinate error was equal to $4'$).
For 203 sources identifications were found in the catalog, and 20 sources
had more than one identification. The authors believed that since no
identifications in the given window were found for five of the sources,
and the number of false detections from the simulation modeling was also
equal to $5\pm4$, then, probably, the remaining five sources were
also spurious.
Thus, no evidence is available that an unknown population of radio sources
exists in the range of WMAP survey frequencies at a flux level of about
1\,Jy. The mean spectral index of point sources of the catalog determined
only from the measurements of the WMAP survey is equal to zero, with an
error of about 0.5 (Paper 2).

An all-sky survey at high frequencies ($>$20\,GHz) has been made for the
first time and a catalog of sources has been compiled. For this reason
it would be interesting to investigate those sources in detail, to
understand which physical objects represent this sample of sources, what
kind of radio spectra and structure they have.  To solve the problem,
SAO RAS has a convenient and  powerful instrument for investigation --- the
generally accessible radio astronomical database CATS
(Verkhodanov et al. 1997), which includes the relevant procedures of
search  and cross-identification of samples of objects.
Besides, the hard disk of the CATS server ({\tt cats.sao.ru}) contains all
the main radio catalogs, catalogs of galaxies, quasars and the data of long-term
monitoring programs. The CGI-procedures of the radio spectrum plotting
``on-line'' require only small changes to include the WMAP catalog.

\section{Cross-identification of the WMAP sources in the CATS}

First, following the authors of Paper 2, we performed the procedures of
searching for identifications in the catalogs GB6 (Gregory et al. 1996)
and PMN. The coordinates of the sources from these catalogs are much more accurate
than those of the initial  WMAP catalog. On the average, the differences of
the WMAP coordinates and identifications are $4-6'$.
At this stage it became clear that the earlier unidentified source
WMAPJ2356+4952 is undoubtedly the bright source J2355+4950, although the
discrepancy of their coordinates is about $12.5'$. The source
WMAP1633+8226 is obviously the known Seyfert galaxy NGC~6251 (J1632+8232),
the difference in the coordinates here is about $6'$.

The general summary of cross-identification of the WMAP sources with the
largest radio surveys looks as follows (in brackets are indicated the
frequencies of the surveys and the main reference): 12 sources in 8C
(38~MHz;  Hales et. al. 1995);
16 sources in 4MASS  (74~MHz; Cohen et al. 2002);
21 sources in 3C/3CR (159~MHz; Edge et al.  1959);
41 sources in 4C    (178~MHz;  Pilkington et al. 1965);
45 sources in 6C    (151~MHz;  Hales el at. 1993);
24 sources in 7C    (151~MHz;  Pooley et al. 1998);
51 sources in WENSS   (325/352~MHz; Rengelink et al. 1997);
105 sources in TEXAS  (365~MHz;  Douglas et al. 1996);
48 sources in SUMSS  (843~MHz; Bock et al. 1999);
92 sources in  WB92  (1400~MHz; White \& Becker 1992);
168 sources in NVSS (1400~MHz;  Condon et al. 1998);
46 sources in FIRST (1400~MHz, White et al. 1997);
162 sources from compiled 1-Jy survey (K\"uhr et al. 1979, 1981);
102 sources in \mbox{VLBIS} (2290~MHz; Preston et al. 1985);
127 sources in PKSCAT90 (2700/8400~MHz;  Wright \& Otrupcek 1990);
58 sources   Z2  (3900~MHz; Amirkhanyan et al. 1992);
100 sources  GB6 (4850~MHz; Gregory et al. 1996);
137 sources  PMN (4850~MHz; Wright et al. 1994, 1996; Griffith et al. 1994,
1995);
73 sources  MITGB (4850~MHz; Bennett et al. 1986; Griffith et al. 1990);
124 sources  VLBApls (5000~MHz; Fomalont et al. 2000);
83 sources  JVAS (8400~MHz; Wilkinson et al. 1998);
66 sources  VLBI-2cm  (15000~MHz; Kellermann et al. 1998).
108 sources are from linear polarization data catalog (Tabara \& Inoue 1980).
Seven of eleven WMAP sources of the northern sky (Decl(1950)$>70^\circ$)
showed a variability on a scale less than  24 hours at a frequency of
2700~MHz (Heeschen et al. 1987).
34 southern WMAP sources are classified as sources
with intra-day variability (IDV-sources; Kedziora-Chudczer et al. 2001).

Then the list of WMAP sources with more accurate coordinates was searched for
with the CATS database with a smaller window ($100''$). In the course of
this procedure it became clear that many identified WMAP sources are
included in the long-time multi-frequency monitoring programs of bright
variable extragalactic radio sources: Naval Research Lab and Green Bank
Interferometer\footnote{The Green Bank Interferometer is a facility of the
National Science Foundation operated by  the National Radio Astronomy
Observatory in support of NASA High Energy Astrophysics programs and
under contract to the US Naval Observatory and Naval Research Laboratory.}
monitoring program (1979--2001) (Fiedler et al. 1987; Waltman et al. 1991;
Lazio et al. 2001; http://ese.nrl.navy.mil/GBI/;
ftp.gb.nrao.edu/pub/fghigo/gbidata/gdata/),
University Michigan RAO\footnote{University of Michigan Radio Astronomy
Observatory is supported by funds from the University of Michigan.}
monitoring (1985--1999)
(http://www.astro.lsa.umich.edu/obs/ra\-dio\-tel/um\-rao.html;
Aller \& Aller 1996), Mets\"ahovi monitoring program (1985--2000)
(Ter\"asranta et al. 1998). Data on similar programs carried out at
RATAN-600 (Kovalev et al. 1999, Mingaliev et al. 1998, 2001;
Kiikov et al. 2002) and data of monitoring at other radio telescopes
(Quiniento et al. 1993; Reuter et al. 1997;  Ghosh et al. 1994;
Tornikoski et al. 1996; Dallacasa et al. 2000) are also available
in the CATS database. The data of these programs increased considerably
the number of flux density measurements of the identified WMAP sources
required for the plotting of integral spectra.

Besides, it should be noted that the CATS database incorporates the data
of the compiled catalogs of radio sources brighter than 1\,Jy at 5 GHz
(K\"uhr et al. 1979, 1981).  This complete sample of
576 bright sources
was studied purposefully in the optical range for identification and
measurement of the redshift (Stickel et al. 1994). However, a considerable
number (26) of unidentified sources or ones with unclear classification
still remains. Nevertheless, we may say that the identified WMAP sources
can not be referred to those poorly studied.

In the end we obtained a catalog of identifications which comprises more
than 206000 flux density measurements in the range from 10\,MHz to
245\,GHz. Since the search for identifications goes on, the catalog is
being replenished. The updated spectra can be plotted ``on-line''
{\tt http://cats.sao.ru/cgi-bin/wmap.cgi} changing at choice the fitting
parameters: a fitting law, a frequency range, an optional
account of the error measurements.

We also searched for identifications of WMAP sources in the optical and
X-ray ranges. 141 WMAP sources are identified with the ROSAT all-sky survey
sources (Voges et al. 1999). 87 of them are bright X-ray sources.
Among 1550 identified in the NVSS bright X-ray sources of the ROSAT survey,
in the so-called sample RBSC-NVSS (Bauer et al. 2000) there are
39 WMAP sources, 10 of them are active galaxies, six --- quasars and the rest
objects are of BL Lac type, or blazars. The redshift is measured for
38 of them.
Seven WMAP sources are gamma-ray sources from the Third EGRET catalog
(Hartman et al. 1999).

Table\,\ref{t1} presents a list of WMAP sources with accurate (the error is
better $5''$) coordinates.  The fourth column gives the type of the
source from three catalogs:
Veron-Cetty \& Veron 2001 (herein VV10, in the table \ref{t1} V),
Stickel et al. 1994 (herein Stickel94, in the table \ref{t1} S),
Fomalont et al. 2000 (herein VLBApls, in the table \ref{t1} F).

\begin{onecolumn}
\hspace{16cm}
\topcaption{\large Parameters of WMAP sources}
\tabletail{\hline}
\tablehead{%
 \hline
 Source  & RA 2000  &DEC 2000  & Type  &  3C   &         &$S_{30}$& $\alpha_{20}$& F-R  \\
 name    &hh~mm~ss.s&~dd~mm~ss  & VSF  & 3CR   &   z     & Jy  &{\normalsize $S_\nu\sim S^\alpha$}      & type  \\
\hline
}
\label{t1}
\begin{supertabular}{|llrlllrrl|}
J0006$-$0623&00 06 13.3&$-$06 23 13& BQG&  3    & 0.347  & 2.07&$-$0.067&II IDV  \\
J0013$-$3954&00 12 59.9&$-$39 54 25& G  &       &        & 1.17&   0.063&   IDV  \\
J0047$-$2517&00 47 33.1&$-$25 17 17& AG &       & 0.00086& 1.12&$-$0.248&        \\
J0050$-$5738&00 49 59.5&$-$57 38 27& QQ &       & 1.797  & 1.42&$-$0.075&        \\
J0050$-$0928&00 50 41.3&$-$09 29 06& BBB&       &        & 1.21&$-$0.233&        \\
J0106$-$4034&01 06 45.1&$-$40 34 20&Q~~Q&       & 0.584  & 2.56&$-$0.018&        \\
J0108+1319 &01 08 50.6& +13 18 31  & A  &       &        & 1.21&$-$0.721&        \\
J0108+0135 &01 08 38.7& +01 34 59  & QQQ&  33   & 2.107  & 1.79&$-$0.267&  II    \\
J0125$-$0005&01 25 28.9&$-$00 05 55& QQQ&       & 1.070  & 1.10&$-$0.253&        \\
J0132$-$1654&01 32 43.4&$-$16 54 48& QQQ&       & 1.022  & 0.55&$-$0.071&        \\
J0133$-$5159&01 33 05.8&$-$52 00 04& QQ &       &        & 0.78&$-$0.025&  IDV   \\
J0136+4751 &01 36 58.6& +47 51 29  & QQQ&       & 0.859  & 2.07&   0.028&        \\
J0204+1514 &02 04 50.4& +15 14 10  & AQG&       & 0.405  & 2.20&$-$0.229&        \\
J0205+3212 &02 05 04.9& +32 12 31  & QQQ&       & 1.466  & 0.81&$-$0.054&        \\
J0210$-$5101&02 10 46.2&$-$51 01 02& BQ &       & 1.003  & 2.77&$-$0.114&        \\
J0217+0144 &02 17 48.9& +01 44 50  &Q~~Q&       & 1.715  & 1.15&$-$0.197&  IDV   \\
J0222$-$3441&02 22 56.4&$-$34 41 28& Q  &       &        & 0.74&$-$0.002&  IDV   \\
J0223+4259 &02 23 12.1& +43 00 03  & BG &  66.0 & 0.0215 & 0.93&$-$0.509&  I     \\
J0237+2848 &02 37 52.3& +28 48 10  & QQQ&       & 1.210  & 2.11&   0.025&        \\
J0253$-$5441&02 53 31.7&$-$54 41 52& Q  &       &        & 2.43&   0.584&  IDV   \\
J0303$-$6211&03 03 50.6&$-$62 11 25& QQ &       &        & 1.46&$-$0.069&  IDV   \\
J0308+0406 &03 08 26.3& +04 02 40  & A  &       &        & 1.23&$-$0.573&        \\
J0309$-$6058&03 09 56.0&$-$61 58 39& Q  &       &        & 1.49&   0.198&        \\
J0311$-$7651&03 11 55.6&$-$76 51 49& Q  &       &        & 1.12&   0.317&        \\
J0319+4130 &03 19 48.2& +41 30 42  & AGG&  84.0 & 0.0172 &19.15&$-$0.376&  I     \\
J0321$-$3711&03 22 36.5&$-$37 14 21& G  &       &        &13.11&$-$0.749&        \\
J0329$-$2357&03 29 54.1&$-$23 57 09& Q  &       &        & 1.32&   0.253&        \\
J0334$-$4008&03 34 13.6&$-$40 08 25& BQQ&       & 1.445  & 1.57&$-$0.166&        \\
J0339$-$0146&03 39 30.9&$-$01 46 35& QQQ&       & 0.852  & 2.08&$-$0.070&  IDV   \\
J0348$-$2749&03 48 38.2&$-$27 49 14&Q~~Q&       & 0.987  & 1.52&   0.296&  IDV   \\
J0403$-$3605&04 03 53.7&$-$36 05 02& QQQ&       & 1.417  & 2.74&   0.208&        \\
J0405$-$1308&04 05 34.0&$-$13 08 16& QQQ&       & 0.571  & 1.00&$-$0.400&        \\
J0406$-$3826&04 06 59.0&$-$38 26 28& QQQ&       & 1.285  & 1.13&$-$0.085&  IDV   \\
J0411+7656 &04 10 45.0&  76 56 47  & EG &       & 0.5985 & 0.87&$-$0.715&        \\
J0423$-$0120&04 23 15.8&$-$01 20 33& QQQ&       & 0.915  & 4.10&$-$0.025&        \\
J0424$-$3756&04 24 42.2&$-$37 56 20&Q~~Q&       & 0.782  & 0.43&$-$0.205&        \\
J0424+0036 &04 24 46.8&  00 36 07  & B~B&       & 0.310  & 1.38&$-$0.035&  IDV   \\
J0433+0521 &04 33 11.1&  05 21 15  & AGG&  120  & 0.0331 & 2.64&$-$0.246&  I     \\
J0440$-$4332&04 40 17.0&$-$43 33 08& QQQ&       & 2.852  & 1.93&$-$0.508&        \\
J0450$-$8100&04 50 05.4&$-$81 01 01& AQ &       & 0.444  & 1.79&   0.181&  IDV   \\
J0453$-$2807&04 53 14.5&$-$28 07 44& QQQ&       & 2.559  & 1.70&$-$0.092&        \\
J0455$-$4616&04 55 50.7&$-$46 15 59& QQ &       & 0.858  & 1.19&$-$0.284&        \\
J0457$-$2324&04 57 03.2&$-$23 24 52& QQB&       & 1.003  & 1.41&$-$0.277&        \\
J0506$-$6109&05 06 44.0&$-$61 09 41& QQ &       & 1.093  & 1.24&$-$0.117&        \\
J0513$-$2159&05 13 49.1&$-$21 49 16& Q  &       &        & 0.66&$-$0.334&        \\
J0518$-$0546&05 18 54.6&$-$05 46 10& ?  &       &        & 1.30&   0.343&        \\
J0519$-$4546&05 19 44.3&$-$45 46 41& AG &       & 0.0342 & 4.61&$-$0.710&  II    \\
J0522$-$3628&05 22 57.2&$-$36 27 36& GGB&       & 0.055  & 4.42&$-$0.329&  I     \\
J0527$-$1241&05 27 28.2&$-$12 41 50& P  &       &        & 1.37&$-$0.236&        \\
J0538$-$4405&05 38 50.3&$-$44 05 09& QB &       & 0.896  & 4.85&   0.044&        \\
J0540$-$5418&05 40 45.9&$-$54 18 22& G  &       &        & 1.26&   0.848&        \\
J0542+4951 &05 42 36.1&  49 51 07  & QQQ&  147.0& 0.545  & 1.38&$-$0.984&        \\
J0546$-$6415&05 46 42.0&$-$64 15 23& Q  &       &        & 0.61&   0.622&        \\
J0550$-$5732&05 50 09.6&$-$57 32 24& Q  &       &        & 1.36&   0.812&        \\
J0555+3948 &05 55 30.8& +39 48 49  &Q~~Q&       & 2.365  & 4.36&$-$0.307&        \\
J0607+6720 &06 07 52.7& +67 20 55  & QQQ&       & 1.970  & 0.88&   0.083&        \\
J0609$-$1542&06 09 40.3&$-$15 41 41& QQQ&       & 0.324  & 7.55&   0.169&  IDV   \\
J0629$-$1959&06 29 23.8&$-$19 59 20& B  &       &        & 1.30&   0.254&        \\
J0633$-$2223&06 33 26.7&$-$22 23 21& A  &       &        & 1.05&   0.349&        \\
J0635$-$7516&06 35 46.5&$-$75 16 17& QQ &       & 0.654  & 4.94&$-$0.136&        \\
J0636$-$2041&06 36 31.0&$-$20 31 52& G  &       &        & 0.23&$-$1.387&        \\
J0639+7324 &06 39 21.9& +73 24 58  & Q  &       &        & 0.80&$-$0.018&        \\
J0646+4451 &06 46 32.0& +44 51 17  &Q~~Q&       & 3.408  & 2.03&   0.279&        \\
J0738+1742 &07 38 07.4& +17 42 19  & BBB&       &>0.424  & 1.73&   0.026&        \\
J0739+0136 &07 39 18.0& +01 37 04  & QQQ&       & 0.191  & 1.64&$-$0.083& IDV    \\
J0741+3112 &07 41 10.7& +31 12 00  & QQQ&       & 0.631  & 1.27&$-$0.349&        \\
J0743-6726 &07 43 31.5&$-$67 26 26 & QQ &       & 1.511  & 0.70&$-$0.579&        \\
J0745+1011 &07 45 33.0& +10 11 13  & EEE&       &        & 0.95&$-$0.922&        \\
J0750+1231 &07 50 52.0& +12 31 04  & QQQ&       & 0.889  & 2.60&   0.201&        \\
J0757+0956 &07 57 06.6& +09 56 34  & B  &       &        & 1.48&$-$0.039&        \\
J0808$-$0751&08 08 15.5&$-$07 51 10& QQQ&       & 1.837  & 1.73&   0.120&        \\
J0816$-$2421&08 16 40.4&$-$24 21 05& G  &       &        & 1.41&   0.728&        \\
J0825+0309 &08 25 50.3& +03 09 24  & BBB&       & 0.506  & 1.41&   0.002&        \\
J0830+2410 &08 30 52.1& +24 10 59  & Q  &       &        & 1.06&   0.281&        \\
J0836$-$2017&08 36 39.1&$-$20 16 59& QQQ&       & 2.752  & 1.32&$-$0.375&  IDV  \\
J0840+1312 &08 40 47.6& +13 12 24  & QQQ&  207  & 0.684  & 0.78&$-$0.354&  II    \\
J0841+7053 &08 41 24.3& +70 53 41  & QQQ&       & 2.172  & 1.19&$-$0.379&  II    \\
J0854+2006 &08 54 48.8& +20 06 31  & BBB&       & 0.306  & 3.30&   0.052&        \\
J0909+0121 &09 09 10.0& +01 21 42  & QQQ&       & 1.018  & 1.80&   0.537&        \\
J0918$-$1205&09 18 06.5&$-$12 05 28& AG &  218.0& 0.0547 & 2.34&$-$0.954&  I     \\
J0927+3902 &09 27 03.0& +39 02 21  & QQQ&       & 0.698  & 7.29&$-$0.235&  II    \\
J0948+4039 &09 48 55.3& +40 39 44  & QQQ&       & 1.252  & 1.77&   0.133&  I     \\
J0955+6940 &09 55 51.7& +69 40 46  & GG &  231.0& 0.0009 & 1.13&$-$0.640&  I IDV \\
J0958+4725 &09 58 19.7& +47 25 08  &Q~~Q&       & 1.873  & 0.99&   0.008&        \\
J1014+2301 &10 14 47.0& +23 01 16  &Q~~Q&       & 0.565  & 1.05&$-$0.027&        \\
J1033+4115 &10 33 03.7& +41 16 06  & QQQ&       & 1.120  & 0.79&   0.026&        \\
J1038+0512 &10 38 46.8& +05 12 29  & G  &       &        & 0.90&   0.307&        \\
J1041$-$4740&10 41 44.7&$-$47 38 38& G  &       &        & 0.91&$-$0.684&        \\
J1048+7143 &10 47 27.6& +71 43 35  & Q  &       &        & 1.11&   0.124&        \\
J1058+0133 &10 58 29.5& +01 33 58  & QQQ&       & 0.892  & 4.06&   0.211&        \\
J1058$-$8003&10 58 43.3&$-$80 03 54& QQ &       &        & 2.52&   0.221&  IDV   \\
J1107$-$4449&11 07 08.7&$-$44 49 07& QQ &       & 1.598  & 1.76&$-$0.208&        \\
J1127$-$1857&11 27 04.4&$-$18 57 18&Q~~Q&       & 1.048  & 2.05&   0.306&        \\
J1130$-$1449&11 30 07.0&$-$14 49 27& QQQ&       & 1.187  & 1.67&$-$0.460&  IDV   \\
J1130+3815 &11 30 52.6& +38 15 18  & Q  &       &        & 1.08&   0.127&        \\
J1147$-$3812&11 47 01.4&$-$38 12 11& QBQ&       & 1.048  & 1.87&   0.211&  IDV   \\
J1153+8058 &11 53 12.4& +80 58 29  & QQQ&       & 1.250  & 1.25&   0.255&        \\
J1153+4931 &11 53 24.4& +49 31 09  & QQQ&       & 0.334  & 1.44&   0.048&  II    \\
J1159+2914 &11 59 31.8& +29 14 43  &Q~~Q&       & 0.729  & 1.44&$-$0.030&        \\
J1209$-$2406&12 09 02.5&$-$24 06 20&B~~B&       &        & 1.08&   0.252&        \\
J1215$-$1731&12 15 46.7&$-$17 31 45& GGG&       &        & 1.11&$-$0.181&        \\
J1219+0549 &12 19 24.0& +05 49 21  & GG &  270.0& 0.0073 & 2.96&$-$0.571&  I/II  \\
J1224$-$8312&12 24 54.6&$-$83 13 10& G  &       &        & 1.20&   0.328&        \\
J1229+0202 &12 29 06.6&  02 03 09  & QQQ&  273.0& 0.158  &29.64&$-$0.167&        \\
J1230+1223 &12 30 49.4&  12 23 28  & AGG&  274.0& 0.0043 &16.70&$-$0.803&  I     \\
J1246$-$2547&12 46 46.8&$-$25 47 49& QQQ&       & 0.633  & 1.45&   0.028&        \\
J1256$-$0547&12 56 11.2&$-$05 47 22& QQQ&  279.0& 0.536  &17.87&   0.108&        \\
J1257$-$3154&12 57 59.0&$-$31 55 17& QQQ&       & 1.924  & 1.50&   0.104&  IDV   \\
J1310+3220 &13 10 28.6&  32 20 43  & QBQ&       & 0.997  & 3.02&$-$0.042&        \\
J1316$-$3339&13 16 07.9&$-$33 38 59& QQQ&       & 1.210  & 1.66&   0.120&        \\
J1329+3154 &13 29 52.8&  31 54 11  & Q  &       &        & 1.12&   0.259&        \\
J1331+3030 &13 31 08.3&  30 30 33  & QQQ&  286  & 0.846  & 2.34&$-$0.633&        \\
J1336$-$3358&13 36 37.3&$-$33 58 01& GG &       & 0.0129 & 0.99&$-$0.499&  I/II  \\
J1337$-$1257&13 37 39.8&$-$12 57 24& QQQ&       & 0.539  & 5.71&   0.134&  IDV   \\
J1354$-$1041&13 54 46.5&$-$10 41 03& QQQ&       & 0.332  & 1.22&   0.362&        \\
J1357+1919 &13 57 04.4&  19 19 08  & QQQ&       & 0.720  & 0.89&$-$0.230&  II    \\
J1408$-$0752&14 08 56.4&$-$07 52 26& QQQ&       & 1.494  & 0.95&$-$0.075&  IDV   \\
J1419+3822 &14 19 46.4&  38 21 48  & Q  &       &        & 1.05&   0.171&        \\
J1427$-$3306&14 27 38.0&$-$33 05 00& G  &       &        & 1.23&   0.776&        \\
J1427$-$4206&14 27 56.2&$-$42 06 17& QQQ&       & 1.522  & 1.71&$-$0.191&        \\
J1459+7140 &14 59 07.6&  71 40 20  & QQ &  309.1& 0.905  & 1.29&$-$0.589&  II    \\
J1504+1029 &15 04 24.9&  10 29 39  & QQQ&       & 0.563  & 1.14&$-$0.323&        \\
J1512$-$0906&15 12 50.4&$-$09 06 00& QQQ&       & 0.361  & 2.42&   0.063&        \\
J1516+0015 &15 16 40.2&  00 15 11  & AGG&       & 0.0518 & 1.19&$-$0.166&  II    \\
J1517$-$2422&15 17 41.5&$-$24 22 23& BBB&       & 0.0486 & 2.30&$-$0.100&        \\
J1549+0237 &15 49 28.4&  02 36 49  & QQQ&       & 0.412  & 1.99&   0.106&        \\
J1550+0527 &15 50 35.2&  05 27 10  & QQQ&       & 1.422  & 1.66&$-$0.111&        \\
J1608+1029 &16 08 45.0&  10 26 41  & Q  &       &        & 1.09&$-$0.085&        \\
J1613+3412 &16 13 41.3&  34 12 48  & QQQ&       & 1.401  & 2.50&$-$0.084&        \\
J1617$-$7717&16 17 46.2&$-$77 17 19& QQ &       & 1.710  & 1.85&$-$0.387&  IDV   \\
J1632+8232 &16 32 31. &  82 32 17  & GGG&       & 0.0243 & 0.92&$-$0.149&  I/II  \\
J1635+3808 &16 35 15.5&  38 08 05  & QQQ&       & 1.814  & 2.07&$-$0.103&  II    \\
J1638+5720 &16 38 13.4&  57 20 24  & QQQ&       & 0.750  & 1.64&   0.188&        \\
J1642+6856 &16 42 07.8&  68 56 49  & AQQ&       & 0.751  & 9.28&$-$0.127&  II IDV \\
J1642+3948 &16 42 58.8&  39 48 37  & QQQ&  345.0& 0.594  & 1.07&$-$0.122&  II    \\
J1651+0459 &16 51 08.7&  04 59 30  & GGG&  348.0& 0.154  & 1.55&$-$1.130&  I/II  \\
J1653+3945 &16 53 52.2&  39 45 36  & BBB&       & 0.0337 & 1.09&$-$0.213&  II    \\
J1658+0741 &16 58 09.0&  07 41 27  & QQQ&       & 0.621  & 1.36&$-$0.047&        \\
J1703$-$6212&17 03 36.6&$-$62 12 40& G  &       &        & 1.68&   0.706&        \\
J1723$-$6500&17 23 41.0&$-$65 00 36& AG &       & 0.0145 & 2.03&$-$0.463&  IDV   \\
J1727+4530 &17 27 27.6&  45 30 40  &Q~~Q&       & 0.714  & 1.38&   0.353&        \\
J1734+3857 &17 34 20.6&  38 57 51  & QQQ&       & 0.976  & 0.96&   0.488&        \\
J1733$-$7935&17 33 40.7&$-$79 35 55& G  &       &        & 1.21&   0.284&        \\
J1740+5211 &17 40 36.9&  52 11 44  & QQQ&       & 1.379  & 1.84&   0.013&        \\
J1748+7005 &17 48 32.6&  70 05 51  & BBB&       & 0.770  & 0.75&$-$0.130&  I IDV \\
J1753+2847 &17 53 42.4&  28 48 05  & Q  &       &        & 0.87&   0.161&        \\
J1758+6638 &17 58 33.4&  66 38 01  & G  &       &        & 0.62&$-$0.311&        \\
J1800+7828 &18 00 45.7&  78 28 04  & QBQ&       & 0.684  & 2.26&$-$0.209&  I IDV \\
J1803$-$6507&18 03 22.8&$-$65 07 33&   G&       &        & 1.23&   0.353&  IDV   \\
J1806+6949 &18 06 50.7&  69 49 28  & BBB&  371.0& 0.051  & 1.62&$-$0.166&  II IDV\\
J1819-6345 &18 19 34.3&$-$63 45 59 & AG &       & 0.0627 & 1.13&$-$0.743&        \\
J1824+5650 &18 24 07.1&  56 51 01  & QBQ&       & 0.664  & 2.06&   0.326&  I     \\
J1829+4844 &18 29 31.7&  48 44 47  & QQ &  380.0& 0.691  & 1.92&$-$0.517&  II    \\
J1837$-$7108&18 37 28.7&$-$71 08 43& QQ &       & 1.356  & 1.90&$-$0.008&        \\
J1842+7946 &18 42 17.1&  79 45 56  & QGG&  390.3& 0.0569 & 1.13&   0.158&  II    \\
J1842+6809 &18 42 33.6&  68 09 25  & Q  &       &        & 0.95&$-$0.831&        \\
J1849+6705 &18 49 16.0&  67 05 41  & Q  &       &        & 0.98&   0.208&        \\
J1850+2825 &18 50 27.7&  28 25 09  & Q  &       &        & 1.15&   0.050&        \\
J1902+3159 &19 02 55.9&  31 59 42  & Q~Q&       & 0.635  & 1.07&$-$0.322&        \\
J1923$-$2104&19 23 32.2&$-$21 04 33& ?  &       &        & 1.97&$-$0.363&        \\
J1927+6117 &19 27 30.0&  61 17 34  & B  &       &        & 0.84&   0.161&  IDV   \\
J1927+7357 &19 27 48.5&  73 58 01  & QQQ&       & 0.302  & 2.77&$-$0.332&  I/II  \\
J1955+5131 &19 55 42.6&  51 31 52  & QQQ&       & 1.230  & 1.20&$-$0.052&  I/II  \\
J1957$-$3845&19 58 00.0&$-$38 45 06& QQQ&       & 0.626  & 1.12&$-$0.424&        \\
J2000$-$1748&20 00 57.0&$-$17 48 58& QQQ&       & 0.650  & 1.86&   0.207&  IDV   \\
J2011$-$1546&20 11 15.7&$-$15 46 40& QQQ&       & 1.180  & 1.55&   0.077&        \\
J2022+6137 &20 22 06.7&  61 36 59  & AGG&       & 0.2266 & 1.98&$-$0.256&  IDV   \\
J2024+1718 &20 24 56.6&  17 18 13  & Q  &       &        & 0.73&   0.045&        \\
J2035$-$6846&20 35 48.8&$-$68 46 34& Q  &       &        & 1.03&   0.364&        \\
J2038$-$13  &20 38 36.4&$-$13 11 44& ?  &       &        & 0.80&   0.885&        \\
J2056$-$4714&20 56 16.4&$-$47 14 48& QQ &       & 1.491  & 1.21&$-$0.271&        \\
J2109+3532 &21 09 31.7&  35 32 40  & ?  &       &        & 1.13&$-$0.115&        \\
J2109$-$4110&21 09 33.2&$-$41 10 20& QQQ&       & 1.0547 & 1.91&$-$0.022&  IDV   \\
J2123+0535 &21 23 44.5&  05 35 22  &Q~~Q&       & 1.941  & 1.08&$-$0.332&  IDV   \\
J2131$-$1207&21 31 35.1&$-$12 07 04& QQQ&       & 0.501  & 2.33&$-$0.148&  IDV   \\
J2134$-$0153&21 34 08.7&$-$01 53 13& QBQ&       & 0.557  & 1.35&$-$0.168&        \\
J2136+0041 &21 36 38.6&  00 41 54  & QQQ&       & 1.936  & 3.73&$-$0.749&  IDV   \\
J2139+1423 &21 39 01.3&  14 23 35  & QQQ&       & 2.427  & 1.65&$-$0.034&        \\
J2143+1743 &21 43 35.6&  17 43 49  & Q  &       &        & 0.99&   0.201&        \\
J2148+0657 &21 48 05.5&  06 57 38  & QQQ&       & 0.990  & 1.35&   0.501&        \\
J2146$-$7755&21 46 30.0&$-$77 55 51& G  &       &        & 8.13&   0.016&        \\
J2157$-$6941&21 57 06.0&$-$69 41 24& AG &       & 0.0285 & 3.05&$-$0.750&  I     \\
J2158$-$1501&21 58 06.2&$-$15 01 09& QQQ&       & 0.672  & 1.44&$-$0.279&        \\
J2202+4216 &22 02 43.2&  42 16 40  & BBB&       & 0.0688 & 3.26&$-$0.113&  I     \\
J2203+3145 &22 03 14.6&  31 45 56  & QQQ&       & 0.298  & 1.90&   0.551&  II    \\
J2203+1725 &22 03 26.8&  17 25 48  & Q  &       &        & 2.30&$-$0.097&        \\
J2206$-$1835&22 06 10.3&$-$18 35 39& QQQ&       & 0.619  & 1.57&$-$0.582&        \\
J2212+2355 &22 12 06.0&  23 55 41  &Q~~Q&       &        & 1.24&   0.160&        \\
J2218$-$0335&22 18 51.8&$-$03 35 00& QQQ&       & 0.901  & 1.23&$-$0.154&        \\
J2220+43   &22 20 09.5&  43 35 27  & ?  &       &        & 0.73&   0.840&        \\
J2225$-$0457&22 25 47.2&$-$04 57 01& QQQ&  446.0& 1.404  & 4.06&$-$0.067&        \\
J2229$-$0832&22 29 39.9&$-$08 32 55& QQQ&       & 1.561  & 1.14&$-$0.094&        \\
J2232+1143 &22 32 36.4&  11 43 50  & QQQ&       & 1.037  & 2.56&$-$0.366&  II    \\
J2235$-$4835&22 35 13.2&$-$48 35 59& Q  &       &        & 1.89&   0.414&        \\
J2236+2828 &22 36 22.4&  28 28 57  & QQQ&       & 0.795  & 1.12&$-$0.307&        \\
J2239$-$5701&22 39 12.0&$-$57 01 01& Q  &       &        & 1.10&   0.096&        \\
J2246$-$1206&22 46 17.9&$-$12 06 52& QQQ&       & 0.630  & 2.48&$-$0.067&  IDV   \\
J2253+1608 &22 53 57.7&  16 08 53  & QQQ&  454.3& 0.859  & 9.64&$-$0.082&        \\
J2256$-$2011&22 56 41.2&$-$20 11 41& B  &       &        & 0.90&   0.346&        \\
J2258$-$2758&22 58 05.8&$-$27 58 17& QQQ&       & 0.926  & 3.67&   0.022&        \\
J2315$-$5018&23 15 44.3&$-$50 18 40& G  &       &        & 1.18&   0.497&        \\
J2331$-$1556&23 31 38.7&$-$15 56 57& QQQ&       & 1.155  & 0.91&$-$0.060&        \\
J2336$-$5236&23 36 12. &$-$52 36 21& QE &       &        & 0.83&$-$0.374&  IDV   \\
J2348$-$1631&23 48 02.0&$-$16 31 12& QQQ&       & 0.576  & 2.23&$-$0.131&  IDV   \\
J2354+4553 &23 54 21.5&  45 53 03  & QQQ&       & 1.992  & 0.98&$-$0.181&  II    \\
J2356+4952 &23 55 09.5&  49 50 08  & AGG&       & 0.237  & 0.49&$-$0.671&        \\
J2357$-$5311&23 57 53.2&$-$53 11 14& QQ &       & 1.006  & 1.32&$-$0.093&  IDV   \\
J2358$-$6054&23 58 48.2&$-$60 54 20& AG &       & 0.0958 & 1.77&$-$0.566&  II    \\
\end{supertabular}
\end{onecolumn}
\noindent
The abbreviations denote the following:
G --- galaxy, A --- AGN, Q --- quasar, B --- BL Lacs, P --- a planetary
nebula. The same column defines the availability of identified sources in each of
these catalogs. The table shows that the type of the objects is sometimes
determined ambiguously.

The table gives the names of identifications from 3C catalog, the measured
redshifts (the data from VV10, Stickel94, VLBApls).  Thus, it is clear that
many WMAP sources had been already mapped with the VLBA system in the
program of the preliminary survey of 376 bright VSOP sources
(Hirabayashi et al. 2001; Fomalont et al. 2000). As it is known, the VSOP
survey is being continued at the present time and its processing has not been
completed yet. The next columns  present the flux at 30\,GHz and
the spectral index at the frequency 20\,GHz approximated from the spectra.
The last column give the object type from the
Fanaroff \& Riley (1974) classification from the Stickel94 survey and IDV sources are
also marked.

\section{Observations and reduction}

Pilot observations of 25 sources were carried out within the frames of the
program of investigation of variable sources in long cycles of monitoring
of flare activity in March 2003 (and J2105+3532 was observed in May).
The in-service complex of continuous spectrum radiometers was used.
All the detectors at frequencies 3.9, 7.7, 11.2 and 21.7 GHz were equipped
with closed loop cryogenic systems lowering the temperature of the first
cascades of amplification to 15-20 K.
Low-noise transistor (HEMT) amplifiers were installed
in the radiometer at the frequency  2.3~GHz.

Table\,\ref{r1} presents the realized sensitivity
at different frequencies $\Delta$S of the telescope RATAN-600
(the level $1\sigma$ with optimum smoothing of drift-scans) in one
observation of a source, passing across the immovable beam pattern.

The observations were made on the South Sector (SS) and on the South
Sector with the Flat reflector (SSF) of RATAN-600 in 2003 March daily
in the upper culmination for (SSF) and in the lower culmination for (SS).
In one observation the flux from the source was recorded at all five
frequencies. It should be noted that the telescope resolution was quite
sufficient for reliable detection of sources and measurement of fluxes
from relatively small-extended sources.

Although, not always the sensitivity of the radiometers was realized in the
observations because of interference, we consider, based on daily
observations of calibration sources, that the flux measurement error was
better than 5\,\% at the frequencies 2.3, 3.9 and 11.2 GHz and
6--10\,\% at 21.7 GHz.

\begin{table}
\caption[l]{Flux density sensitivity of RATAN-600}
\begin{tabular}{l|l|l|l|l|l}
\noalign{\smallskip}
\hline
  Wavelength (cm)   &13.0  & ~7.6   &  ~3.9 &  ~2.7 & 1.38 \\
\hline
  Frequency (GHz)      & ~2.3 & ~3.9   & ~7.7  &  11.2 & 21.7  \\
\hline
  $\Delta$S(mJy)(SS) & 10   &  ~~6   &  10  &   10  & 20    \\
  $\Delta$S(mJy)(SSF)& 15   &  ~10   &  20  &   15  & 30    \\
\hline
\end{tabular}
\label{r1}
\end{table}

The flux calibration was performed using daily observations of the sources
NGC\,7027, DR\,21, 1245--19, 1850--001 and 1345+12.
For additional check, the best secondary calibrator 3C\,286 was observed.
The fluxes for them were determined earlier and brought into consistence with the
basic radio astronomical scale of the fluxes of the reference sources of
Baars et al. (1977) and with the renewed flux measurements from the paper
by Ott et al. (1994). The adopted fluxes of the reference sources for the
given set of observations are presented in Table\,\ref{r2}.

\begin{table}
\begin{center}
\caption {Flux densities of calibration sources (Jy)}
\label{r2}
\begin{tabular}{l|r|r|r|r|r}
\noalign{\smallskip}
\hline
Source       & \multicolumn{5}{|c}{Frequency (GHz)} \\
\cline{2-6}
     name    & 2.3~  &    3.9~   &    7.7~   &  11.2~ &  21.7~\\
\hline
 1245-199    &   4.29  &   2.99   &    1.76  &    1.23 &   0.70\\
 1328+30     &   11.57 &   8.57   &    5.53  &    4.27 &   2.57\\
 1345+12     &    4.40 &   3.32   &    2.22  &    1.75 &   1.12\\
 1850-001    &    2.27 &   3.04   &    3.88  &    4.20 &   4.50\\
 2037+42     &   12.70 &   17.3   &   20.7   &   21.1~ &  19.40\\
 2105+42     &   3.04  &    4.9   &    6.33  &   6.03  &   5.90 \\
\hline
\end{tabular}
\end{center}
\end{table}

The recording of data, preliminary reduction and writing on the  computer HDD
were made  with the package of acquisition programs created by
P.\,Tsybulev.
The reduction of the obtained records of observations and reference sources
was done with the aid of standard programs by the methods from the paper
by Aliakberov et al. (1985). The current data-reduction includes the
procedures of subtraction of the background, convolution with the beam
pattern and Gauss analysis.

Thus we made about 350 flux measurements of the selected sources.
The data of these simultaneous measurements are fitted by a flat,
inverse or power spectra. Generally, these data are consistent with the
former measurements, although it is evident that  numerous WMAP sources
have variable radio emission.

\section{Analysis of spectra and discussion}

The spectra of 208 WMAP sources are shown in Appendix at the end of the
paper. For the sources
WMAPJ0518--05, WMAPJ2038--13 and WMAPJ2220+43 we have found only possible
identifications in the catalogs  NVSS and GB6. This is why their radio
spectra are given only for the picture to be complete and they should be
treated with care
\footnote{ {\it Added in proof:} We have observed the fields of
NVSS\,J203809$-$131904 and NVSS\,J222120+433514
at the end of May with SSF and detected the sources with fluxes $135\pm30$\,mJy and
$120\pm25$\,mJy, respectively, which is consistent with their spectra.}.

\begin{figure}[t]
\centerline{\vbox{\psfig{figure=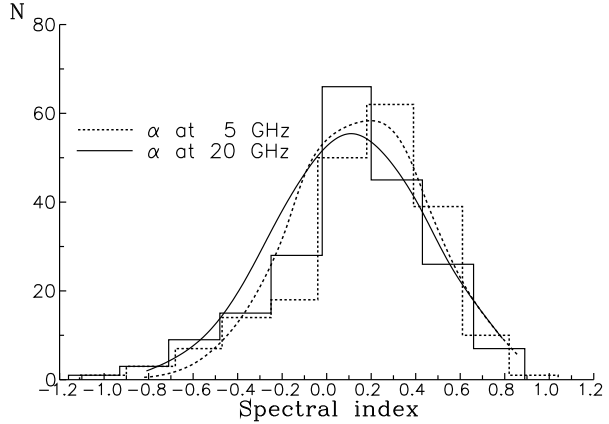,width=8.cm,angle=-90}}}
\caption{Spectral index distribution of the WMAP sources in the ranges
 5 GHz and 20 GHz.}
\label{al}
\end{figure}

In the procedure of spectrum plotting the spectral indices at two
frequencies, 5 and 20 GHz ($\alpha_5$, $\alpha_{20}$),
if the spectrum changes the curvature, or one spectral index for the linear
(power) spectrum are given. We have analyzed these data and obtained
the distribution of the spectral indices throughout the whole sample.
Fig.\,\ref{al} shows the distribution of both indices. They are quite
similar, although  $\alpha_5$ is significantly shifted to the right with
respect to the distribution of  $\alpha_{20}$. When approximating these
distributions by Gaussians, then $\alpha_{5}$ has a maximum near +0.25,
while $\alpha_{20}$ near +0.2. This can be accounted for by the presence
of sources with GHz peak (GPS-sources) in the sample, for which this flux
peak falls at a frequency higher than 5\,GHz. On the other hand, similar
sources will not show a rightward displacement in the distribution of
$\alpha_{20}$, if the flux peaks fall at the frequencies lower than 20\,GHz.
The spectra of numerous WMAP sources are difficult to approximate by a
common simple relation (linear, linear with the addition of the exponential
function, or parabolic) in the total range of frequencies from
10--100 MHz to 100--240 GHz. Probably, it is due to the fact that different
radiating components, extensive halo, extended double components and jets,
make a contribution to the integral spectrum. For this reason, measurements
with antennae with different angular resolution may give discrepant
results. In our case the resolution was the lowest at low frequencies
(for instance, the UTR survey, Braude et al. 2002). This is especially
evident from recent maps of the radio galaxy Virgo\,A at a low frequency,
74 MHz (VLA), where an extended radio structure is present around the
bright and more compact structure near the optical galaxy with a radio and
optical jet.

\begin{onecolumn}
\hspace{16cm}
\begin{center}
\topcaption{WMAP flux density measurements with RATAN-600}
\tablefirsthead{%
\hline
 Source      &Frequency& Flux &  Error  &  Source       &Frequency& Flux & Error      \\
name & ~GHz~ & ~Jy~  &  ~Jy~    &  name & ~GHz~ & ~Jy~  &  ~Jy~        \\
\hline
}
\tablehead{
\hline
 Source & Frequency& Flux &  Error  &   Source      &Frequency&Flux &Error      \\
 name & ~GHz~ &~Jy~  & ~Jy~    &  name& ~GHz~ & ~Jy~  &  ~Jy~\\
\hline
}
\tabletail{\hline}
\begin{supertabular}{|l|r|l|l||l|r|l|l|}
J0329$-$2357&  2300 &  0.600 & 0.08 &  J0816$-$2421& 11200 &  1.05  &  0.10  \\
	    &  3900 &  0.555 & 0.05 &              & 21700 &  0.91  &  0.10  \\
	    &  7700 &  0.960 & 0.07 &  J0830+2410  &  2300 &  0.67  &  0.05  \\
J0403$-$3605&  2300 &  0.922 & 0.1  &              &  3900 &  1.08  &  0.05  \\
	    &  3900 &  1.080 & 0.1  &              &  7700 &  1.09  &  0.08  \\
	    &  7700 &  2.213 & 0.15 &              & 11200 &  1.43  &  0.10  \\
	    & 11200 &  3.718 & 0.20 &              & 21700 &  1.57  &  0.10  \\
	    & 21700 &  4.435 & 0.20 &  J1014+2301  &  2300 &  0.85  &  0.05  \\
J0406$-$3826&  2300 &  0.843 & 0.10 &              &  3900 &  0.975 &  0.05  \\
	    &  3900 &  1.184 & 0.10 &              &  7700 &  0.9   &  0.08  \\
	    &  7700 &  1.390 & 0.10 &              & 11200 &  1.140 &  0.10  \\
	    & 11200 &  1.490 & 0.10 &              & 21700 &  0.980 &  0.10  \\
	    & 21700 &  1.309 & 0.10 &  J1048+7143  &  2300 &  1.45  &  0.10  \\
J0424$-$3756&  2300 &  0.966 & 0.1  &              &  3900 &  1.43  &  0.10  \\
	    &  3900 &  1.294 & 0.1  &              &  7700 &  1.13  &  0.10  \\
	    &  7700 &  1.694 & 0.15 &              & 11200 &  1.24  &  0.10  \\
	    & 11200 &  2.222 & 0.2  &              & 21700 &  0.91  &  0.10  \\
	    & 21700 &  2.072 & 0.2  &  J1127$-$1857&  2300 &  0.900 &  0.05  \\
J0527$-$1241&  2300 &  1.805 & 0.05 &              &  3900 &  1.160 &  0.05  \\
	    &  3900 &  1.791 & 0.05 &              &  7700 &  1.200 &  0.08  \\
	    &  7700 &  1.665 & 0.05 &              & 11200 &  1.660 &  0.06  \\
	    & 11200 &  1.569 & 0.05 &              & 21700 &  1.440 &  0.15  \\
	    & 21700 &  1.354 & 0.10 &  J1153+8058  &  2300 &  1.82  &  0.1   \\
J0607+6720  &  2300 &  0.88  & 0.05 &              &  3900 &  1.72  &  0.1   \\
	    &  3900 &  0.93  & 0.05 &              &  7700 &  1.55  &  0.1   \\
	    &  7700 &  0.87  & 0.05 &              & 11200 &  1.52  &  0.1   \\
	    & 11200 &  0.83  & 0.05 &  J1632+8232  &  3900 &  1.42  &  0.04  \\
	    & 21700 &  0.95  & 0.05 &              &  7700 &  1.45  &  0.04  \\
J0629$-$1959&  2300 &  0.819 & 0.1  &              & 11200 &  1.54  &  0.04  \\
	    &  3900 &  1.131 & 0.1  &              & 21700 &  1.44  &  0.04  \\
	    &  7700 &  1.254 & 0.1  &  J1800+7828  &  2300 &  2.32  &  0.1   \\
	    & 11200 &  1.619 & 0.1  &              &  3900 &  2.37  &  0.12  \\
	    & 21700 &  1.589 & 0.1  &              &  7700 &  2.16  &  0.12  \\
J0633$-$2223&  2300 &  0.732 & 0.073&              & 11200 &  2.15  &  0.12  \\
	    &  3900 &  0.778 & 0.078&              & 21700 &  1.72  &  0.15  \\
	    &  7700 &  0.750 & 0.075&  J1824+5650  &  2300 &  1.32  &  0.1   \\
	    & 11200 &  0.672 & 0.067&              &  3900 &  1.36  &  0.1   \\
	    & 21700 &  0.574 & 0.06 &              &  7700 &  1.55  &  0.1   \\
J0741+3112  &  2300 &  2.600 & 0.26 &              & 11200 &  1.76  &  0.15  \\
	    &  3900 &  2.27  & 0.22 &              & 21700 &  2.35  &  0.15  \\
	    &  7700 &  1.81  & 0.18 &  J1927+7357  &  2300 &  3.60  &  0.25  \\
	    & 11200 &  1.52  & 0.15 &              &  3900 &  3.155 &  0.25  \\
	    & 21700 &  1.55  & 0.15 &              &  7700 &  3.555 &  0.25  \\
J0745+1011  &  2300 &  4.210 & 0.1  &              & 11200 &  3.850 &  0.3   \\
	    &  3900 &  3.834 & 0.1  &              & 21700 &  4.290 &  0.3   \\
	    &  7700 &  2.775 & 0.1  &  J2022+6137  &  2300 &  2.220 &  0.15  \\
	    & 11200 &  2.367 & 0.1  &              &  3900 &  2.690 &  0.15  \\
	    & 21700 &  1.304 & 0.1  &              &  7700 &  3.120 &  0.20  \\
J0750+1231  &  2300 &  1.175 & 0.11 &              & 11200 &  2.810 &  0.20  \\
	    &  3900 &  1.32  & 0.13 &              & 21700 &  2.540 &  0.20  \\
	    &  7700 &  1.80  & 0.18 &  J2109+3532  &  2300 &  1.56  &  0.10  \\
	    & 11200 &  2.44  & 0.23 &              &  3900 &  1.285 &  0.10  \\
	    & 21700 &  2.350 & 0.28 &              &  7700 &  1.315 &  0.10  \\
J0808$-$0751&  2300 &  1.955 & 0.18 &              & 11200 &  1.190 &  0.10  \\
	    &  3900 &  2.120 & 0.20 &              & 21700 &  1.060 &  0.15  \\
	    &  7700 &  1.700 & 0.17 &  J2355+4950  &  2300 &  2.05  &  0.10  \\
	    & 11200 &  2.105 & 0.20 &              &  3900 &  1.63  &  0.10  \\
	    & 21700 &  2.020 & 0.20 &              &  7700 &  1.06  &  0.10  \\
J0816$-$2421&  2300 &  0.27  &  0.05&              & 11200 &  0.80  &  0.10  \\
	    &  3900 &  0.64  &  0.06&              & 21700 &  0.58  &  0.10  \\
	    &  7700 &  0.76  &  0.06&              &       &        &        \\
\end{supertabular}
\end{center}
\end{onecolumn}

\begin{figure*}
\centerline{
 \vbox{
  \hbox{\psfig{figure=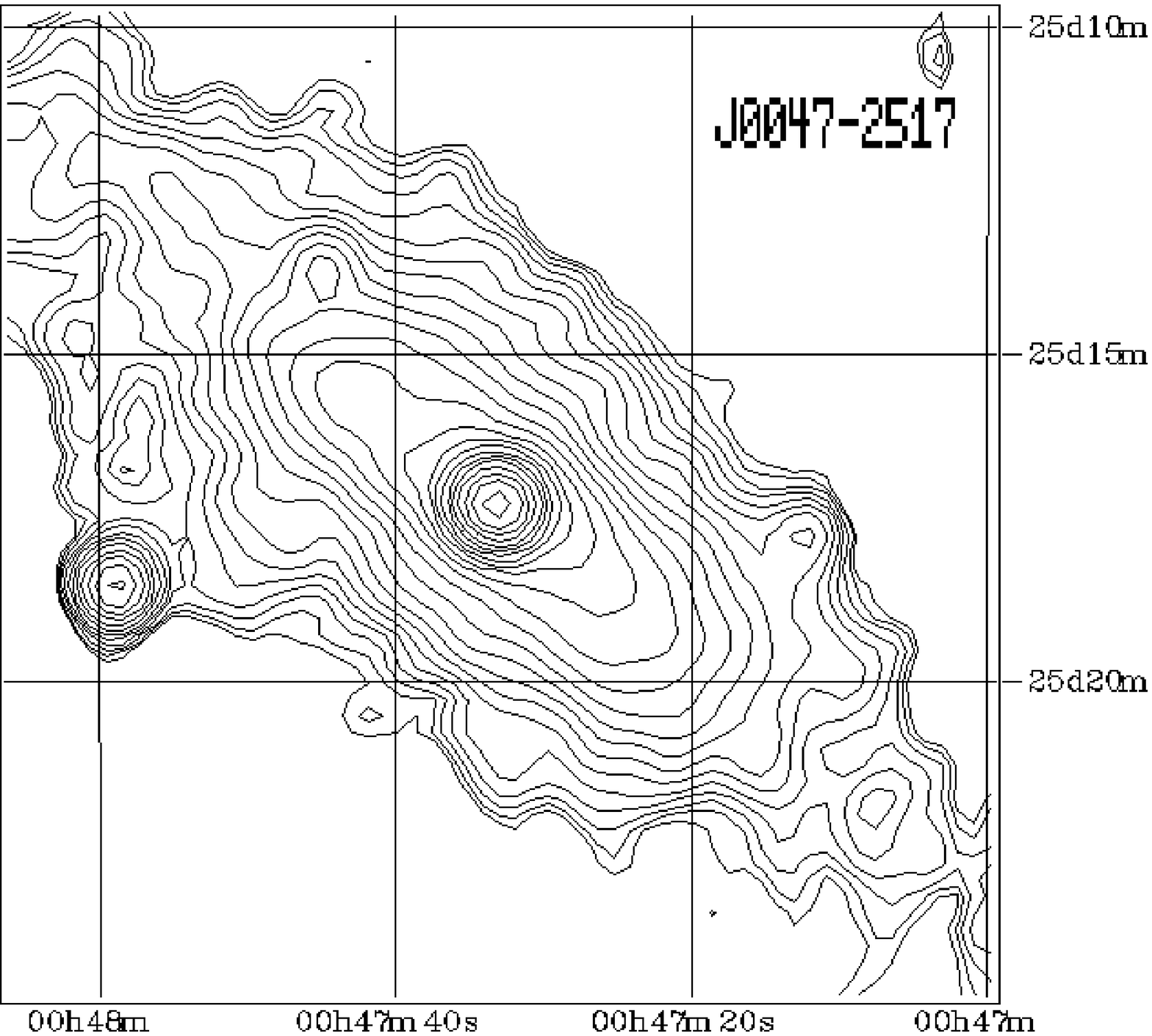,width=7.5cm,angle=0}
	\hspace{0.5cm}
	\psfig{figure=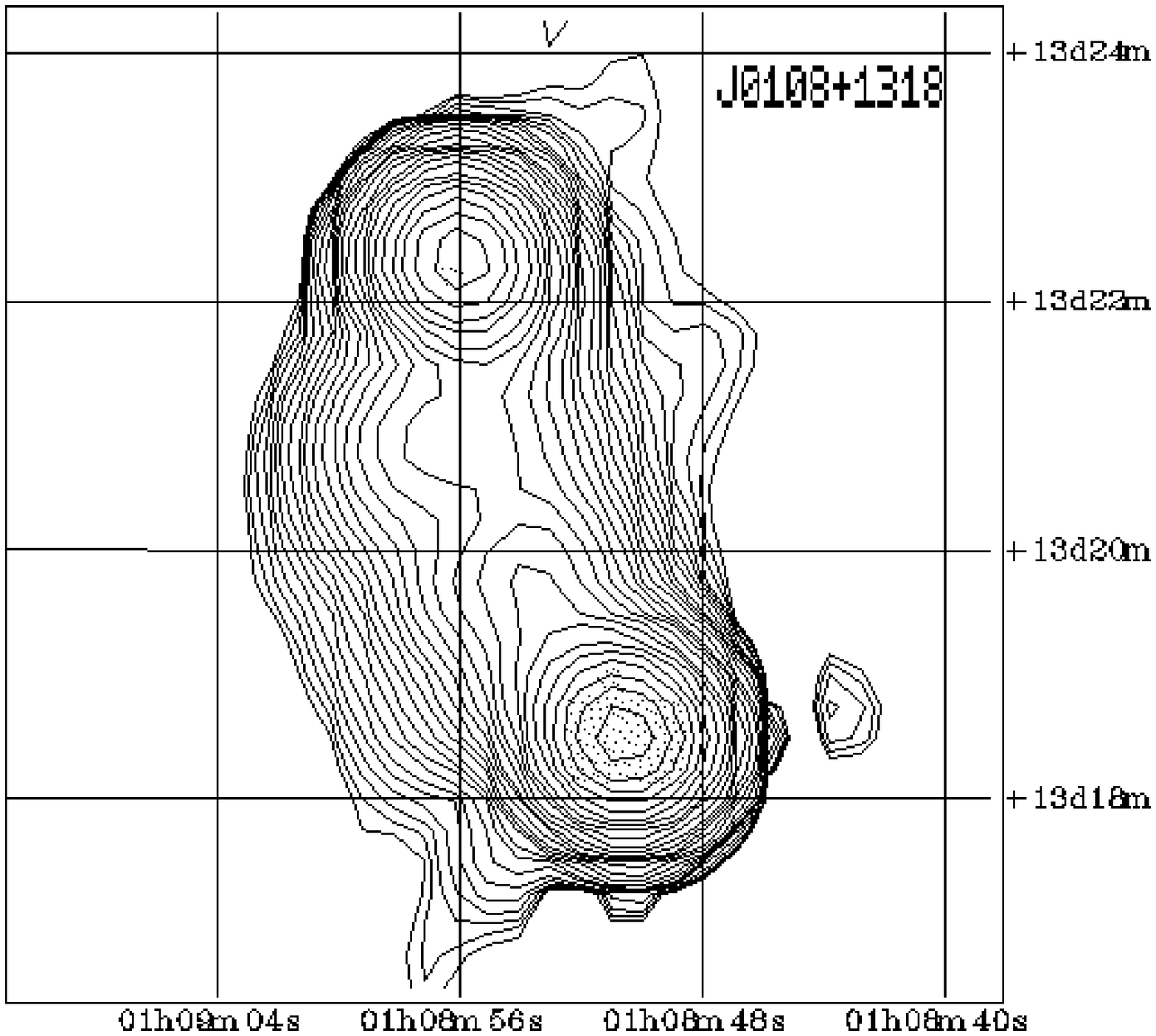,width=8.0cm,angle=0}
  }
 \vspace{0.5cm}
  \hbox{\psfig{figure=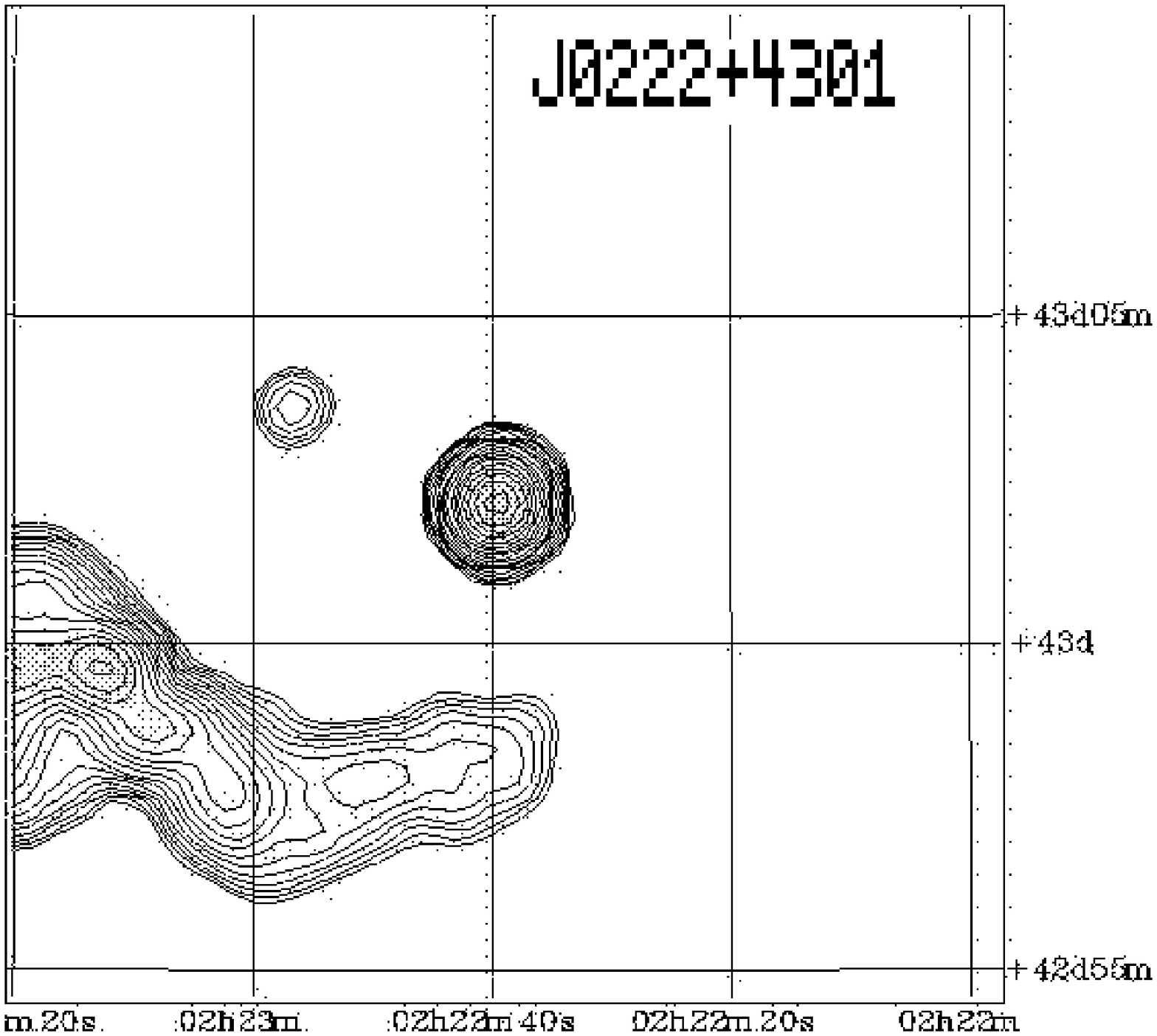,width=8.0cm,angle=0}
	\hspace{0.5cm}
	\psfig{figure=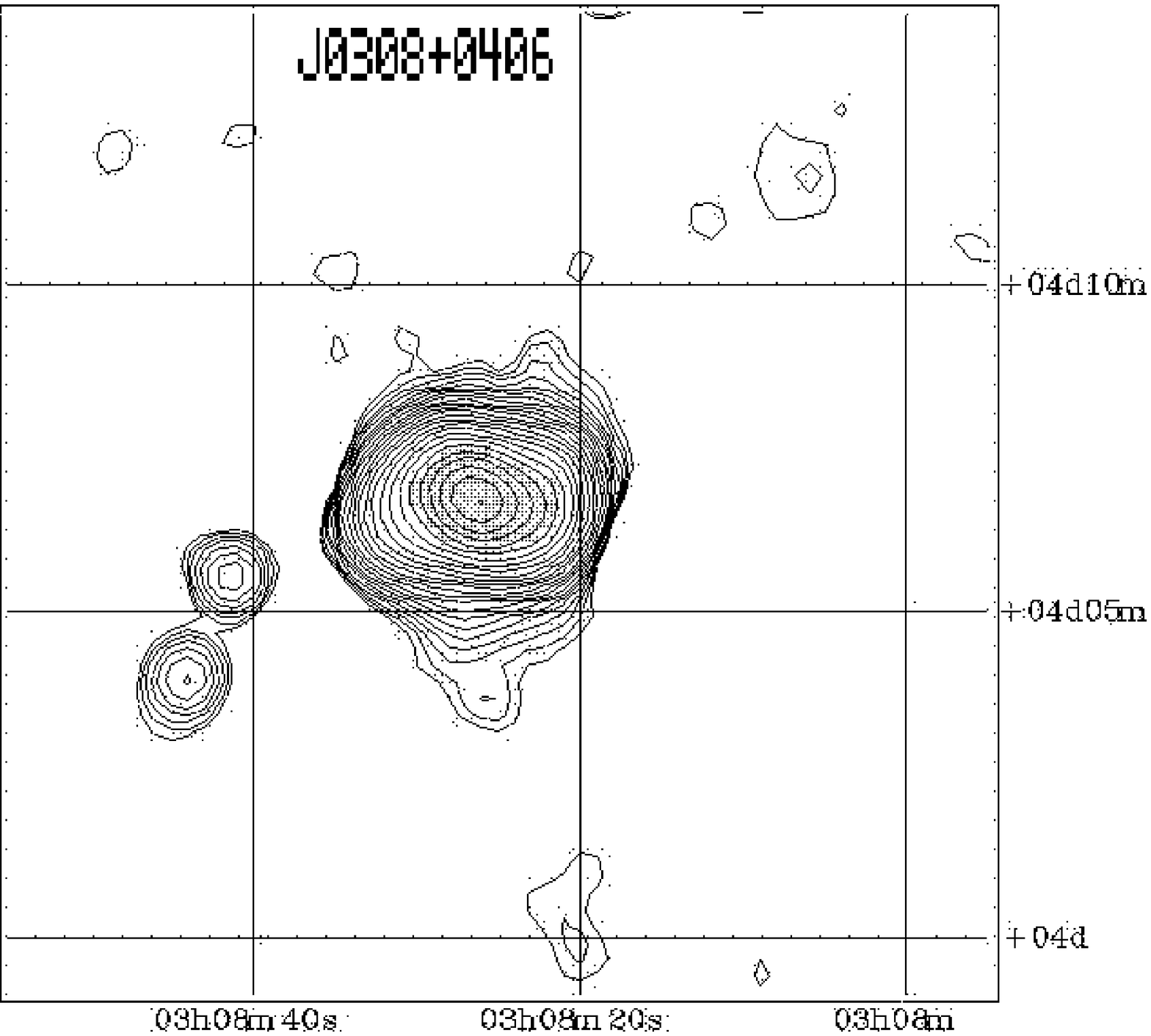,width=8.0cm,angle=0}
  }
 \vspace{0.5cm}
  \hbox{\psfig{figure=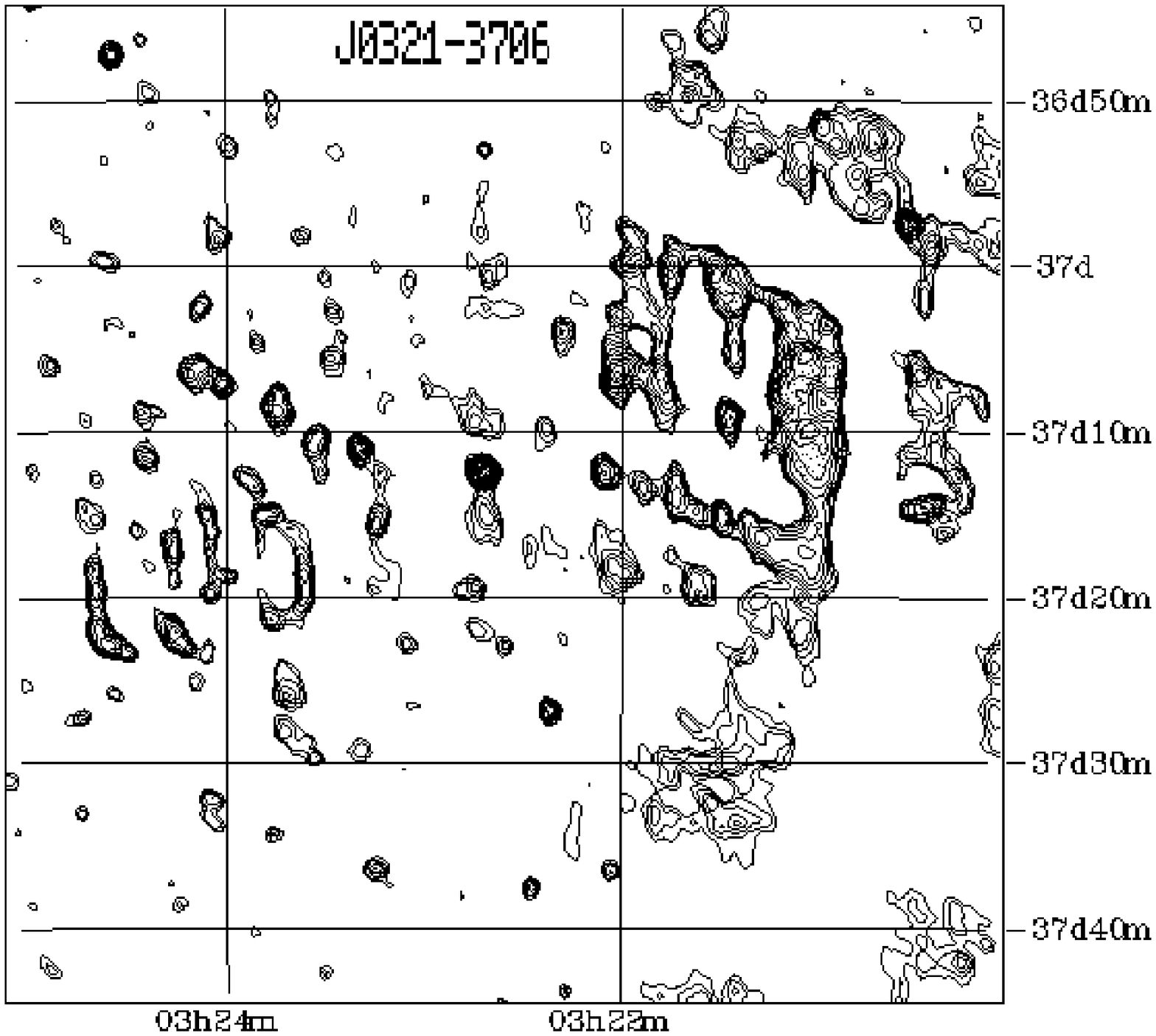,width=8.0cm,angle=0}
	\hspace{0.5cm}
	\psfig{figure=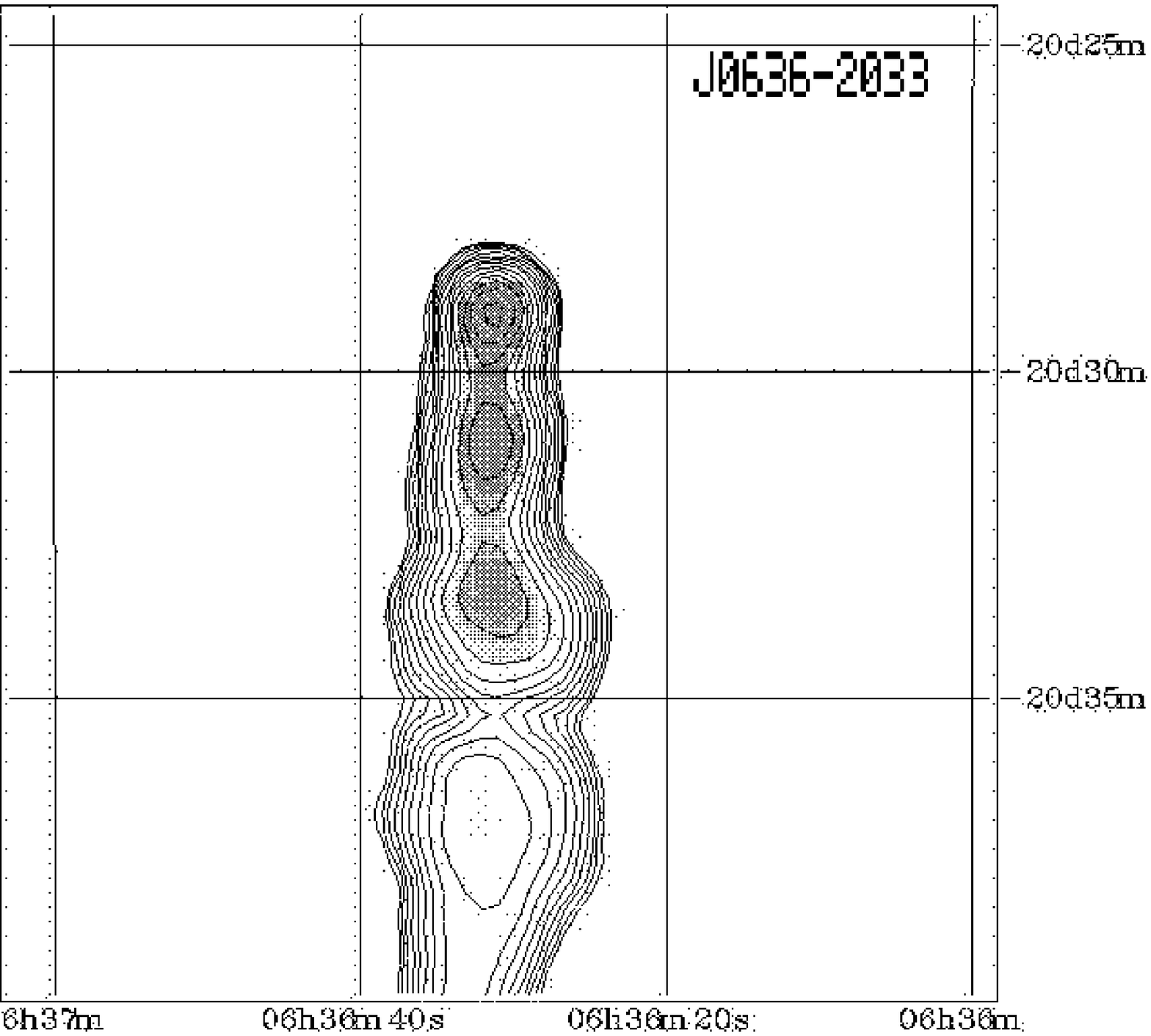,width=8.0cm,angle=0}
  }
 }
}
\caption{ a) NVSS maps of the extended WMAP sources at 1400 MHz.}
\label{m1}
\end{figure*}

\setcounter{figure}{1}
\begin{figure*}
\centerline{
 \vbox{
  \hbox{\psfig{figure=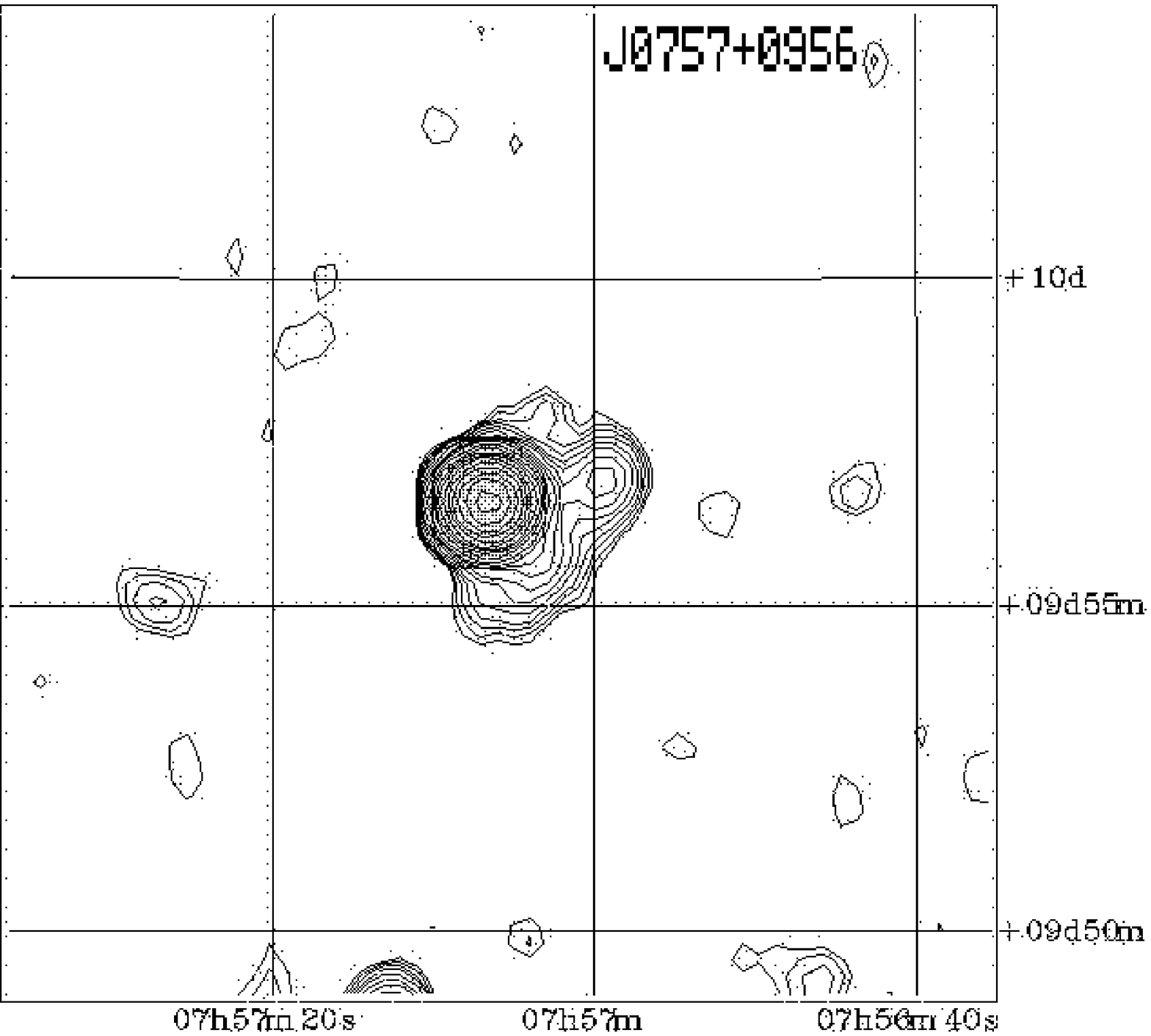,width=8.0cm,angle=0}
       \hspace{0.5cm}
       \psfig{figure=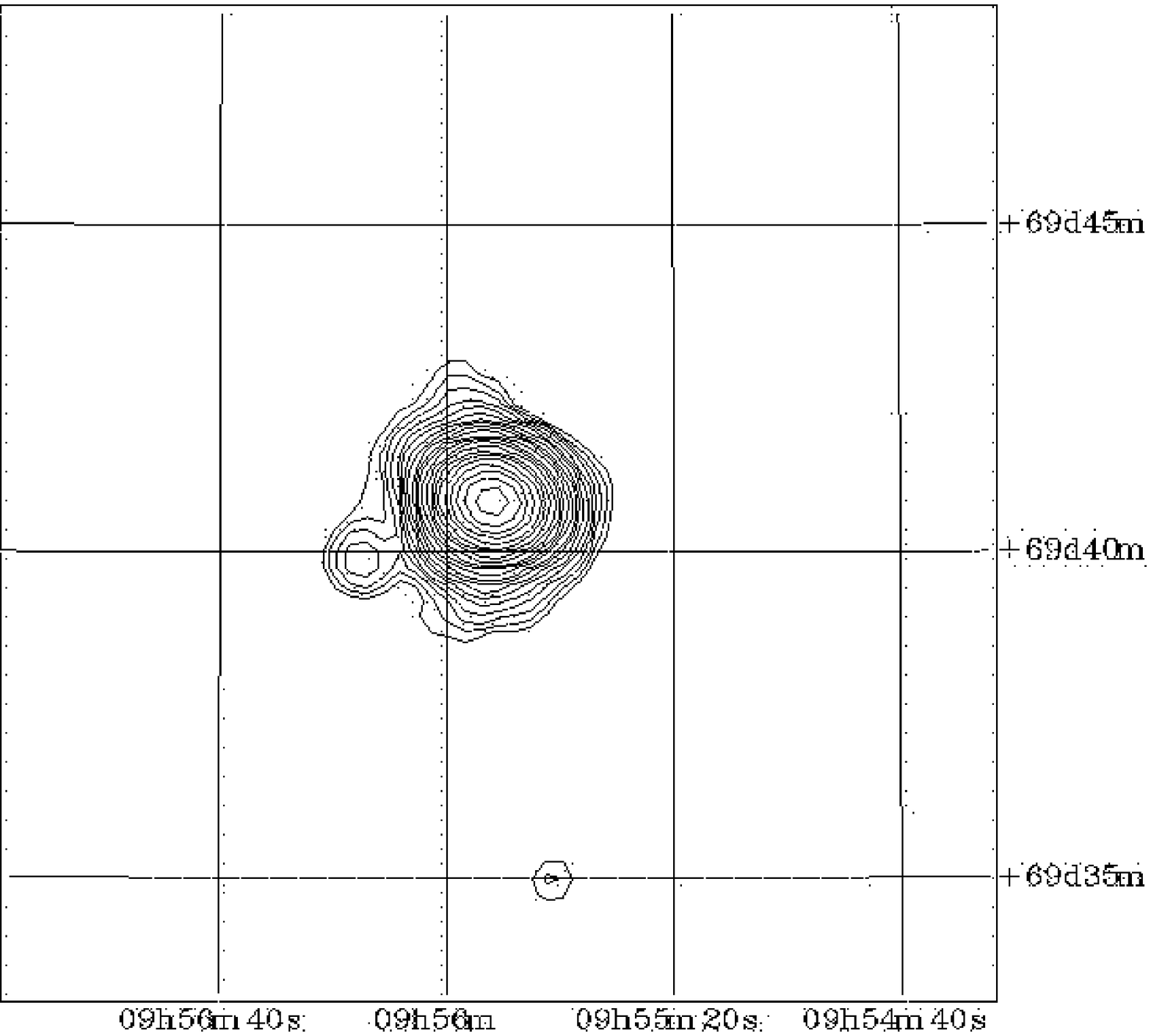,width=8.0cm,angle=0}
 }
 \vspace{0.5cm}
 \hbox{\psfig{figure=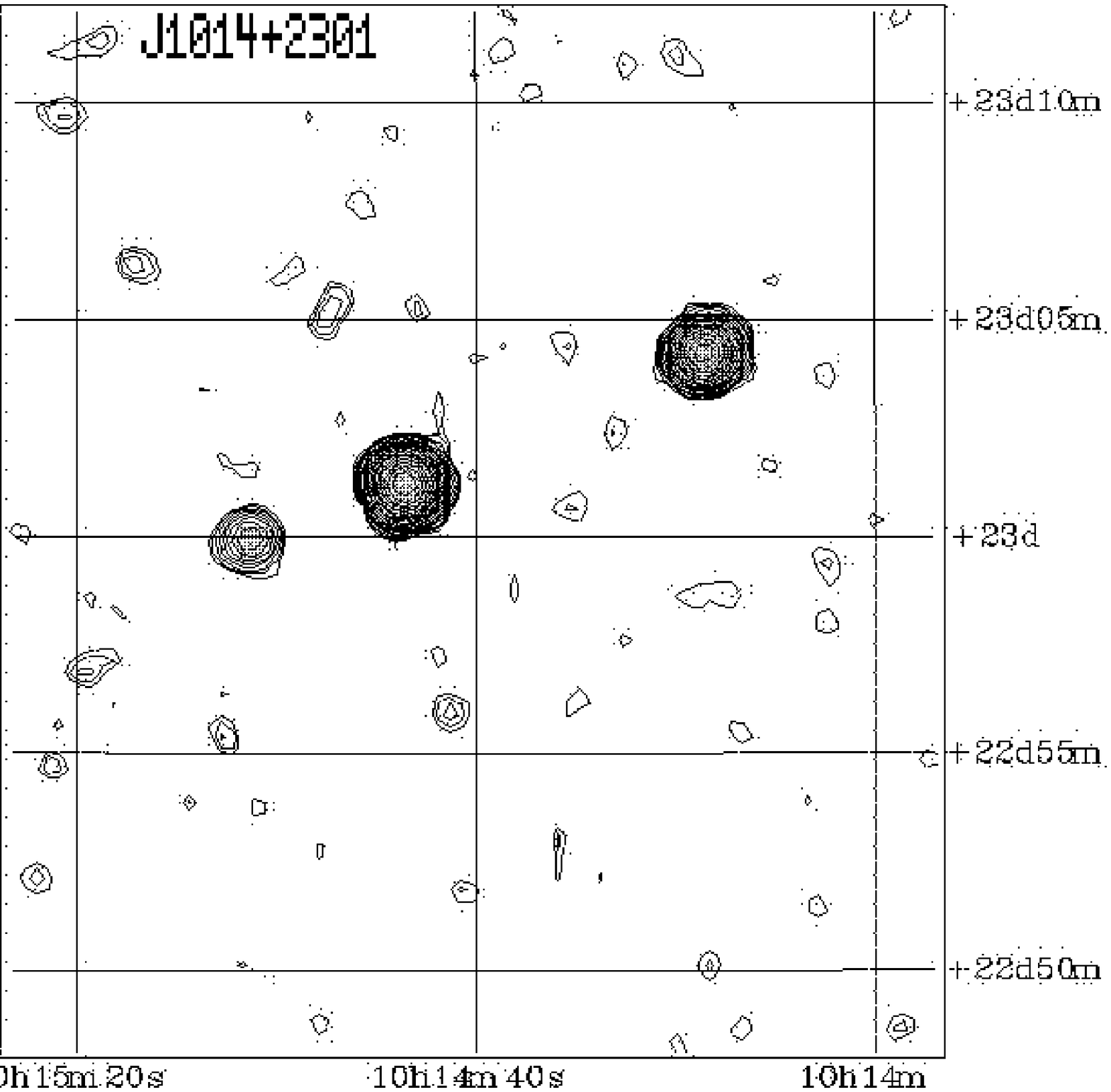,width=8.0cm,angle=0}
	\hspace{0.5cm}
       \psfig{figure=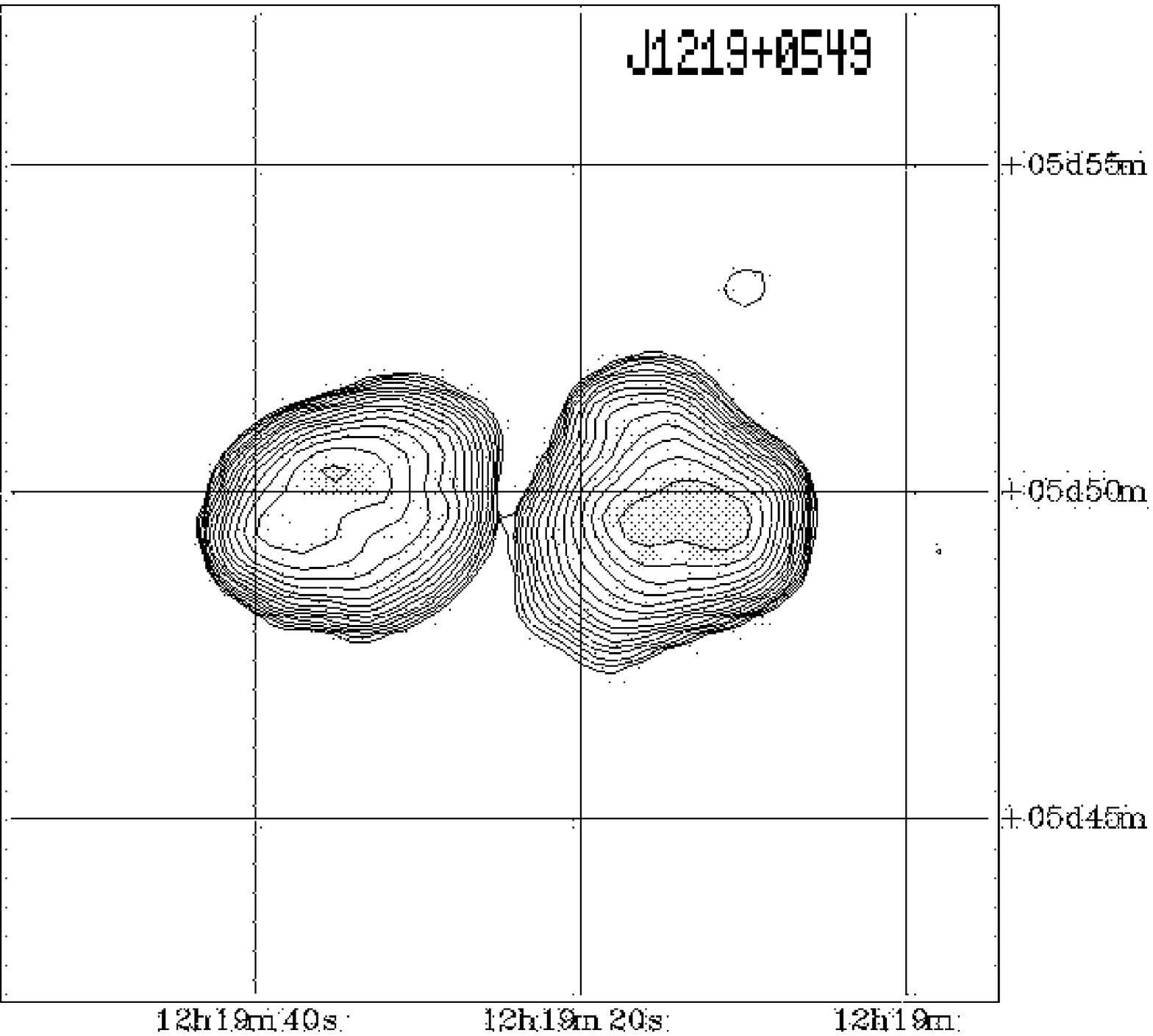,width=8.0cm,angle=0}
 }
 \vspace{0.5cm}
  \hbox{\psfig{figure=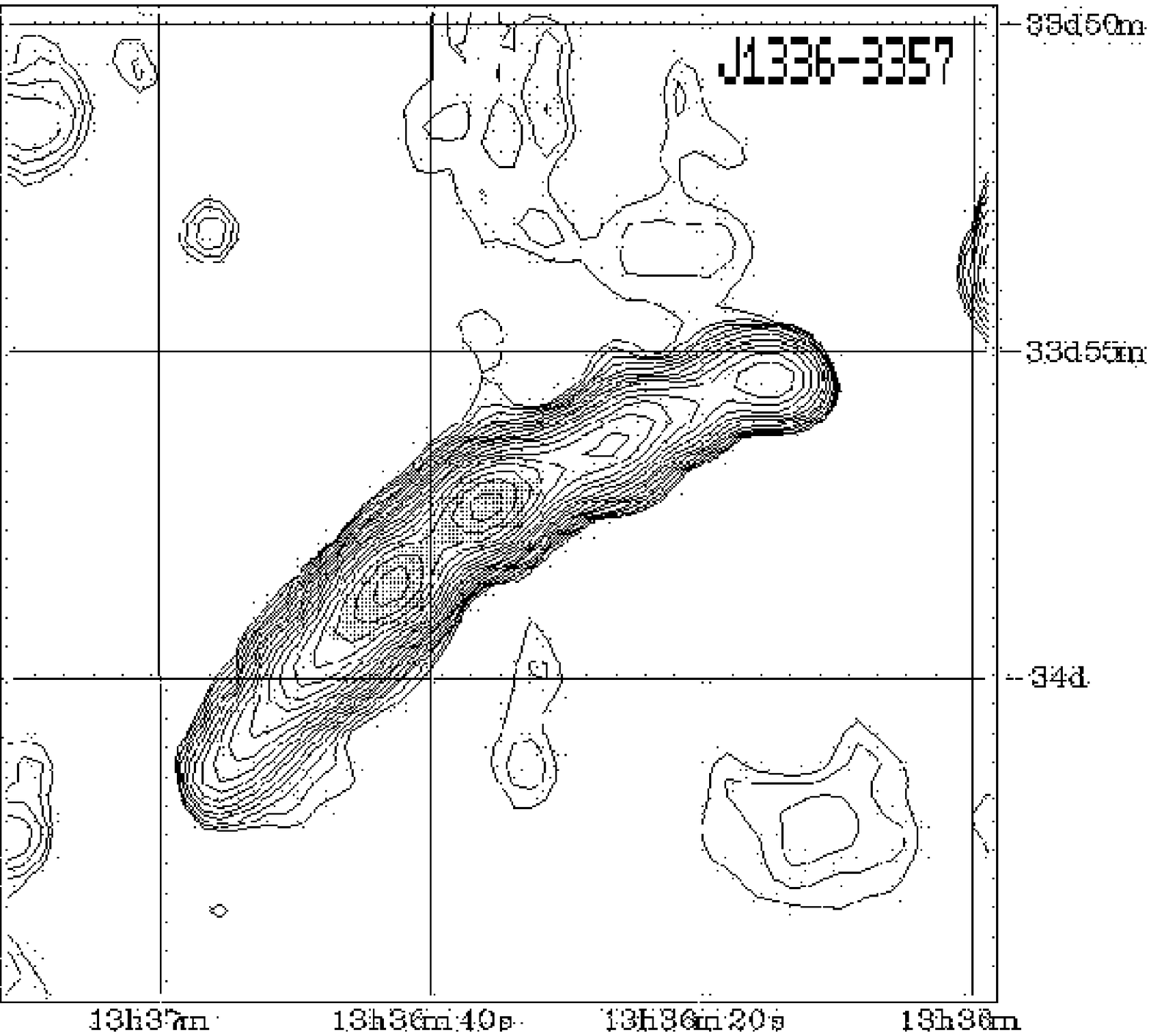,width=8.0cm,angle=0}
	\hspace{0.5cm}
	\psfig{figure=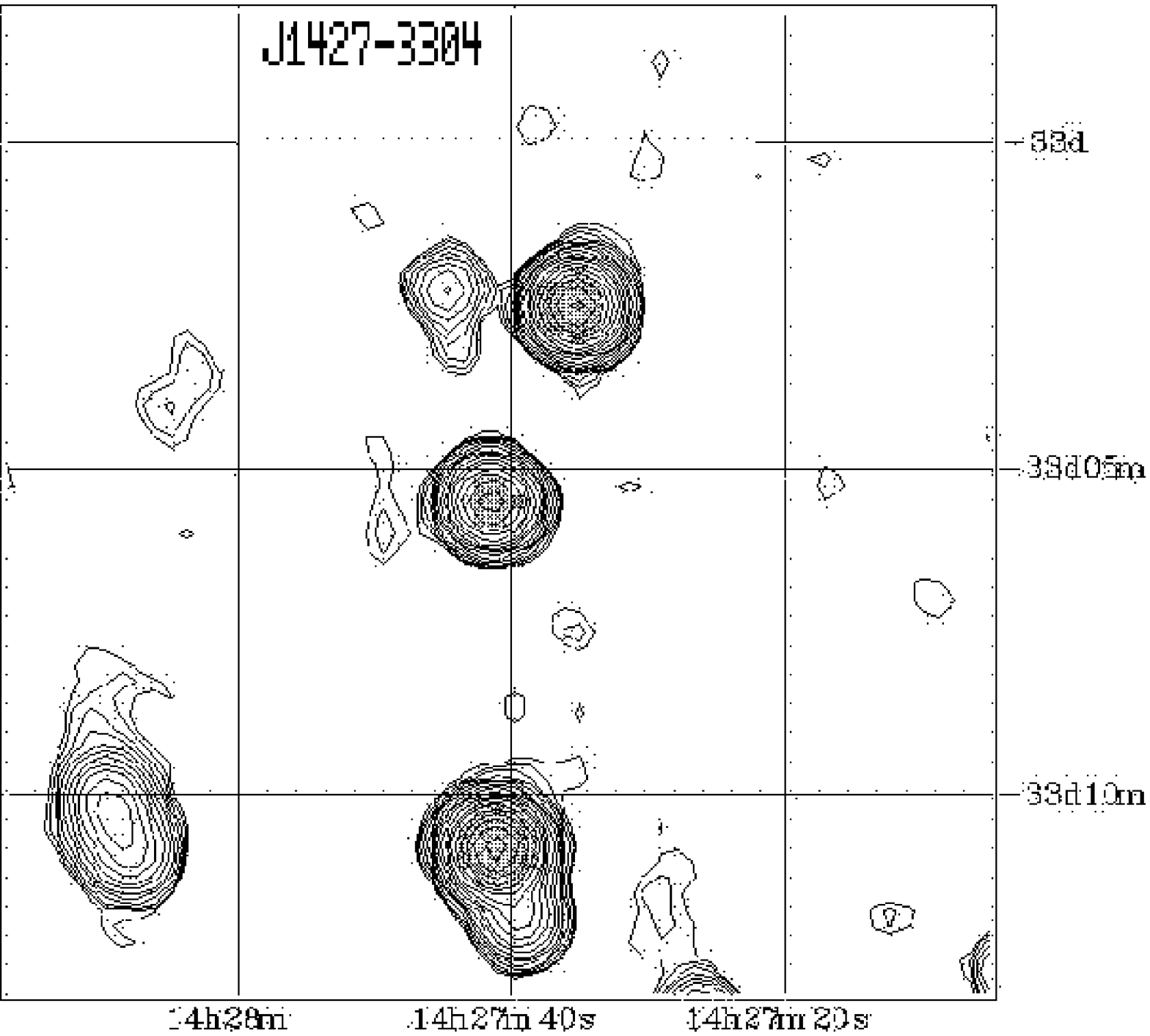,width=8.0cm,angle=0}
  }
 }
}
\caption{b) (continued) NVSS maps of the extended WMAP sources.}
\label{m2}
\end{figure*}

\setcounter{figure}{1}
\begin{figure*}
\centerline{
 \vbox{
  \hbox{\psfig{figure=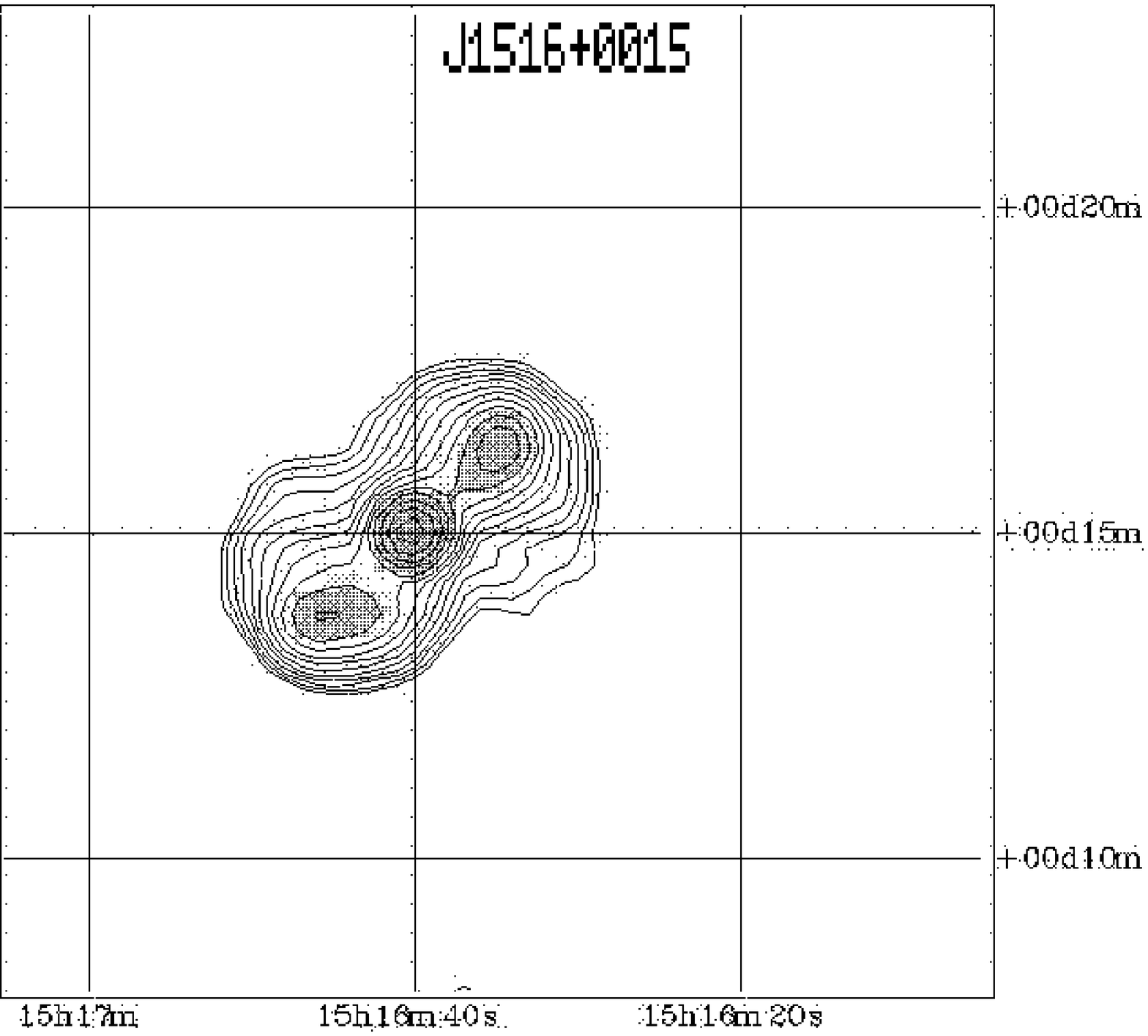,width=8.0cm,angle=0}
	\hspace{0.5cm}
	\psfig{figure=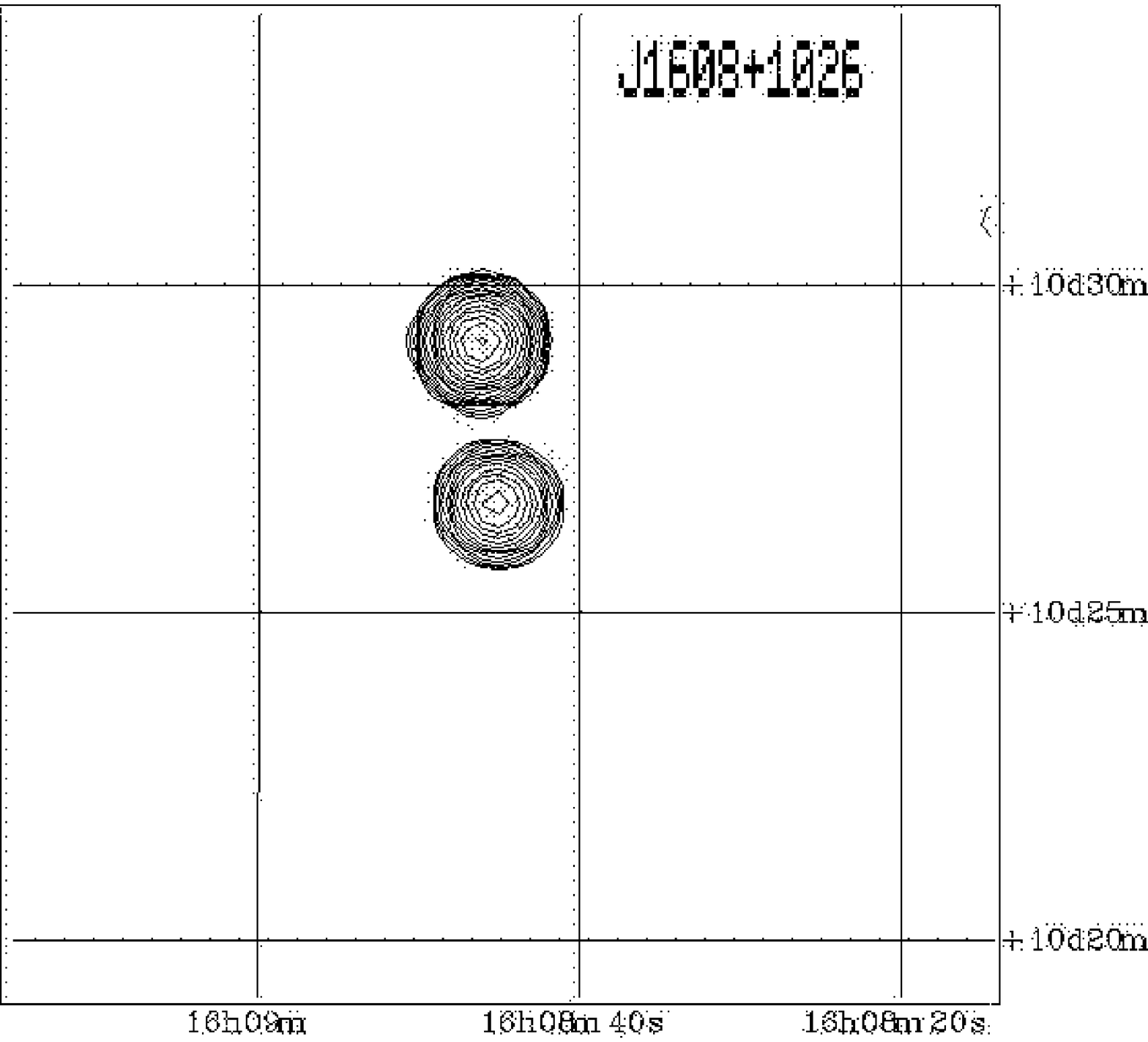,width=8.0cm,angle=0}
  }
 \vspace{0.5cm}
  \hbox{\psfig{figure=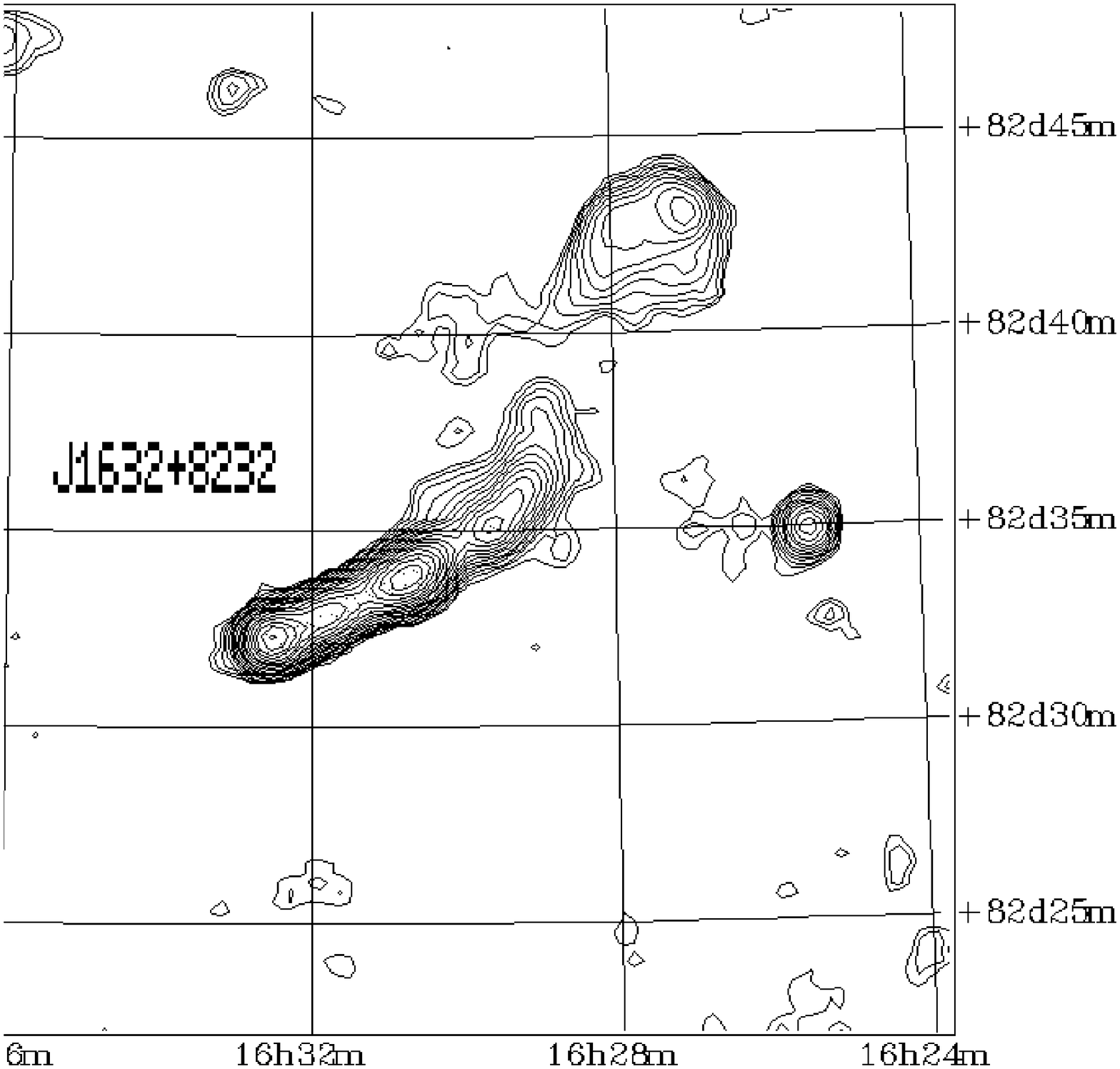,width=8.0cm,angle=0}
	\hspace{0.5cm}
	\psfig{figure=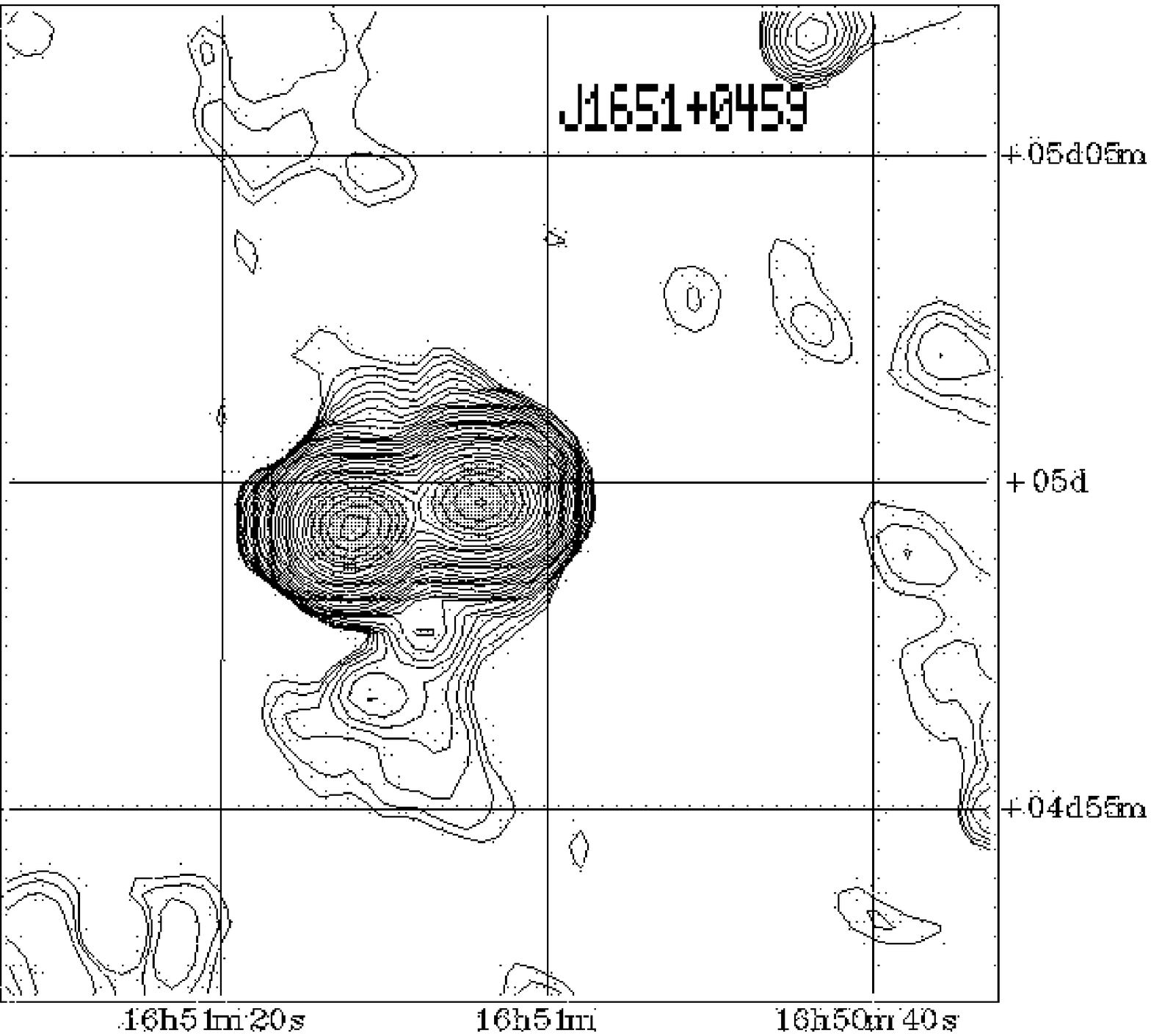,width=8.0cm,angle=0}
  }
 \vspace{0.5cm}
  \hbox{\psfig{figure=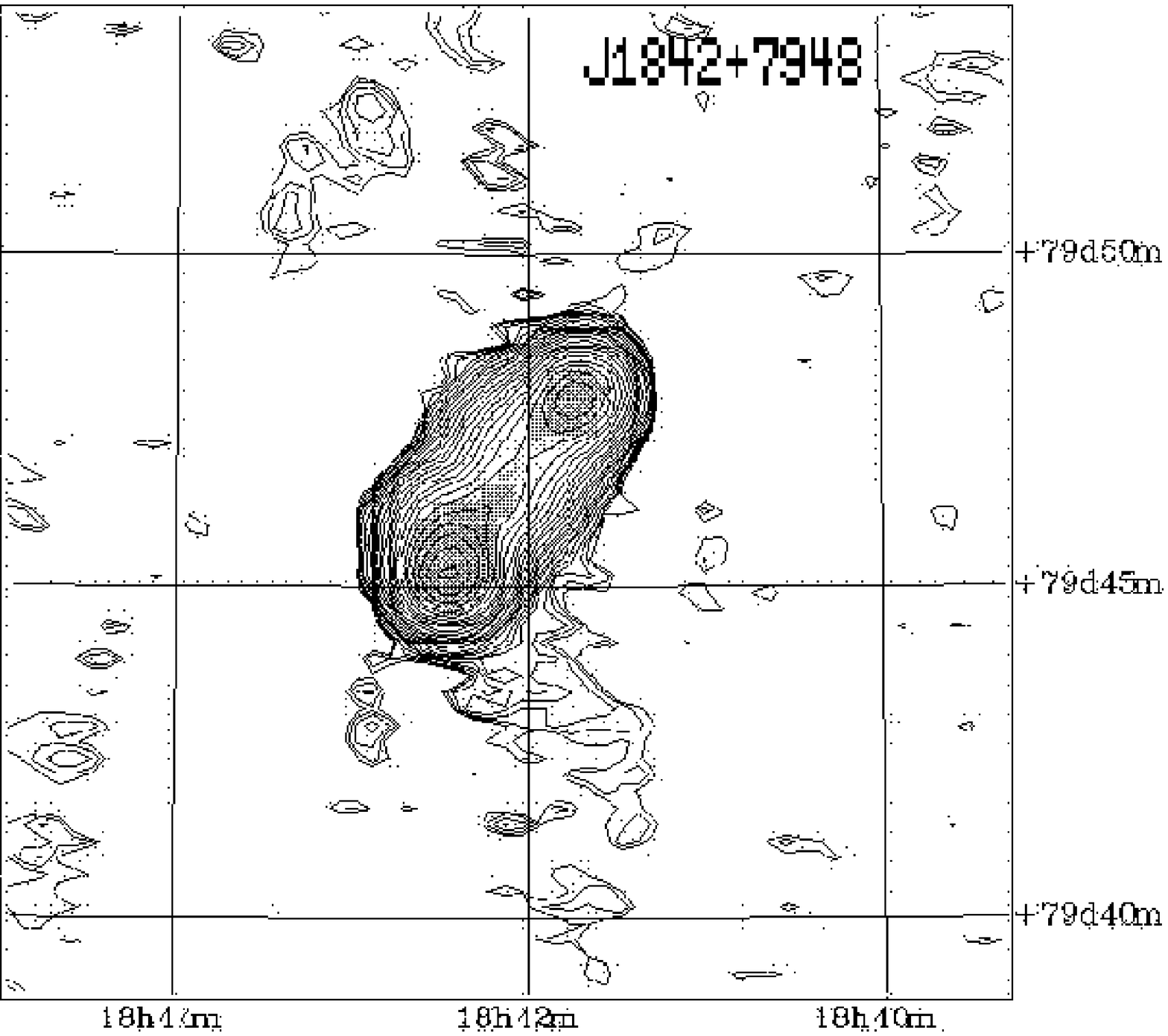,width=8.0cm,angle=0}
	\hspace{0.5cm}
	\psfig{figure=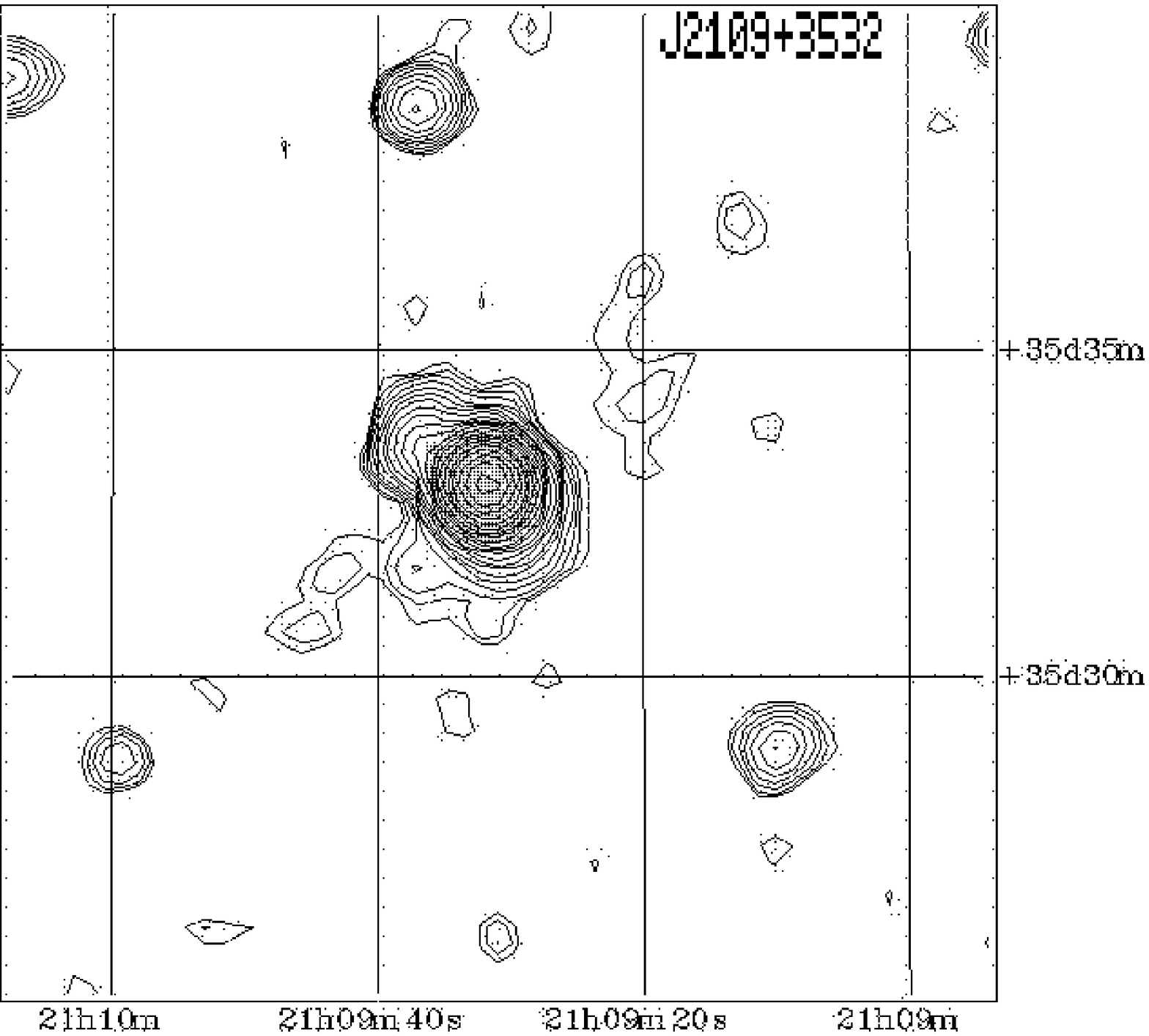,width=8.0cm,angle=0}
  }
 }
}
\caption{c) (continued) NVSS maps of the extended WMAP sources.}
\label{m3}
\end{figure*}

It can be seen from the distribution of the spectral indices  and from the spectra
themselves that the percentage of sources with the GHz peak is large
(27, 13\,\%) and the proportion of sources with the classical power
spectrum is not large (16, 8\,\%).
The great majority of WMAP sources have flat and inverse spectra
($\sim82$, 40\,\%). 15 (7\,\%) sources have combined spectra
(3C\,84 spectrum type) with a power low-frequency component and a flat
(inverse) high-frequency component. A detailed comparison of the spectra
and the structure of the sources needs a separate study.

\subsection{Radio maps of WMAP sources}

Numerous bright WMAP sources are included into the lists of VLBI monitoring
(for instance, Kellermann et al. 1998). This is why, detailed
multi-frequency radio maps with a millisecond angular resolution have
been drawn for them, and after the launching of the 8\,m space telescope
HALSA (satellite VSOP) even with a sub-millisecond resolution.
Most frequently such maps reveal a double structure of sources or a nucleus
with a jet (Fomalont et al. 2000). The super-luminous WMAP source, the
radio galaxy 3C120, was mapped sequentially during 16 months at a frequency
of 43\,GHz (Gomez et al. 2001). For the first time, a detailed investigation
into the inner structure of the jet with a spatial resolution
of about 0.07 pc was carried out.

Fig.\,\ref{m1} displays NVSS maps of 16 extended WMAP sources
at the frequency 1.4\,GHz.
Everywhere the first level of the brightness isophotes
is 0.001\,Jy/beam and the following levels are with a
multiplication factor of 1.41. Everywhere the reference grid corresponds
to the epoch J2000. But for the lacking source
J1230+1223 (Virgo\,A), this is a complete sample of the extended
(on arc min scale) WMAP sources identified with 168 sources of the
NVSS survey, that is, having the structure (or multiple components) of
radio emission more than  45$''$ in extent. Among them there are two galaxies
with a star formation burst, J0047--2517 (NGC253) and J0955+6940 (M82),
the known radio galaxies, Fornax A (J0321--3706, NGC 1316) and Her A
(J1651+0459), already mentioned Seyfert (Sy\,2) galaxy J1632+8232 (NGC6251).

In the window $30''$ 45 WMAP sources  are identified with the sources of
the FIRST catalog  (White et al. 1997).
All of them have angular sizes smaller than 12 arcseconds, but for the
radio galaxy Virgo\,A.
The SUMSS maps at the frequency 843\,MHz of three southern WMAP sources,
having angular sizes of about $1'$, are presented in Fig.\,\ref{sl} on the left.
The first level of the brightness isophotes
is also 0.001\,Jy/beam and the following levels are with a
multiplication factor of 1.41.
Fig.\,\ref{s1} (right) presents the FIRST maps
of three WMAP sources, having angular dimensions more than $5''$,
at the frequency 1.4~GHz. The first level of the brightness isophotes is

\setcounter{figure}{2}
\begin{figure*}
\centerline{
 \vbox{
  \hbox{\psfig{figure=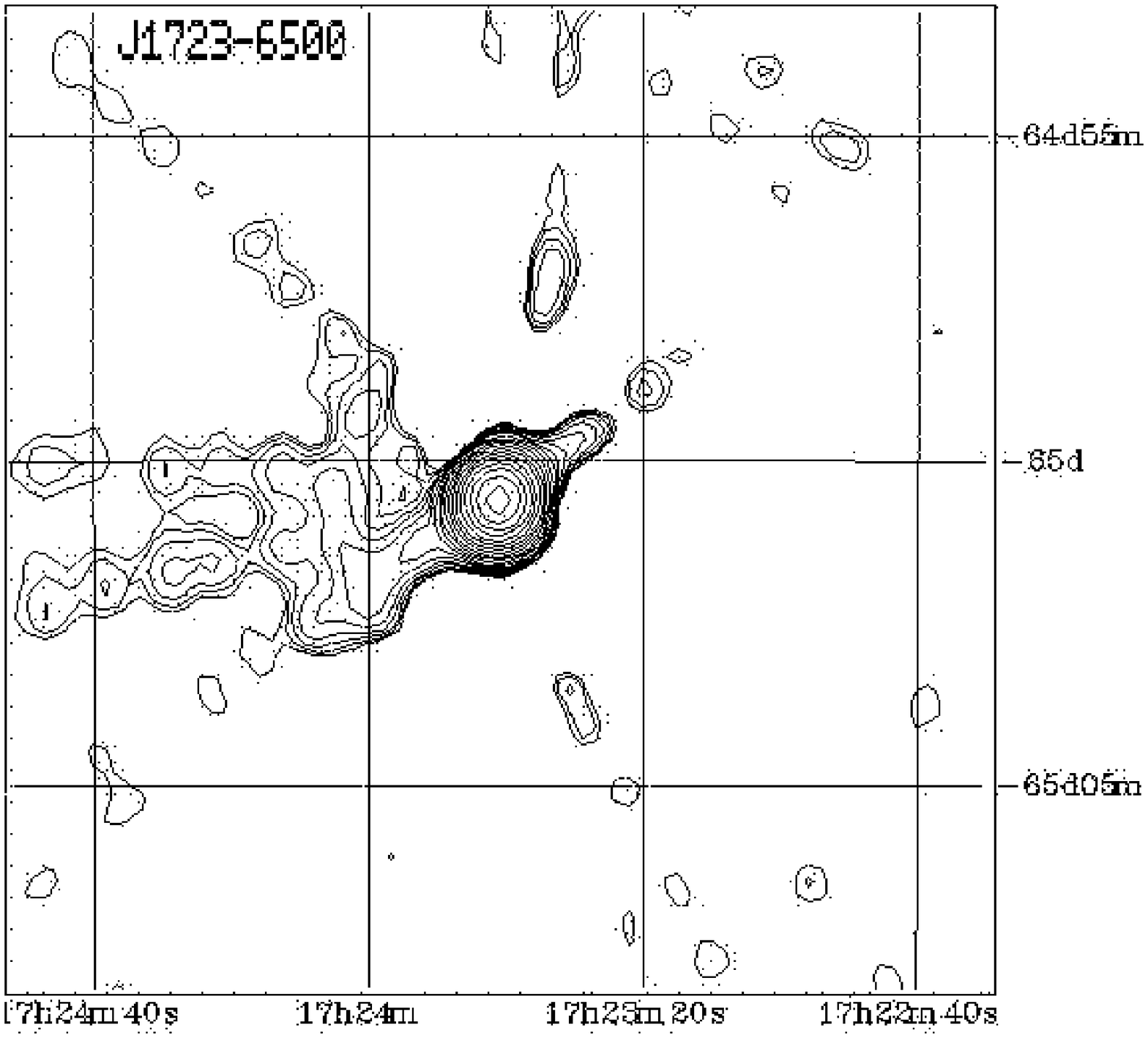,width=8.0cm,angle=0}
	\hspace{0.5cm}
	\psfig{figure=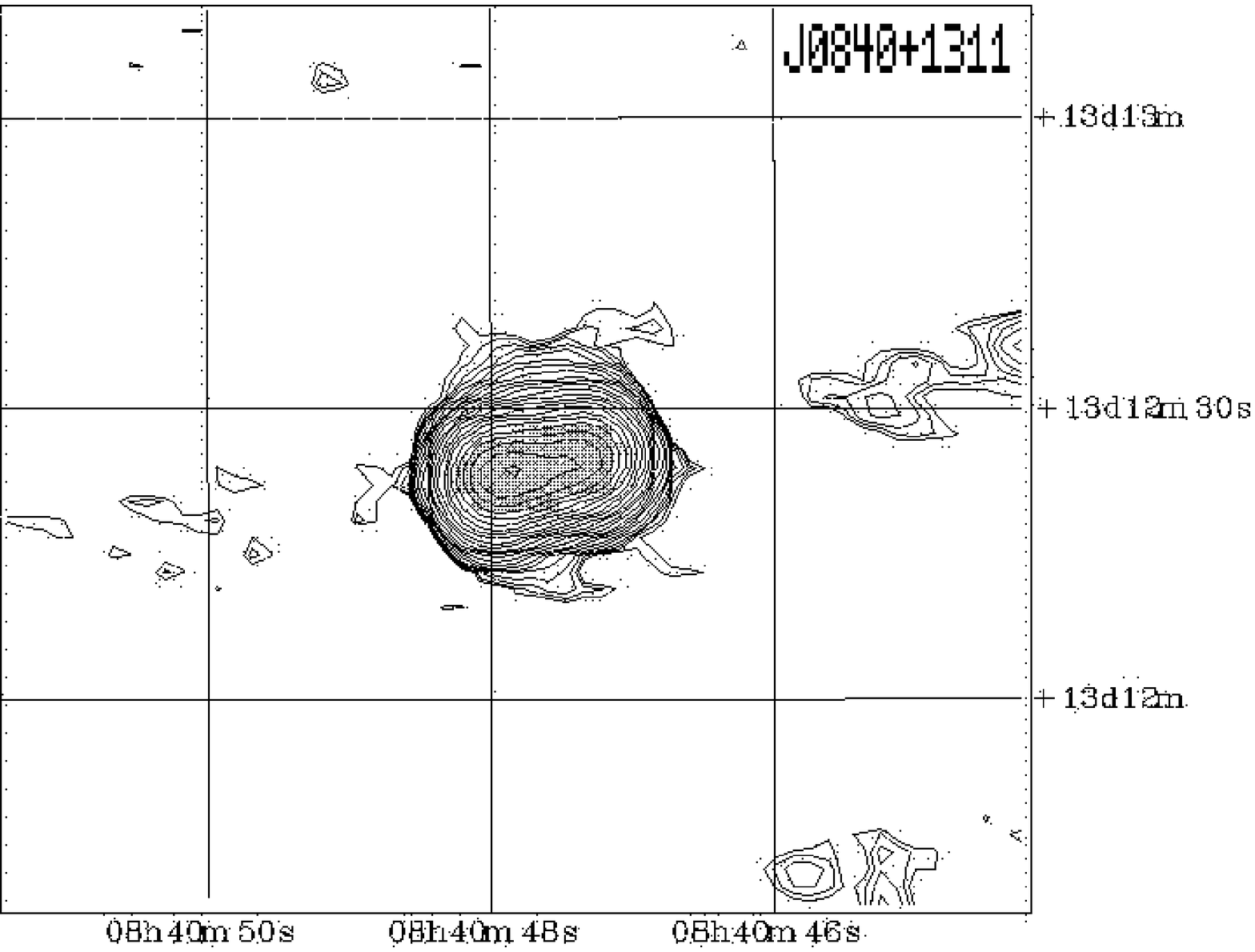,width=8.0cm,angle=0}
  }
 \vspace{0.5cm}
  \hbox{\psfig{figure=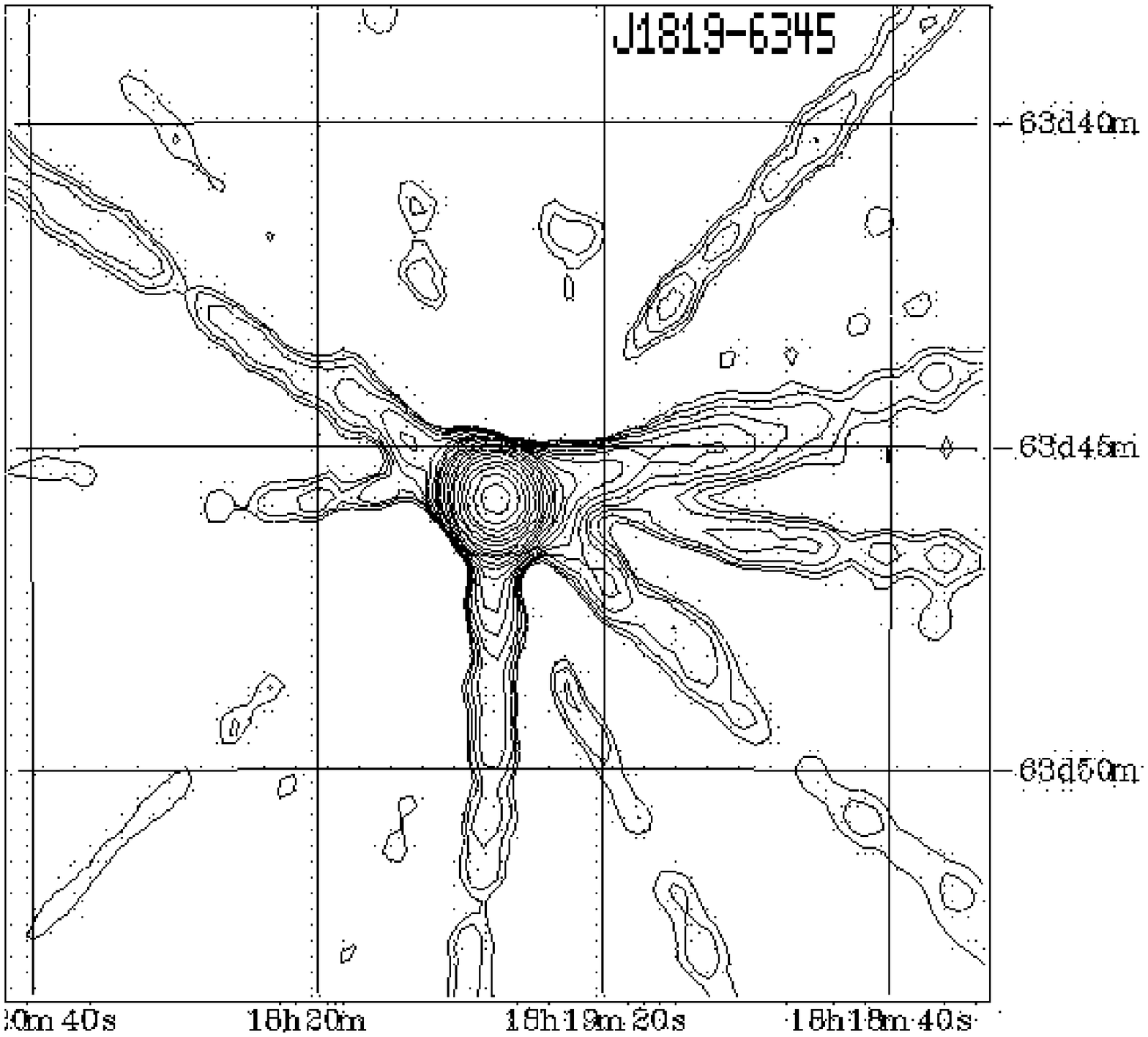,width=8.0cm,angle=0}
	\hspace{0.5cm}
       \psfig{figure=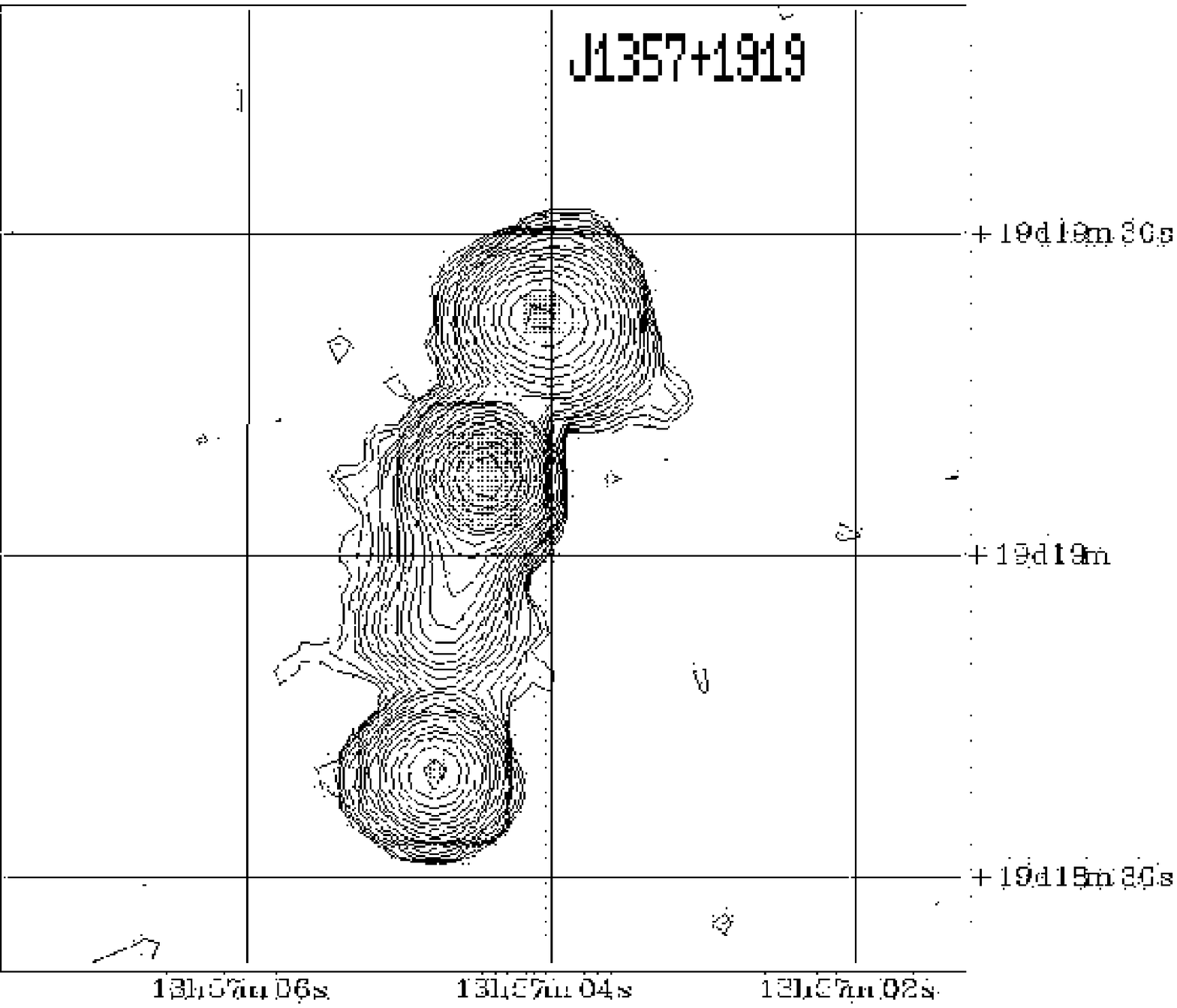,width=8.0cm,angle=0}
  }
 \vspace{0.5cm}
  \hbox{\psfig{figure=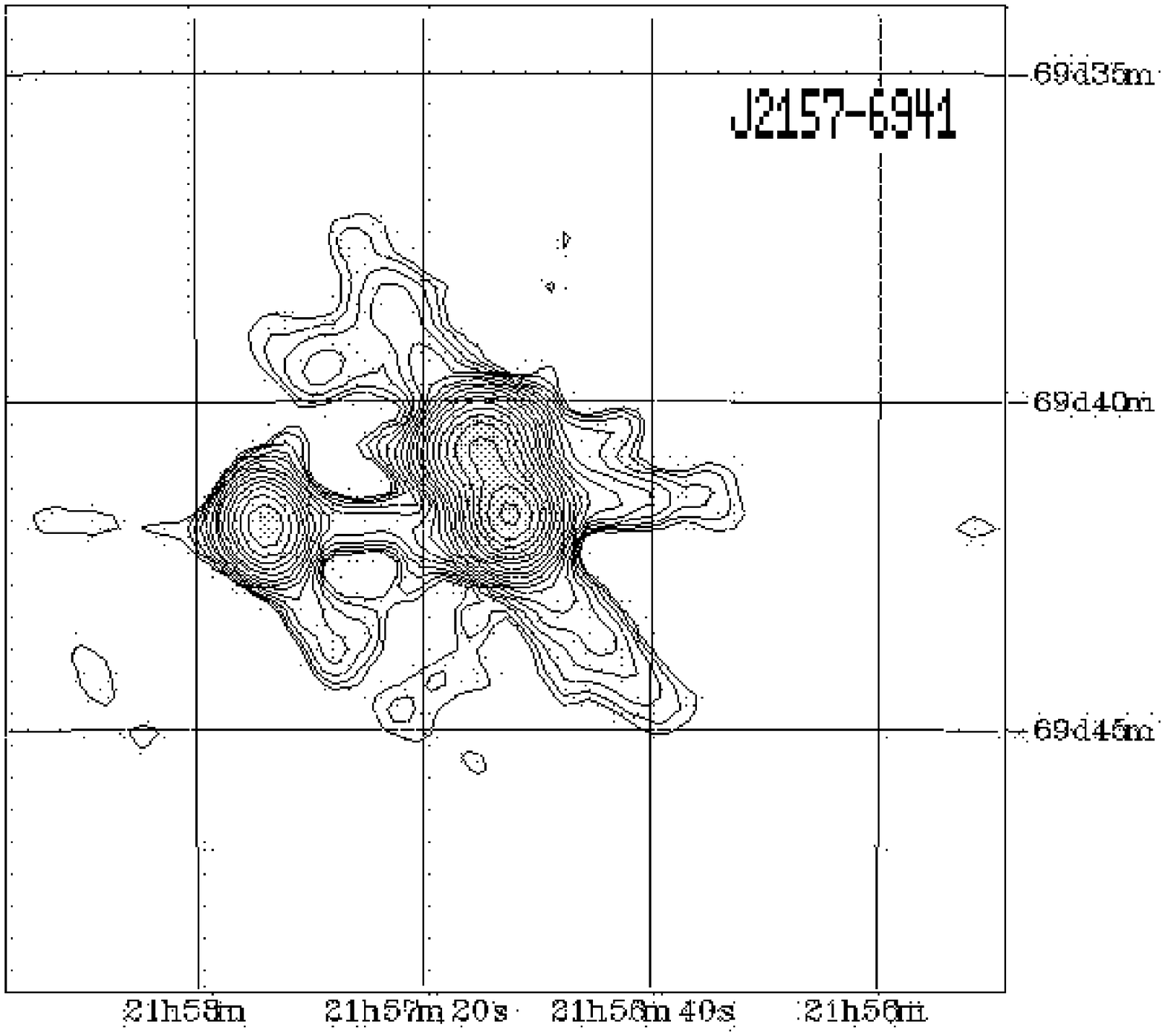,width=8.0cm,angle=0}
	\hspace{0.5cm}
       \psfig{figure=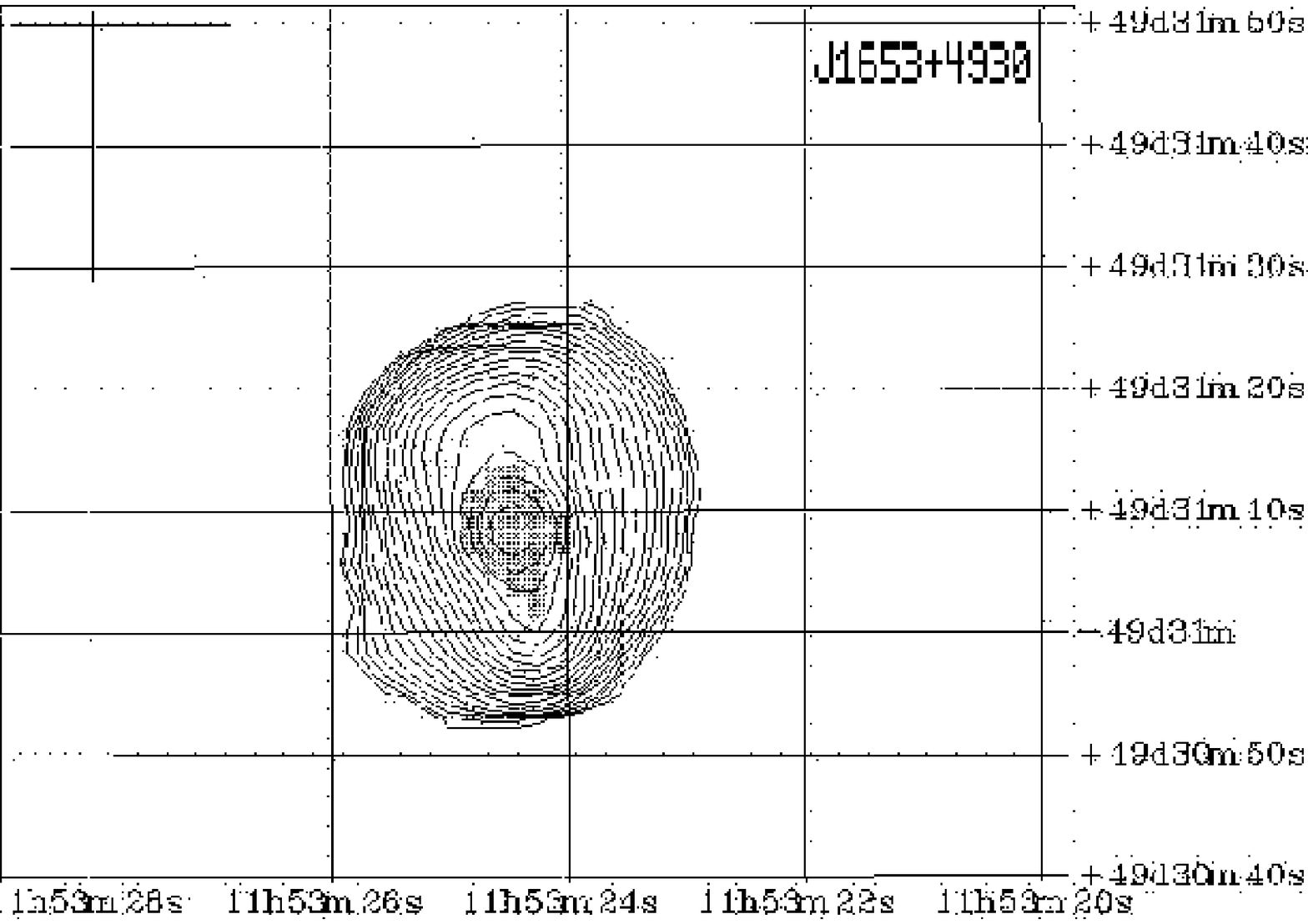,width=8.5cm,angle=0}
  }
 }
}
\caption{Maps of the WMAP sources (left: SUMSS, 843\,MHz; right: FIRST, 1400\,MHz).}
\label{s1}
\end{figure*}
\noindent
also 0.001\,Jy/beam and the following levels are with a
multiplication factor of 1.41.
Everywhere the coordinate grid corresponds to the epoch J2000.

\begin{figure}[t]
\centerline{\vbox{\psfig{figure=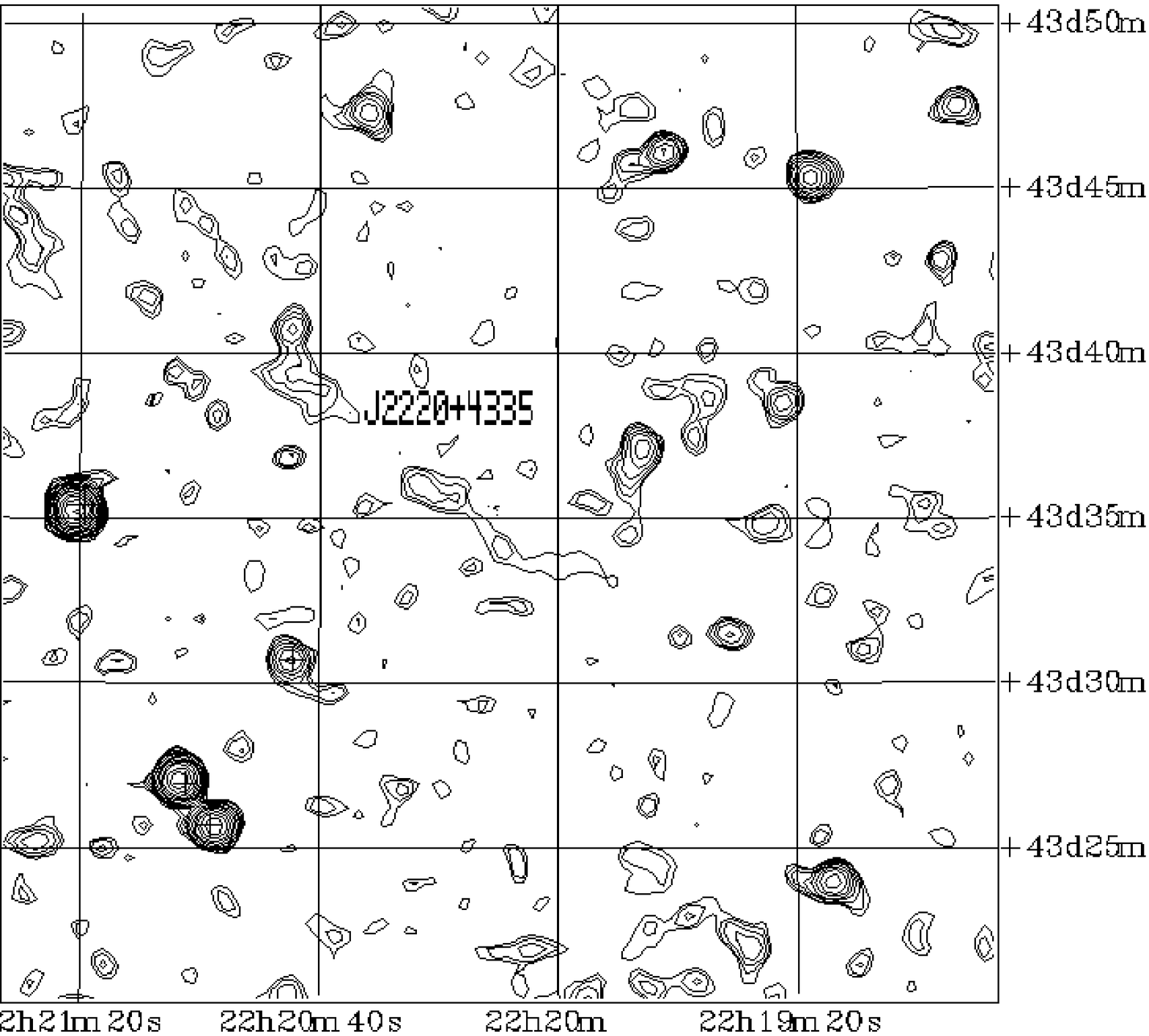,width=8.cm,angle=0}}}
\caption{Radio map of the region around the source WMAPJ2230+43 from the
NVSS survey at 1.4~GHz. The isophotes begin from the map brightness level
0.0005\,Jy/beam and go further to a maximum with a multiplication factor of
1.41. The cross in the middle marks the position of WMAPJ2230+43.
}
\label{w22}
\end{figure}

We have failed to find clear identification for the source WMAPJ2220+43.
Possibly, this is a double source south-east of WMAPJ2220+43 or a source
east of WMAPJ2221+43, which is its double on the NVSS map
(Fig.\,\ref{w22}), although they are located $\sim12'$ away from it.

\section{Conclusions}

We have studied spectral features of the complete sample of WMAP
sources, which is presently the only all-sky survey catalog at frequencies
higher than 20\,GHz. We have shown that the most of WMAP sources are
already known sources of the former more low-frequency
surveys 3C, 4C, GB6, PMN, NVSS, FIRST and others, and
203 sources (97.5\,\%) are identified with optical objects:
141 quasars (68\,\%), 42 galaxies or active galactic nuclei (20\,\%),
19 Bl Lac objects (9\,\%) and one planetary nebula IC418.

Based on our observational data of 26 sources and compilation of $\sim$206000
flux measurements, we have studied the  radio spectra and
constructed the most complete distribution of spectral indices over the
whole sample of WMAP sources. The spectra are frequently plotted in a
frequency interval including 4--5 orders --- from 10\,MHz to 245\,GHz.

A considerable part of WMAP sources  was included in the programs of
long-time monitoring of variability of extragalactic sources at
frequencies from 300 MHz to 245 GHz. First of all, it should be noted that
15--20-year programs of investigations of the identified WMAP sources
were performed with the GBI (NRAO) at two frequencies, and
the 25\,m telescope at UMRAO (USA) at three frequencies, and with the
12\,m telescope in Mets\"ahovi (Finland), and revealed various
variability of many WMAP sources.
Over 40 WMAP sources are the compact IDV sources. Samples of some bright
variable WMAP sources have repeatedly been observed at RATAN-600
(Amirkhanian et al. 1992;  Kovalev et al. 1999; Mingaliev et al. 2001;
Kiikov et al. 2002). We have also carried out observations
of 26 ``spectrally poorest'' sources,  accessible to observations
at RATAN-600.

Immediately after the publication of the WMAP survey data of the
first year (Paper 2) the WMAP catalog was included into the generally
accessible  CATS database (Verkhodanov et al. 1997) for the procedures
of searching, cross-identification of different data and plotting
of radio spectra. Such an extensive catalog can be a basis for performing
statistical studies of complete samples of
extragalactic objects.

We have presented the NVSS radio maps of the brightness distribution for
16 extended WMAP sources and the radio maps of the extended  structure
for other six WMAP sources from the SUMSS and FIRST surveys.

\vspace{0.2cm}

\begin{acknowledgements}

The project of the CATS database was supported through grant
of RFBR in 1996-1999. The author is grateful to his colleagues
O.V.\,Verkhodanov, V.N.\,Chernenkov and H.\,Andernach for productive
activity in creation and support of the CATS database. In the course
of work we made use of the generally accessible data of a few monitoring
programs (THE GBI-NASA MONITORING PROGRAM, UMRAO), to the authors of which
we express the most sincere thanks. We present radio maps from the
generally accessible FITS files of the surveys FIRST, NVSS (NRAO) and
SUMSS (Sydney University Molonglo Sky Survey).
This work was partially supported by grant of the program ``Astronomy''
in 2003.
\end{acknowledgements}


\begin{onecolumn}

\hspace{16cm}

{\large
Integral spectra of first 40 MMAP sources are presented below.
}

\vspace{1cm}
In order to do the complete version of the paper,  you must download
the 208 figures from ftp://cats.sao.ru/WMAP/spectra.tgz.

\end{onecolumn}

\clearpage


\begin{figure}\centerline{\vbox{\psfig{figure=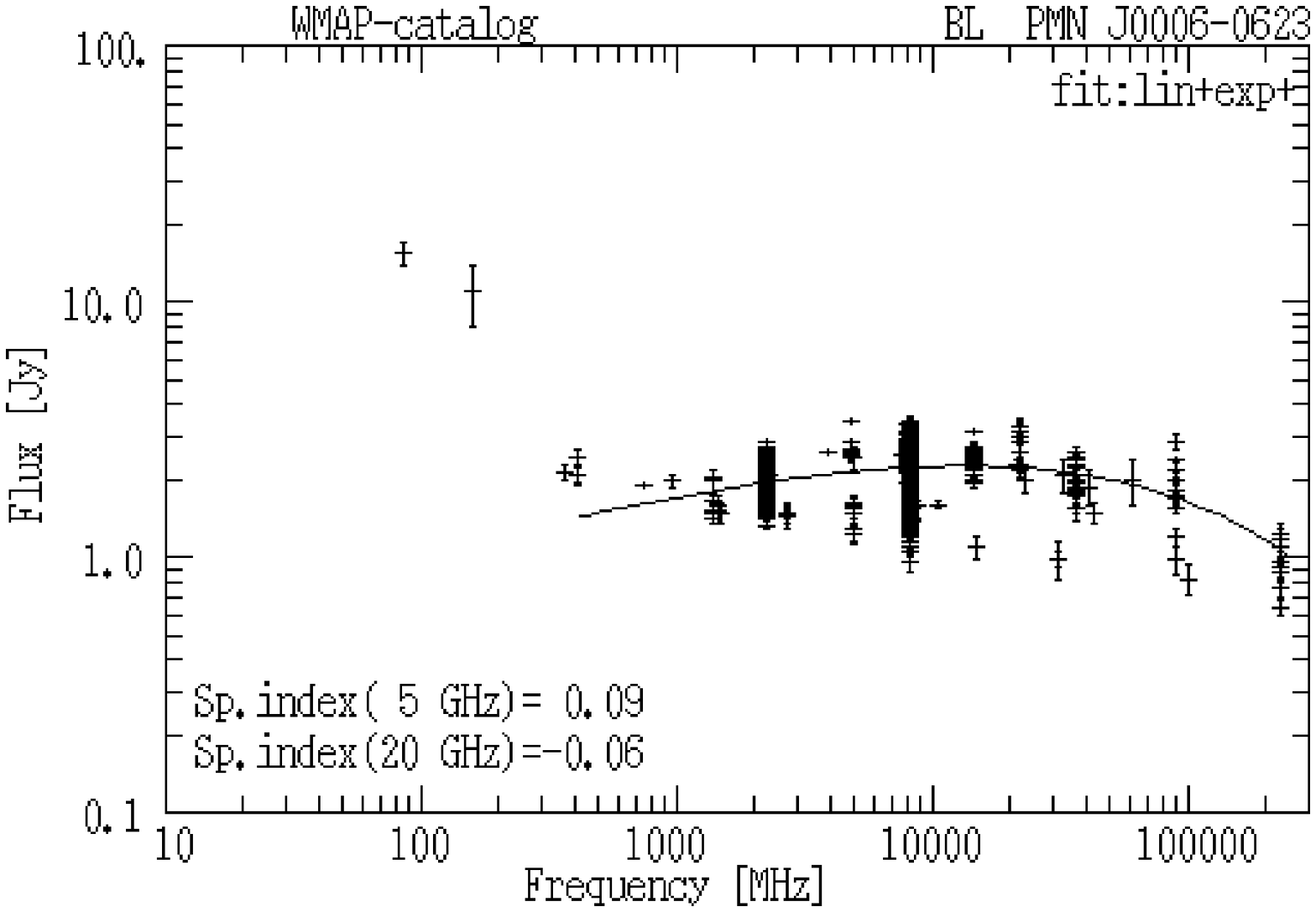,width=8.2cm,angle=0}}}\end{figure}
\begin{figure}\centerline{\vbox{\psfig{figure=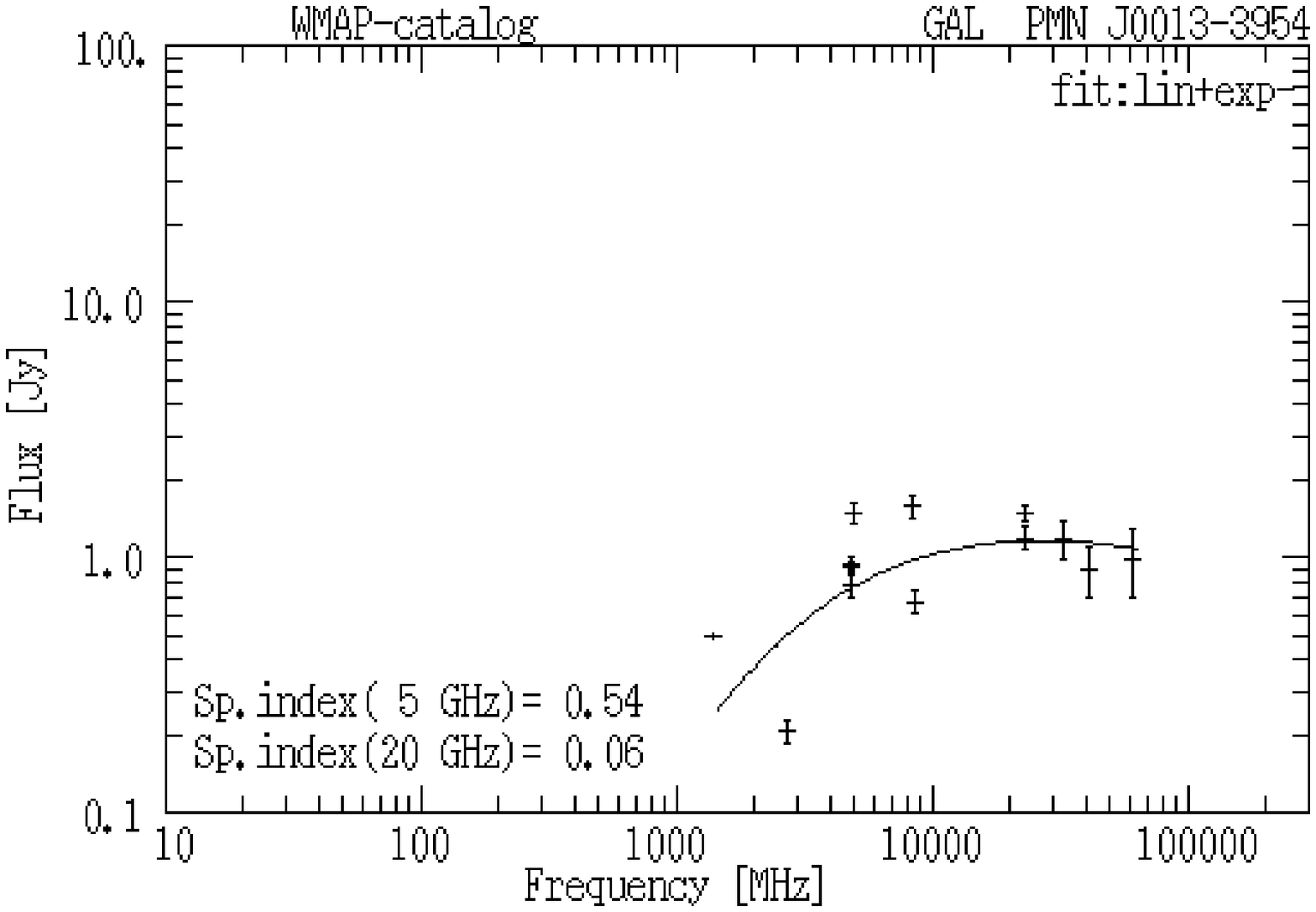,width=8.2cm,angle=0}}}\end{figure}
\begin{figure}\centerline{\vbox{\psfig{figure=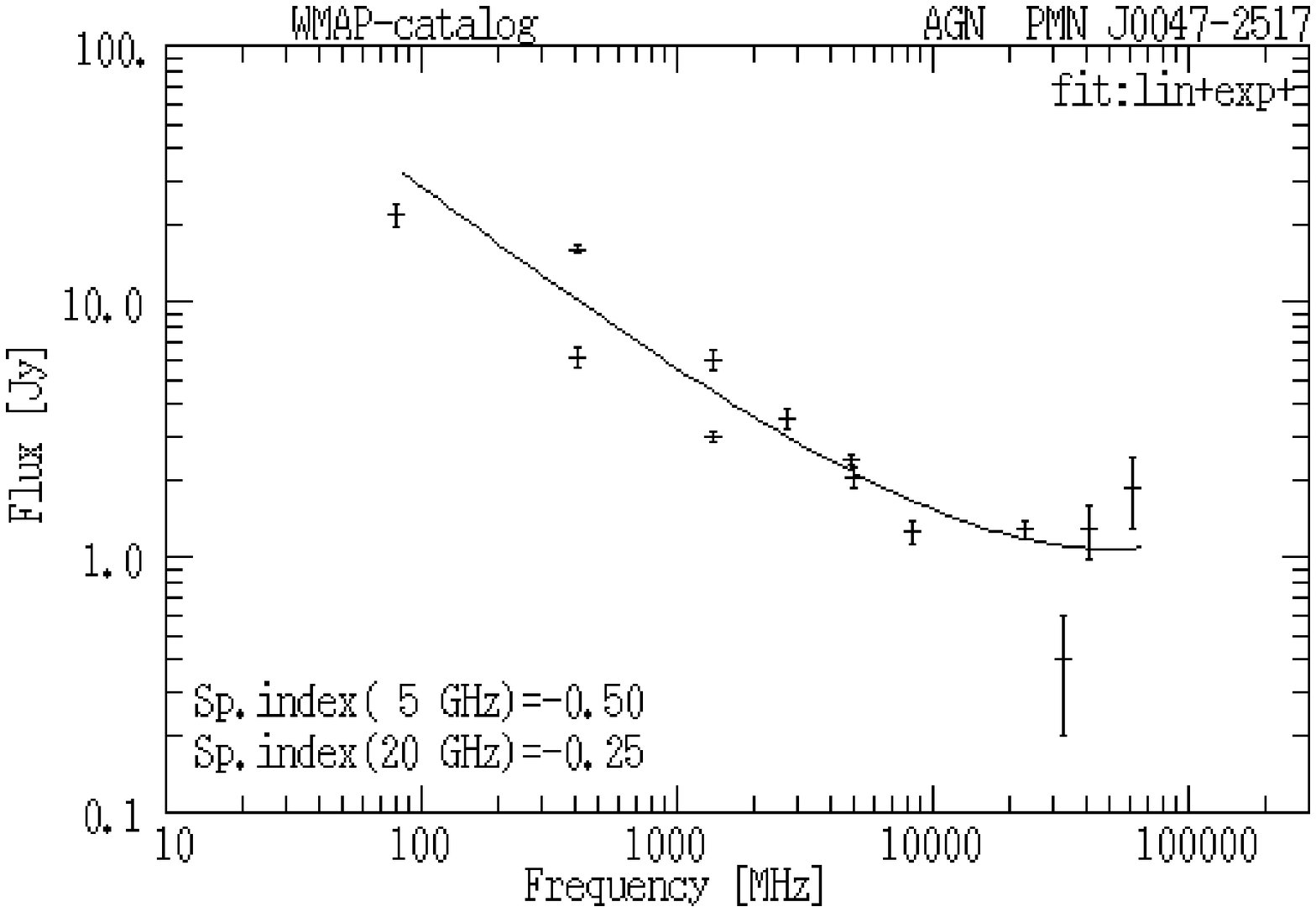,width=8.2cm,angle=0}}}\end{figure}
\begin{figure}\centerline{\vbox{\psfig{figure=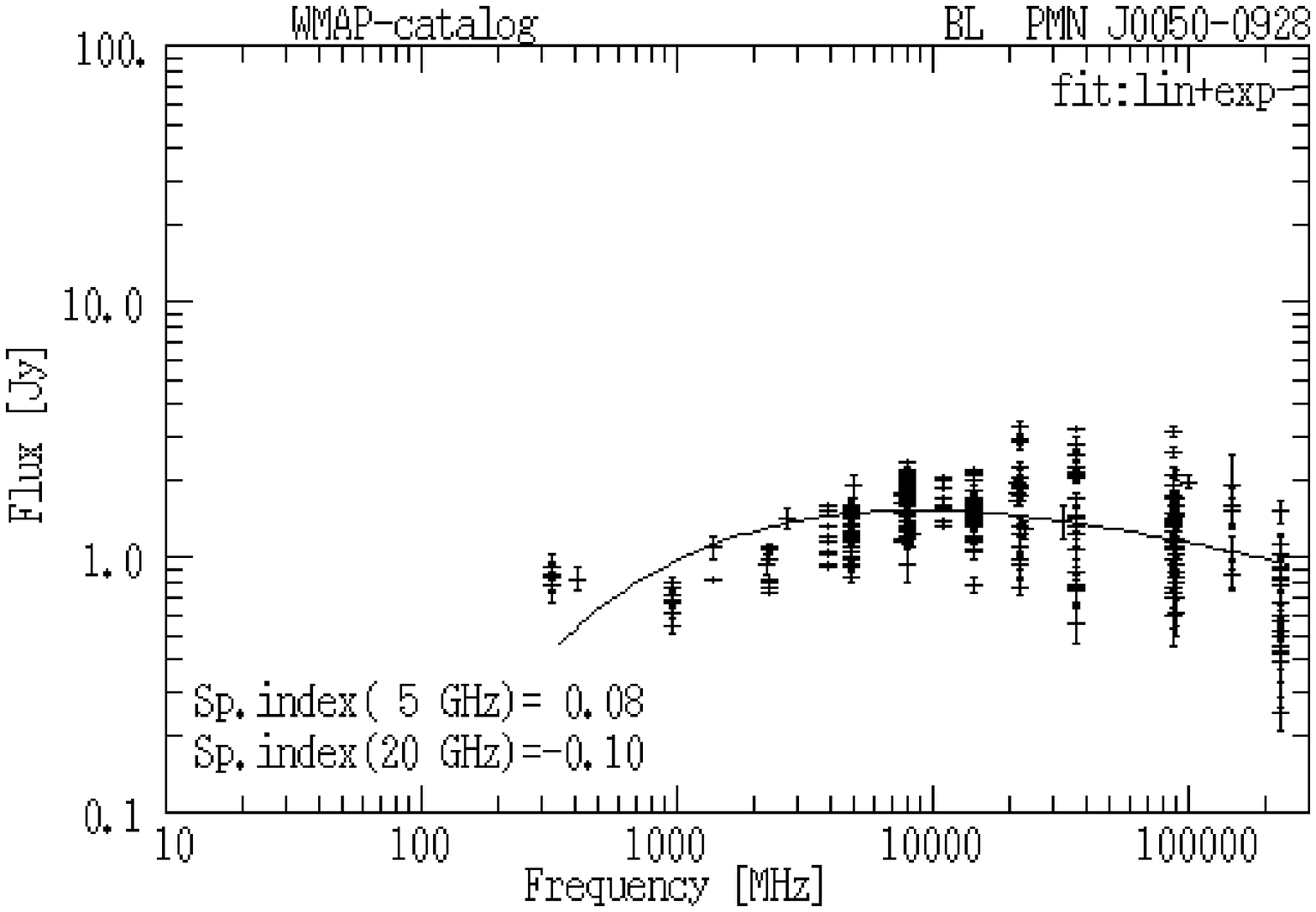,width=8.2cm,angle=0}}}\end{figure}
\begin{figure}\centerline{\vbox{\psfig{figure=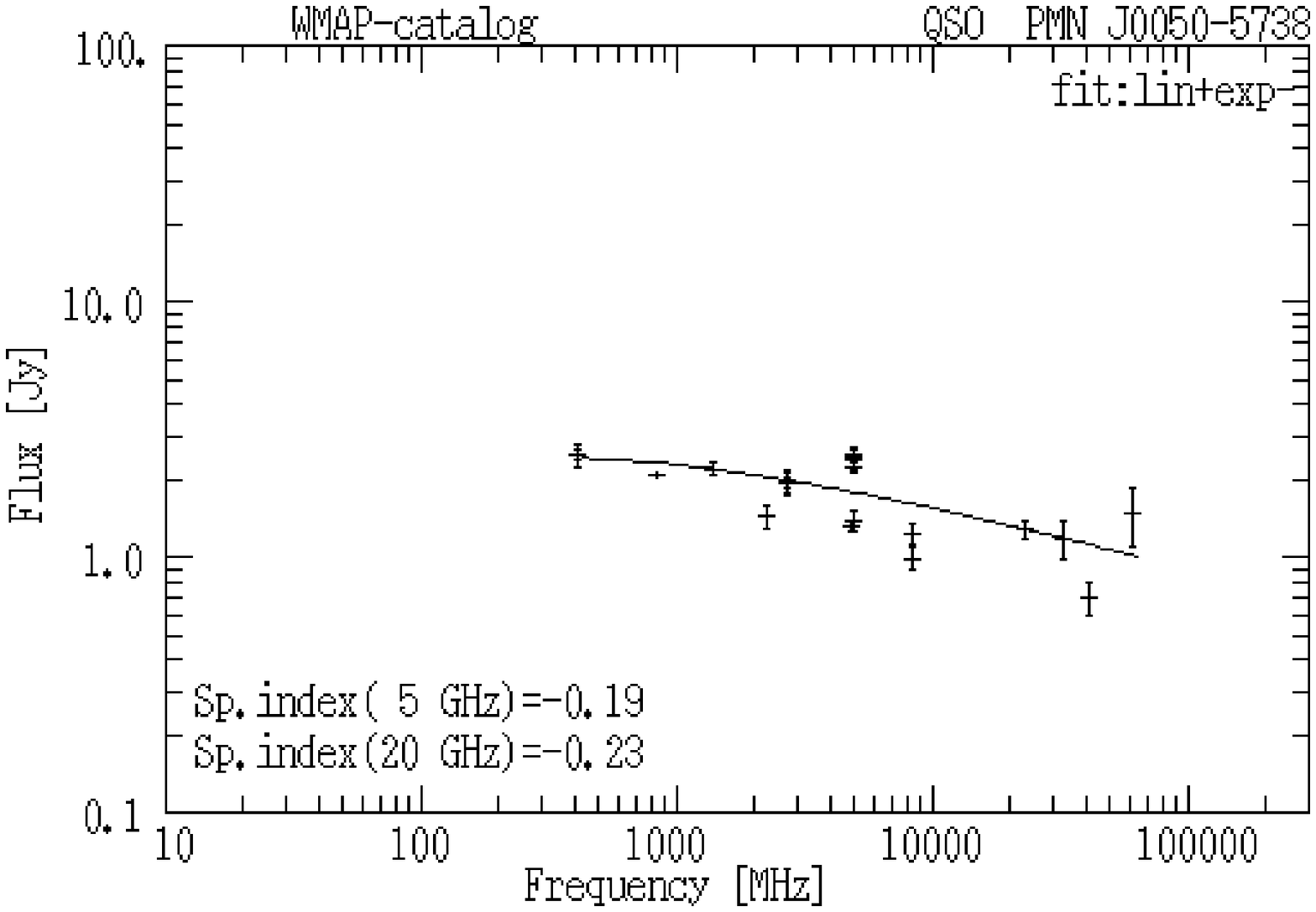,width=8.2cm,angle=0}}}\end{figure}
\begin{figure}\centerline{\vbox{\psfig{figure=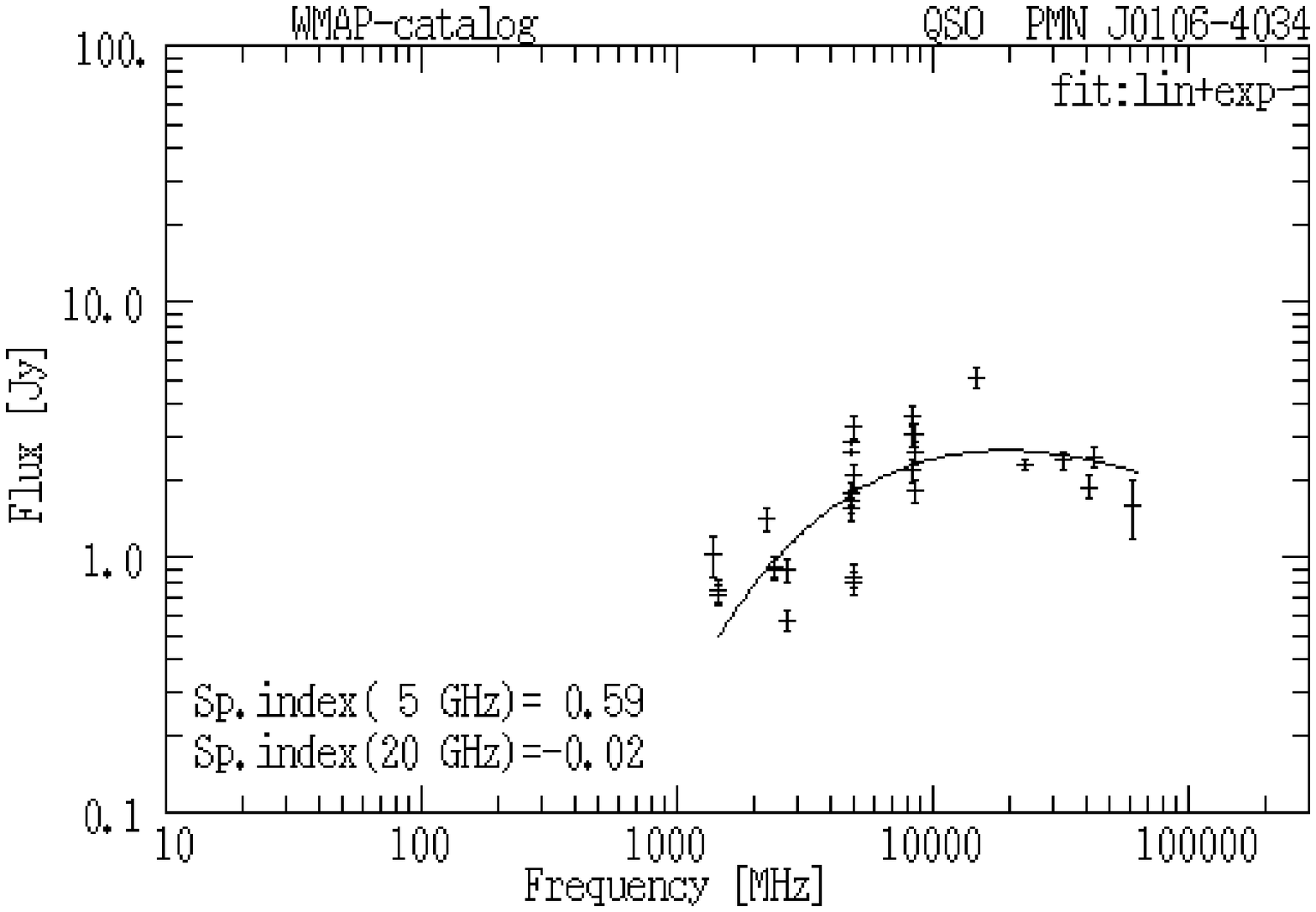,width=8.2cm,angle=0}}}\end{figure}
\begin{figure}\centerline{\vbox{\psfig{figure=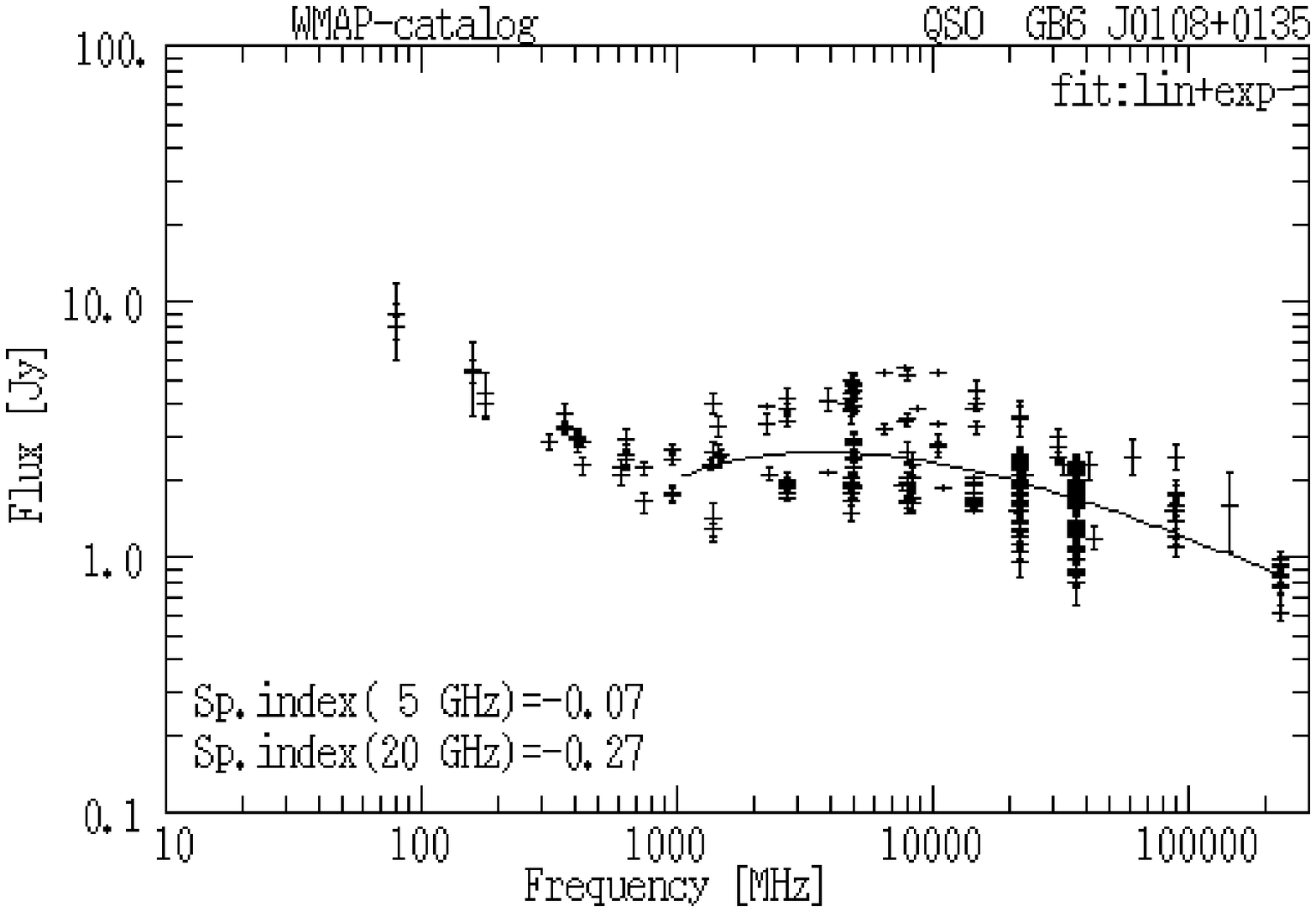,width=8.2cm,angle=0}}}\end{figure}
\begin{figure}\centerline{\vbox{\psfig{figure=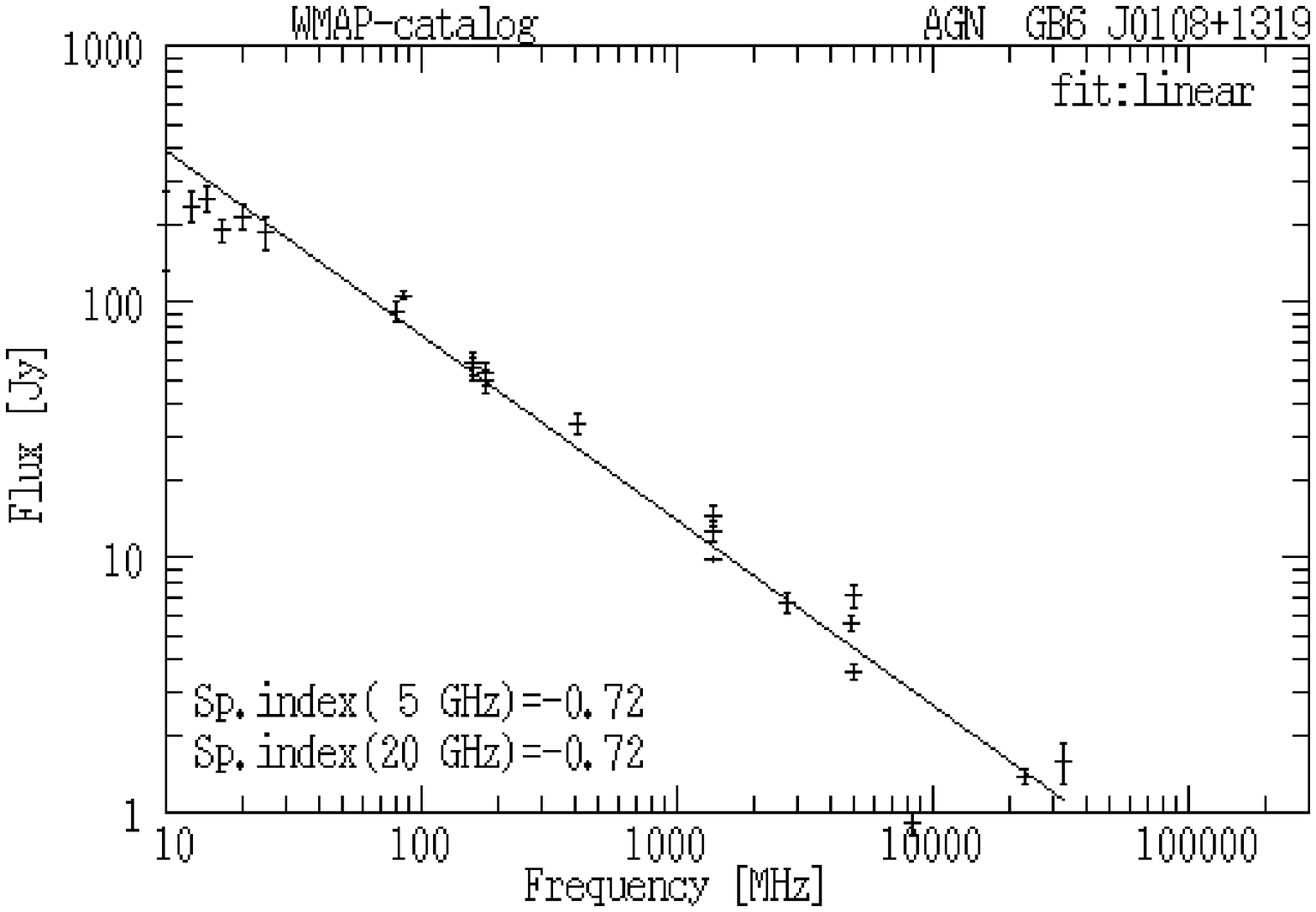,width=8.2cm,angle=0}}}\end{figure}\clearpage
\begin{figure}\centerline{\vbox{\psfig{figure=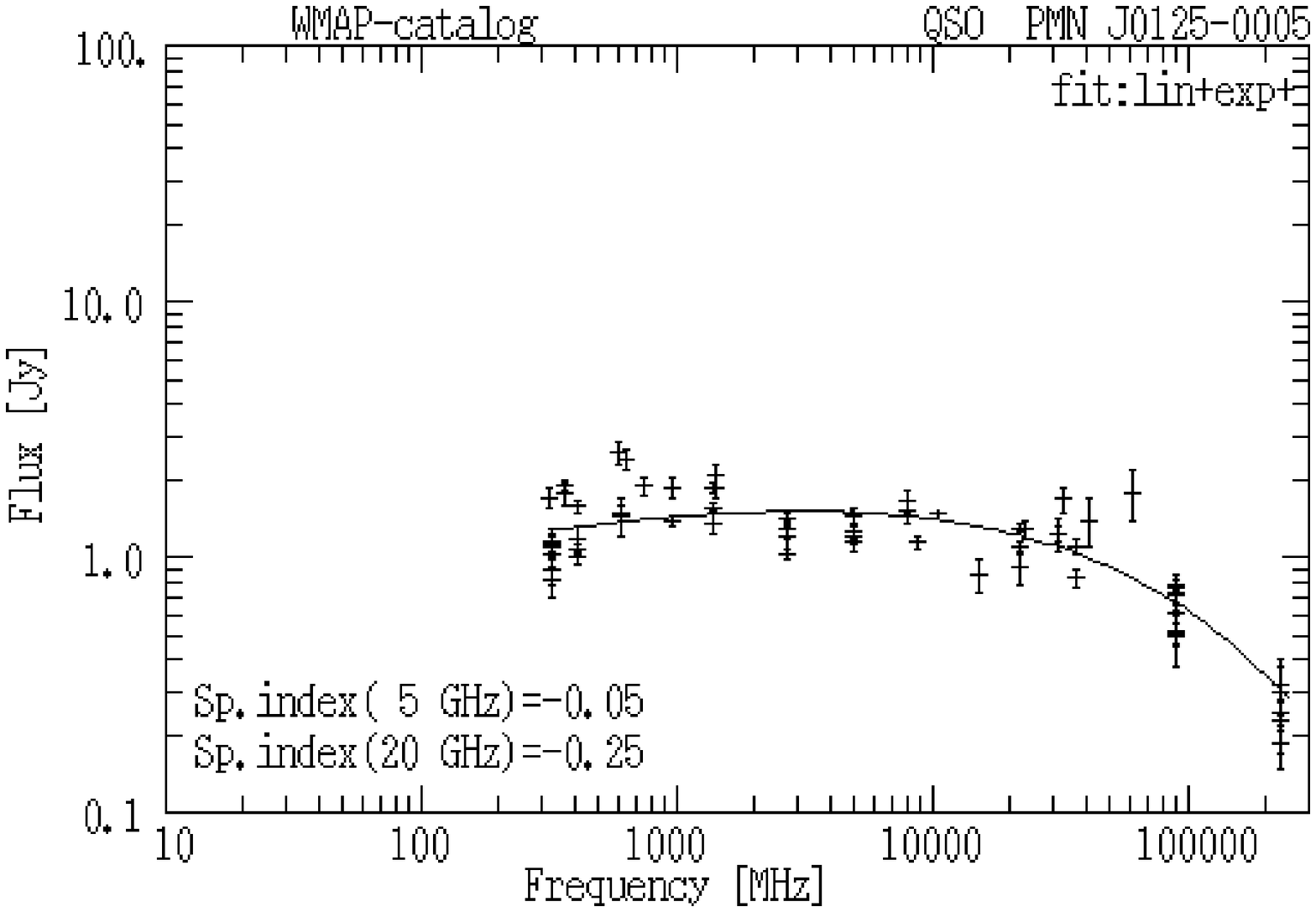,width=8.2cm,angle=0}}}\end{figure}
\begin{figure}\centerline{\vbox{\psfig{figure=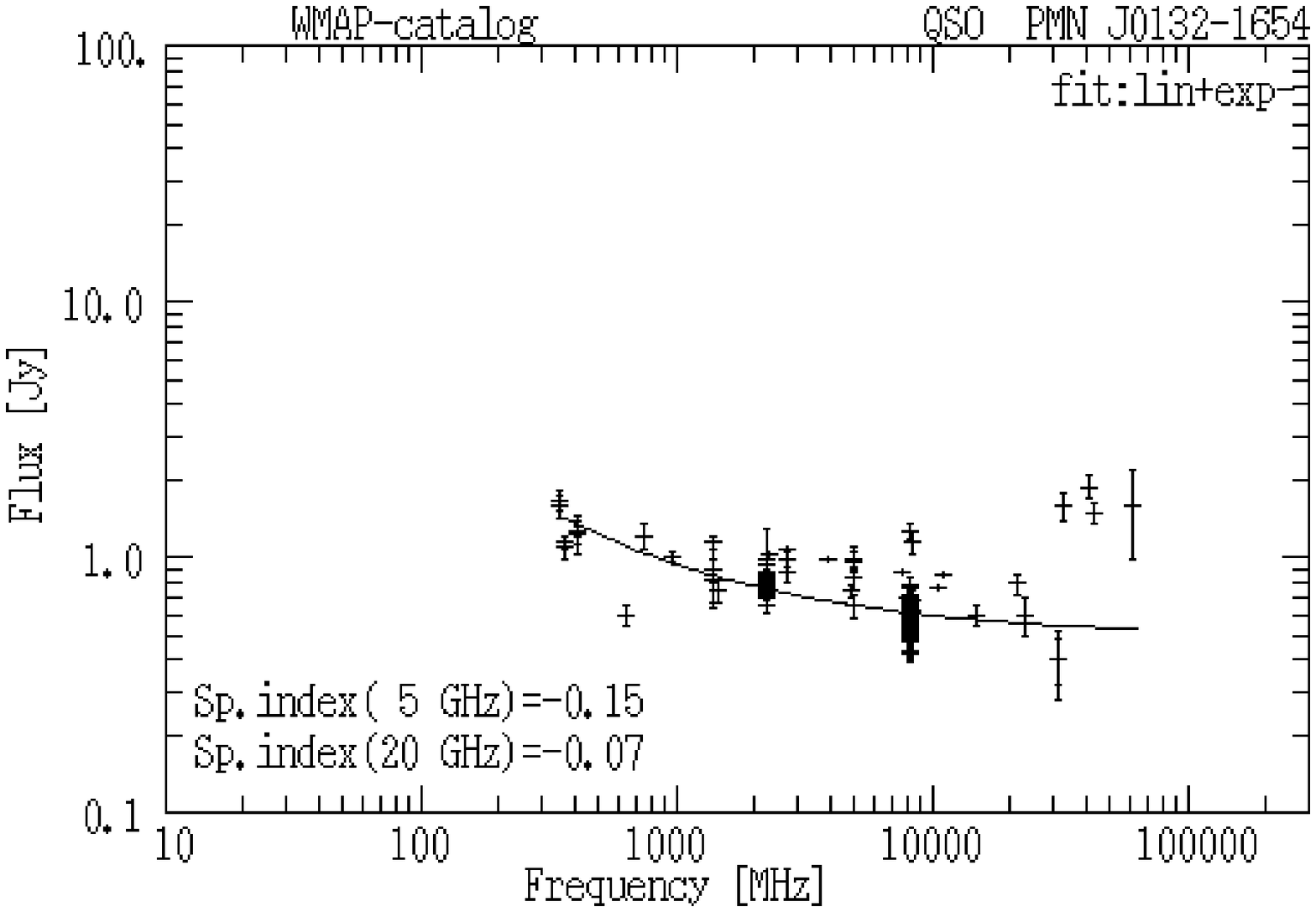,width=8.2cm,angle=0}}}\end{figure}
\begin{figure}\centerline{\vbox{\psfig{figure=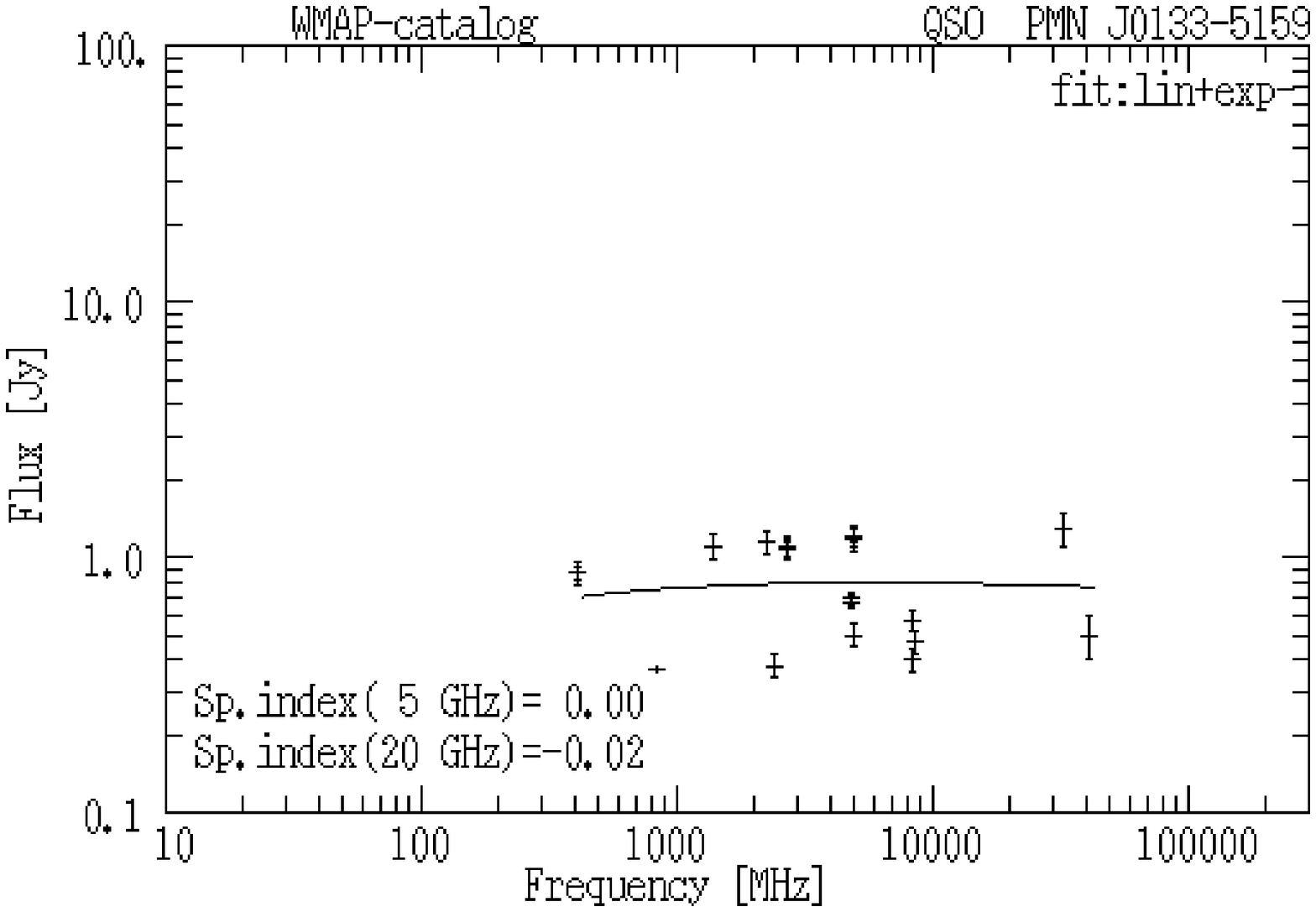,width=8.2cm,angle=0}}}\end{figure}
\begin{figure}\centerline{\vbox{\psfig{figure=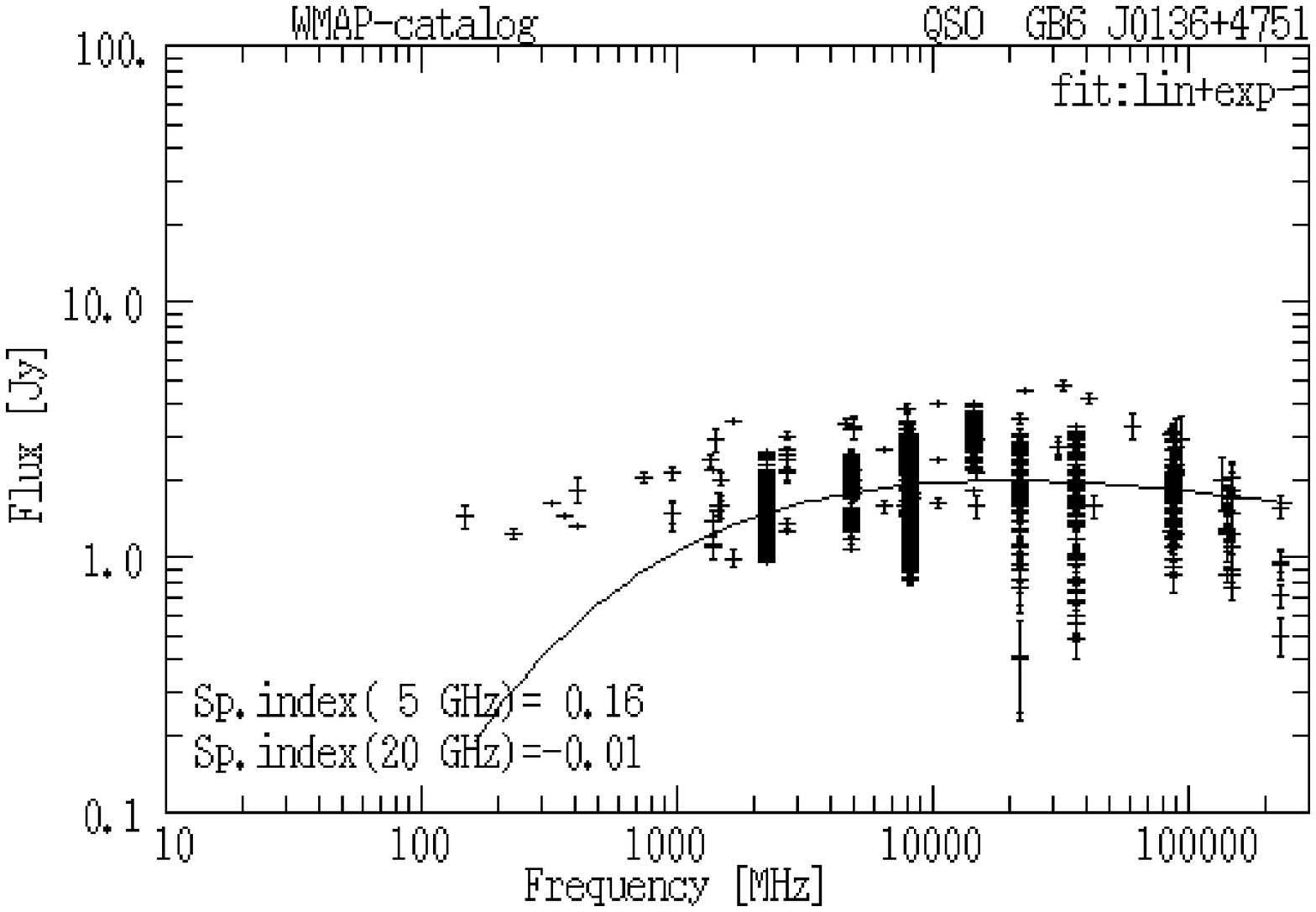,width=8.2cm,angle=0}}}\end{figure}
\begin{figure}\centerline{\vbox{\psfig{figure=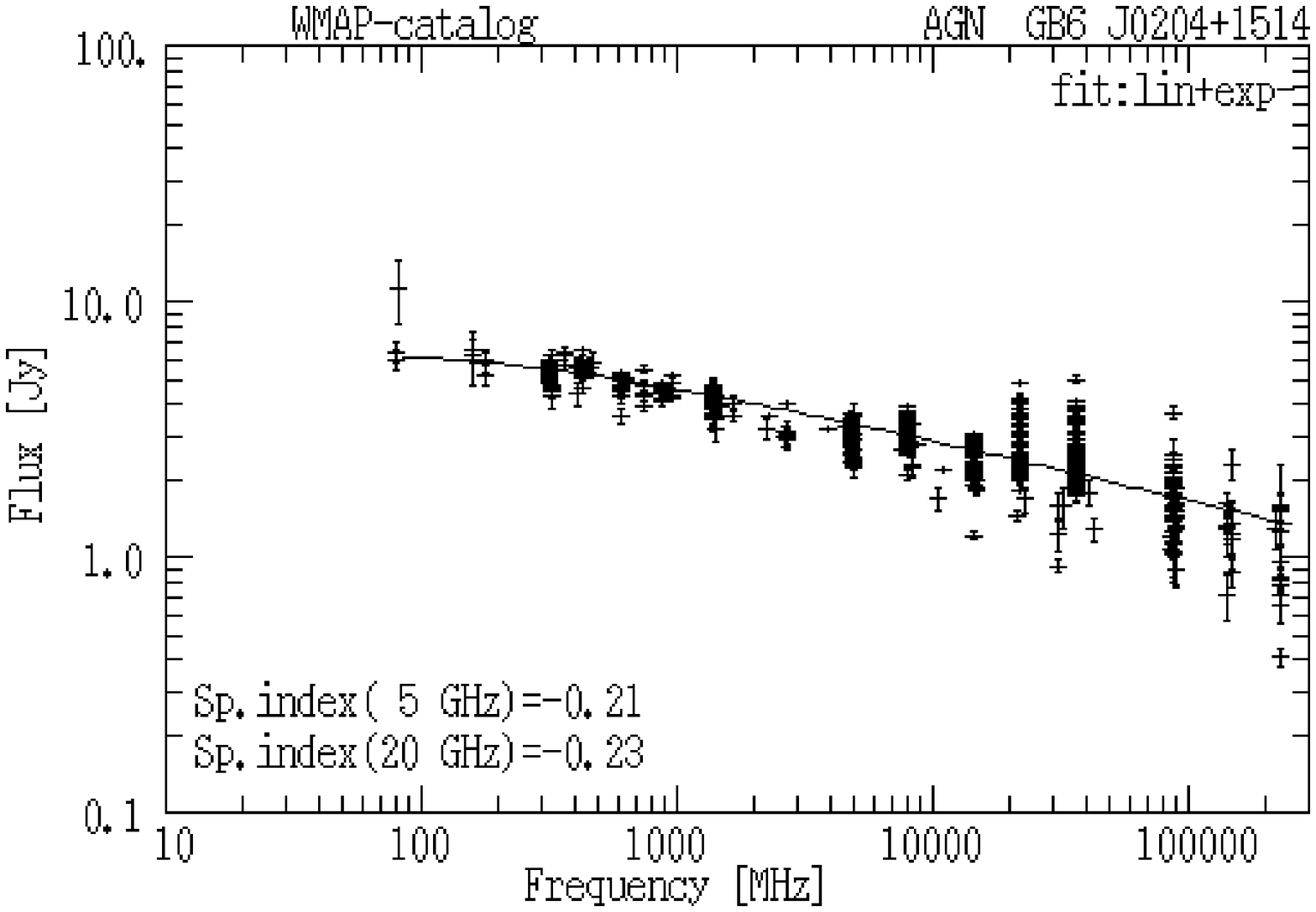,width=8.2cm,angle=0}}}\end{figure}
\begin{figure}\centerline{\vbox{\psfig{figure=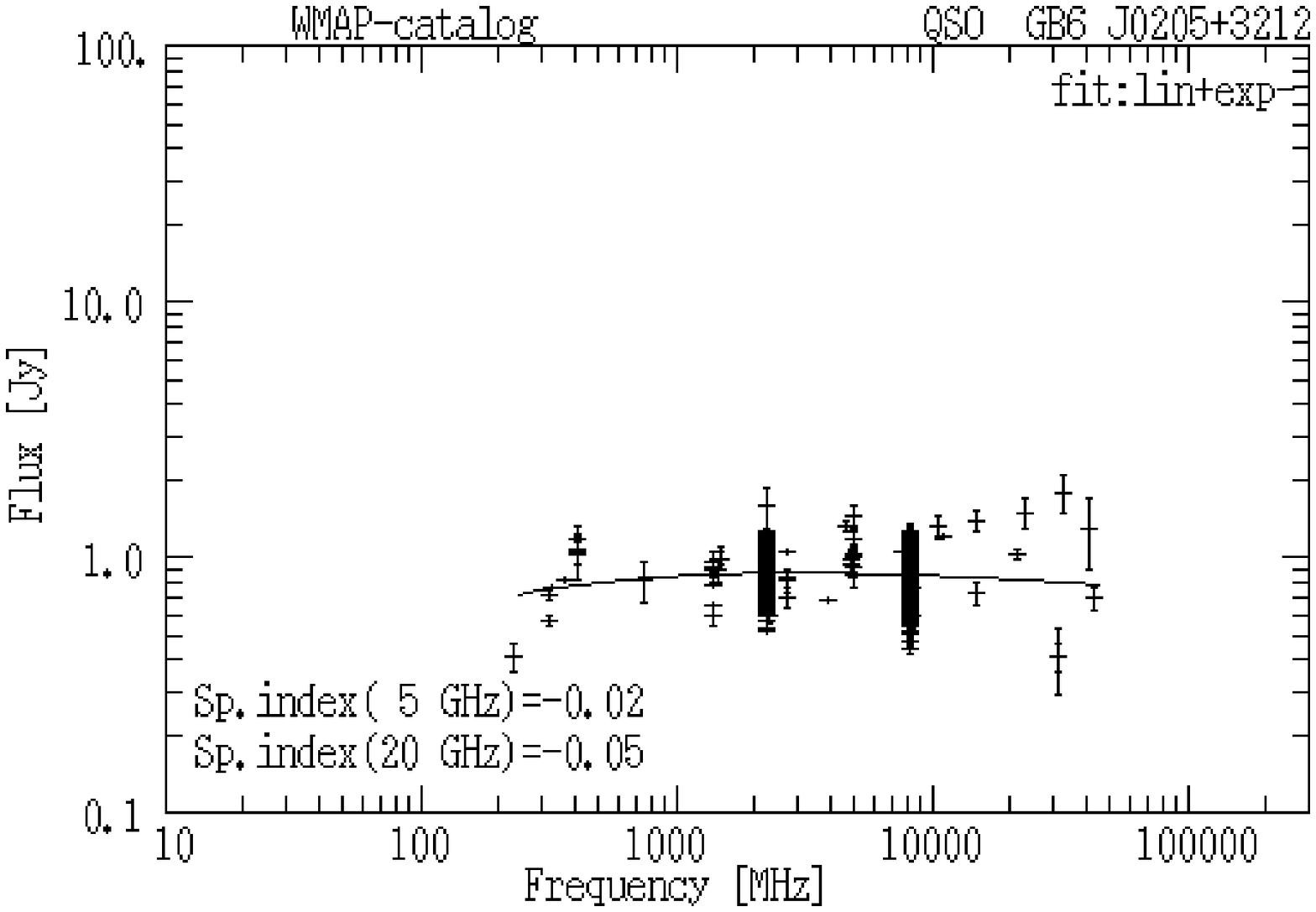,width=8.2cm,angle=0}}}\end{figure}
\begin{figure}\centerline{\vbox{\psfig{figure=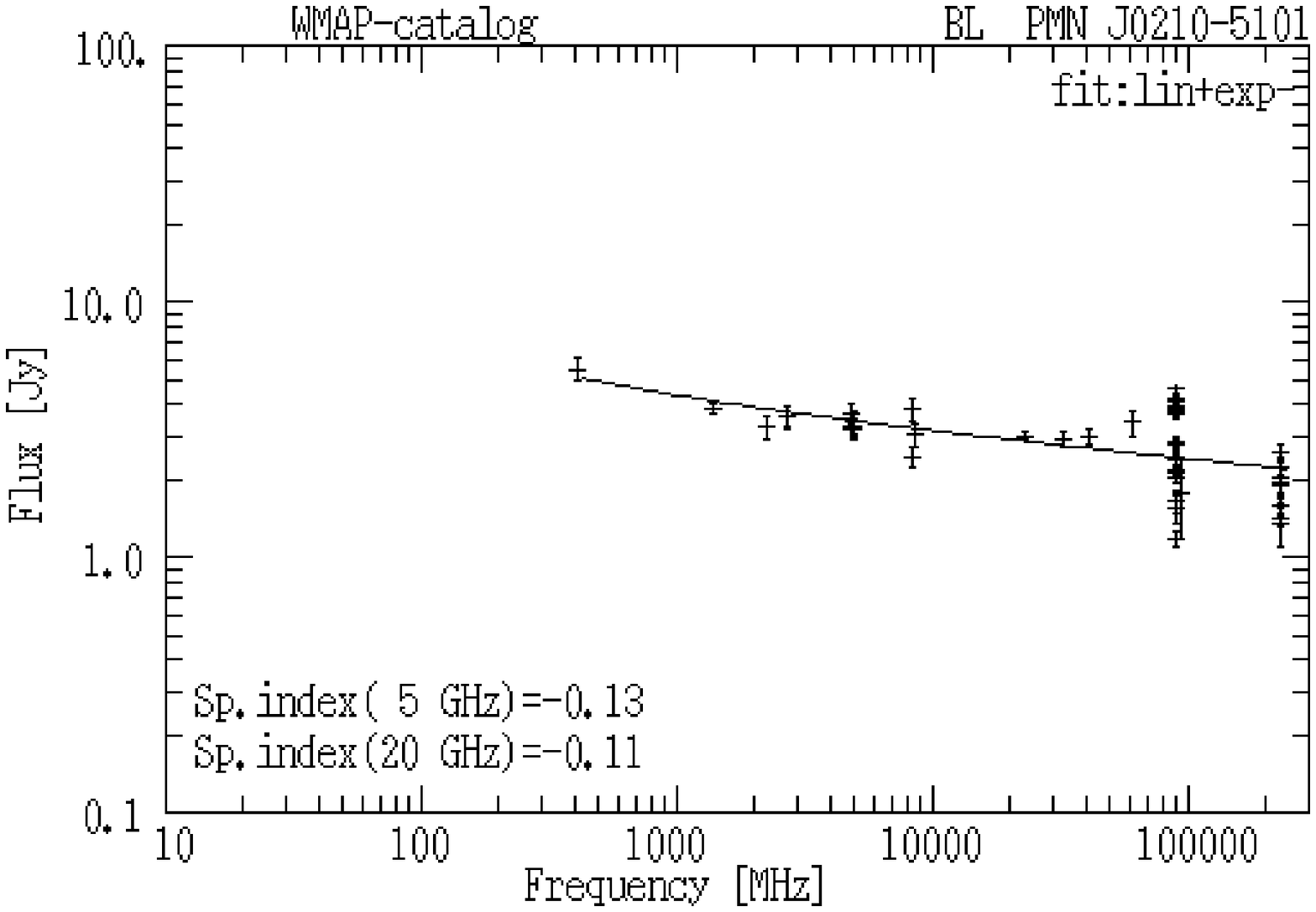,width=8.2cm,angle=0}}}\end{figure}
\begin{figure}\centerline{\vbox{\psfig{figure=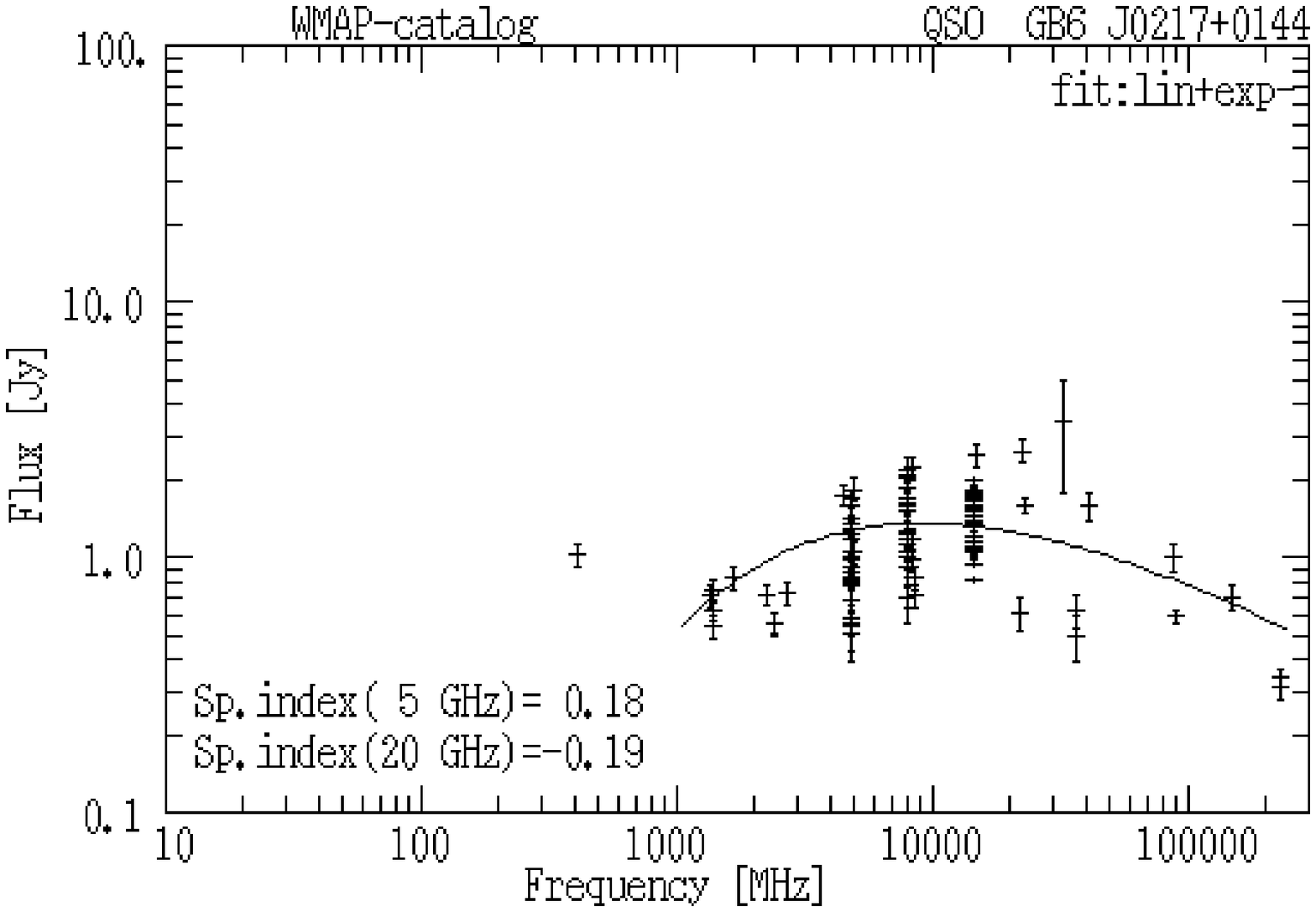,width=8.2cm,angle=0}}}\end{figure}\clearpage
\begin{figure}\centerline{\vbox{\psfig{figure=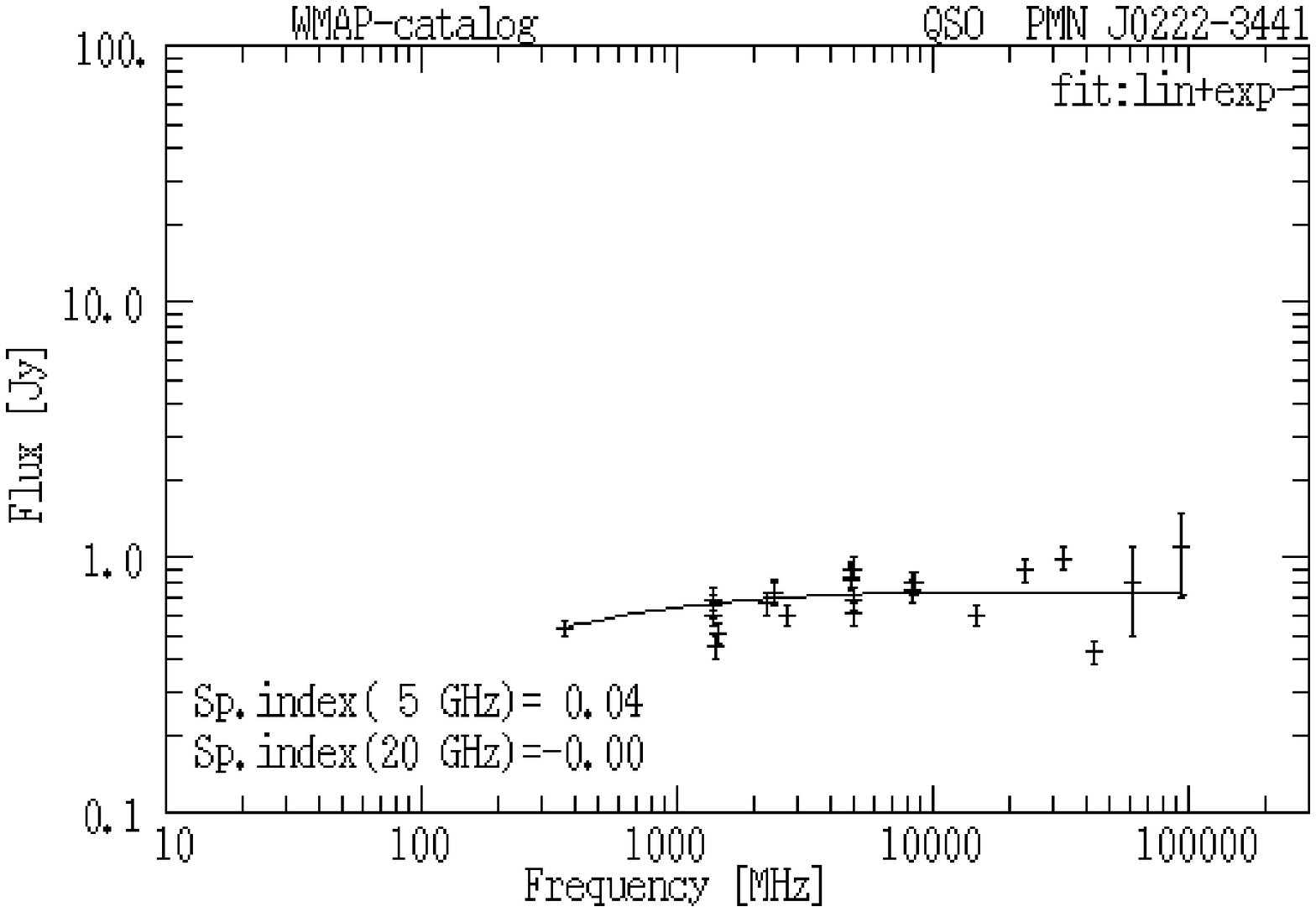,width=8.2cm,angle=0}}}\end{figure}
\begin{figure}\centerline{\vbox{\psfig{figure=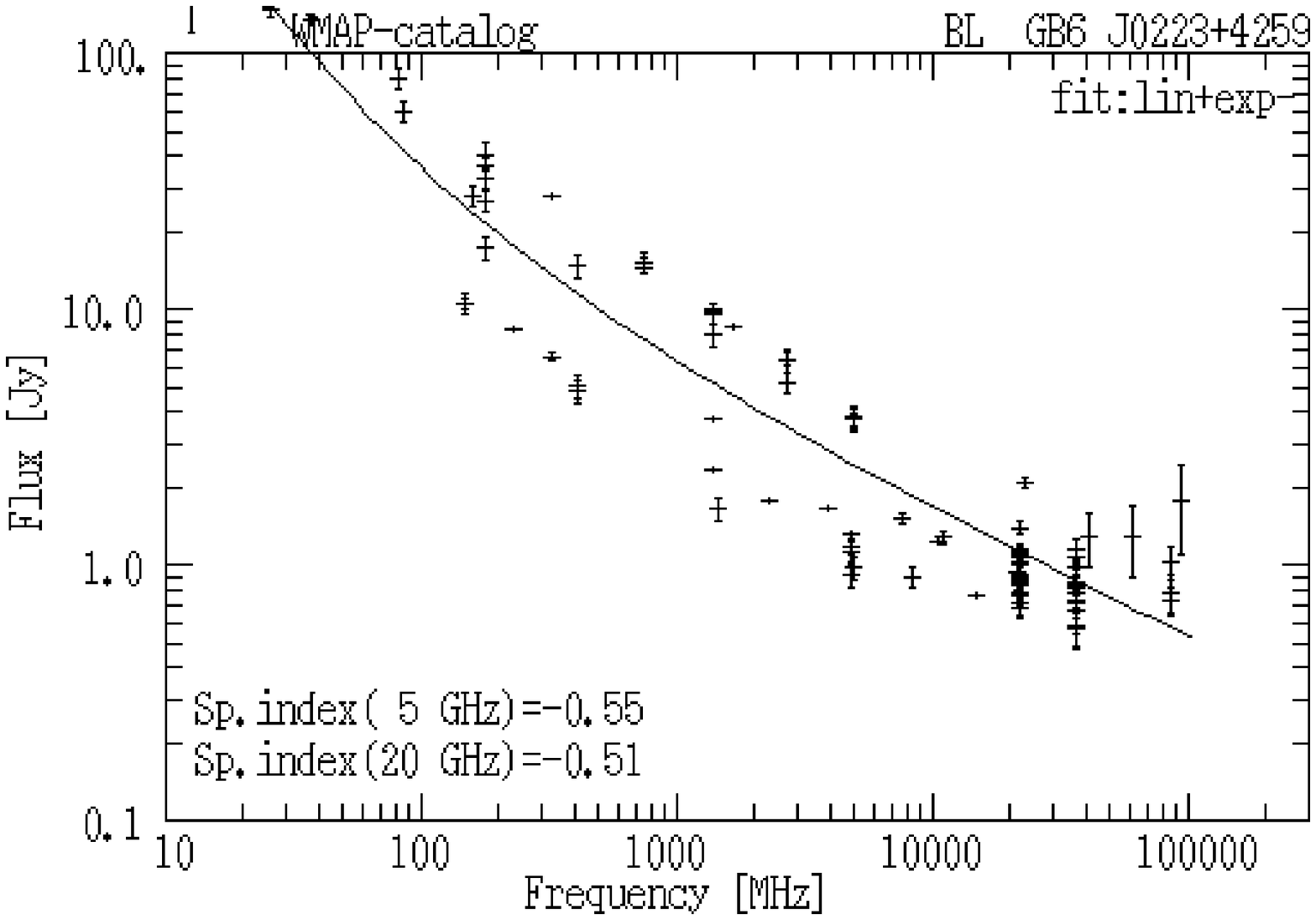,width=8.2cm,angle=0}}}\end{figure}
\begin{figure}\centerline{\vbox{\psfig{figure=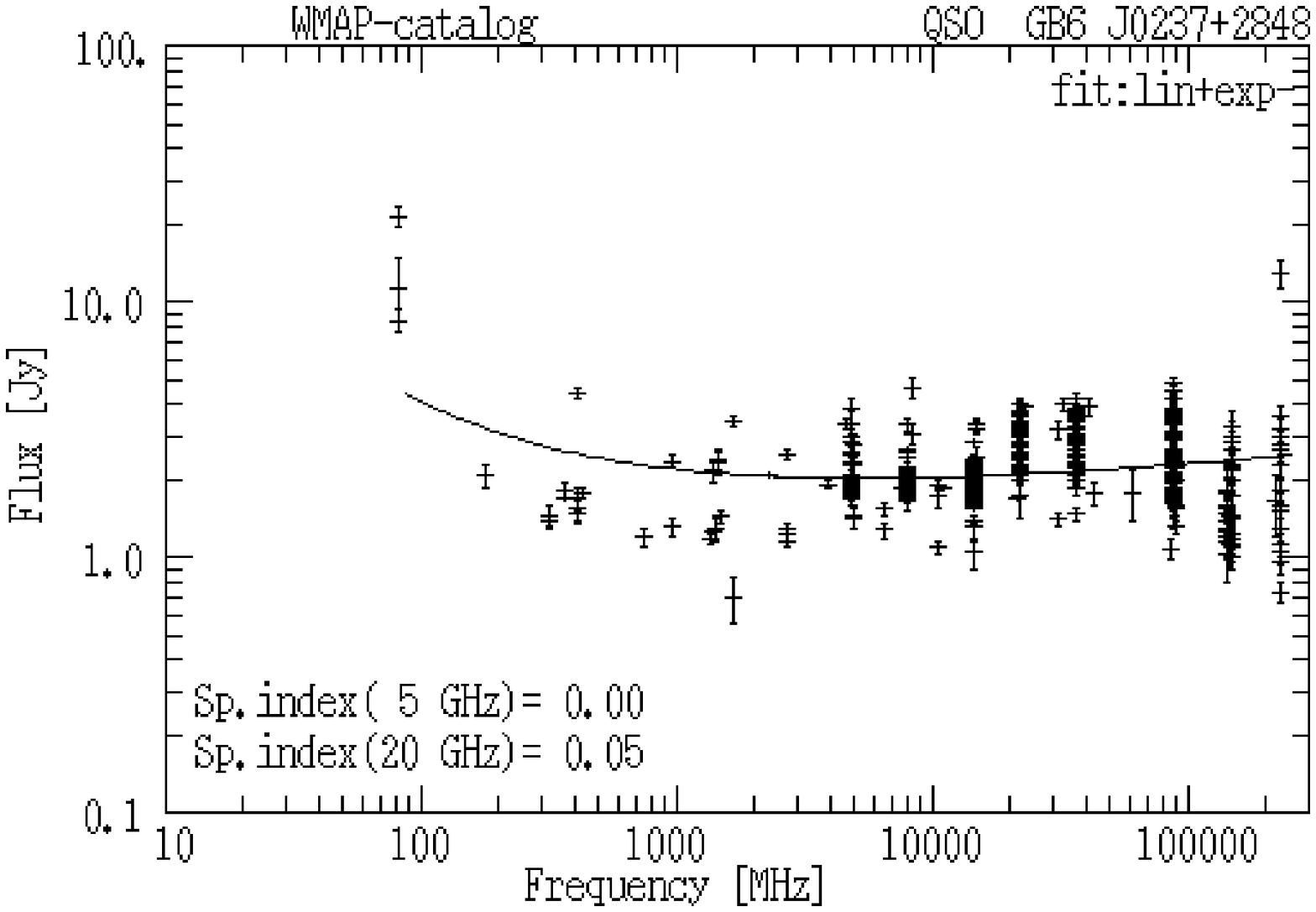,width=8.2cm,angle=0}}}\end{figure}
\begin{figure}\centerline{\vbox{\psfig{figure=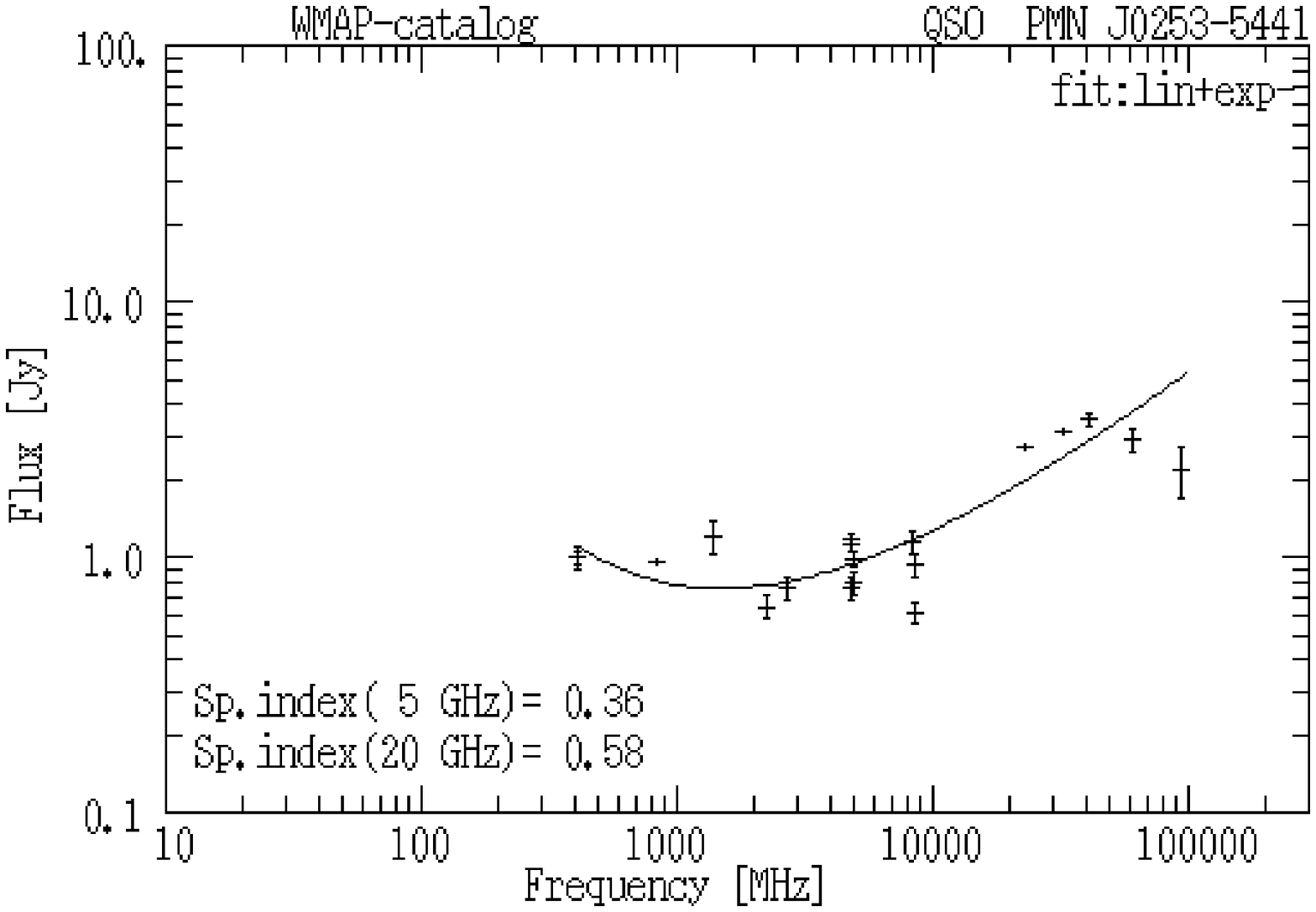,width=8.2cm,angle=0}}}\end{figure}
\begin{figure}\centerline{\vbox{\psfig{figure=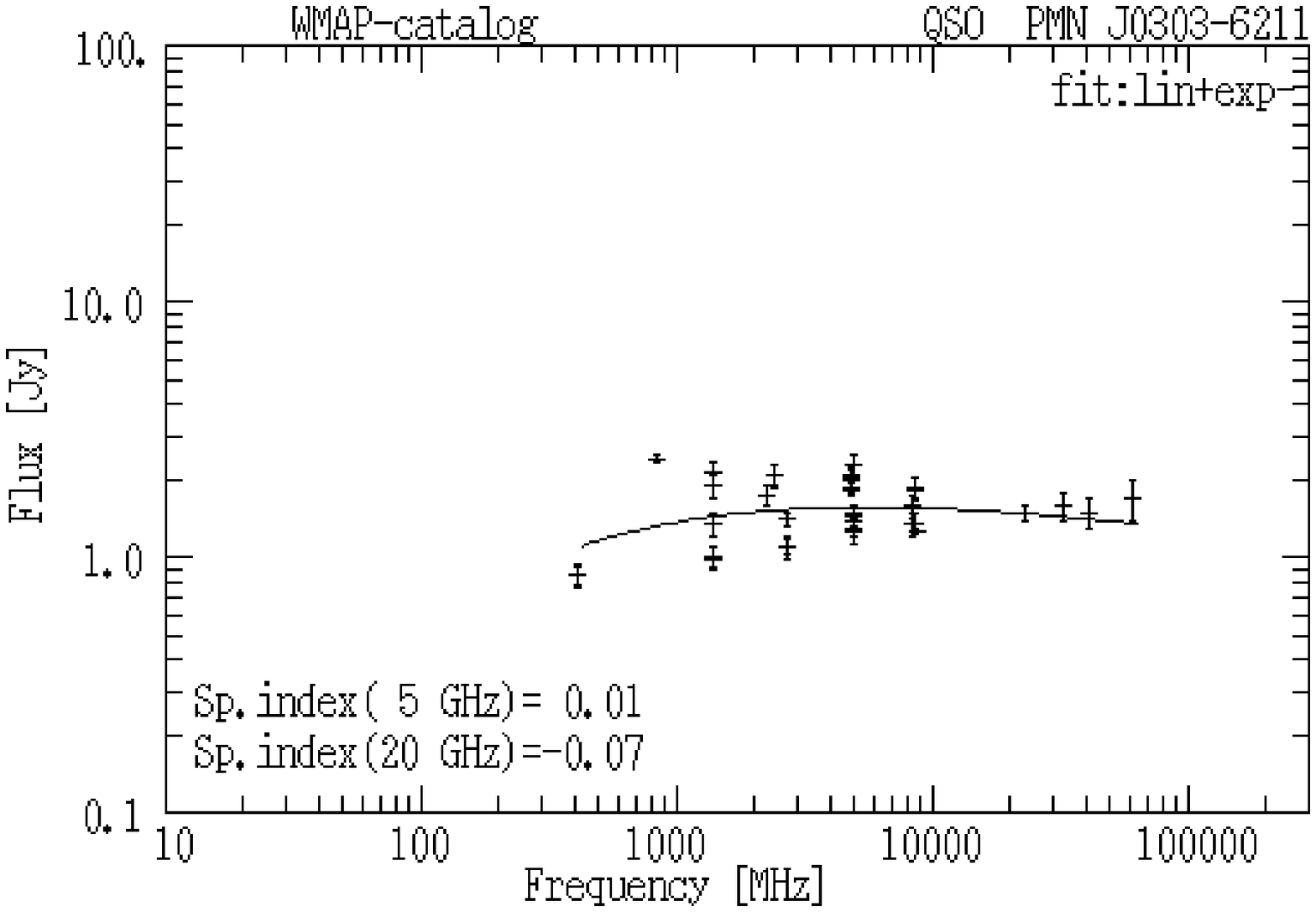,width=8.2cm,angle=0}}}\end{figure}
\begin{figure}\centerline{\vbox{\psfig{figure=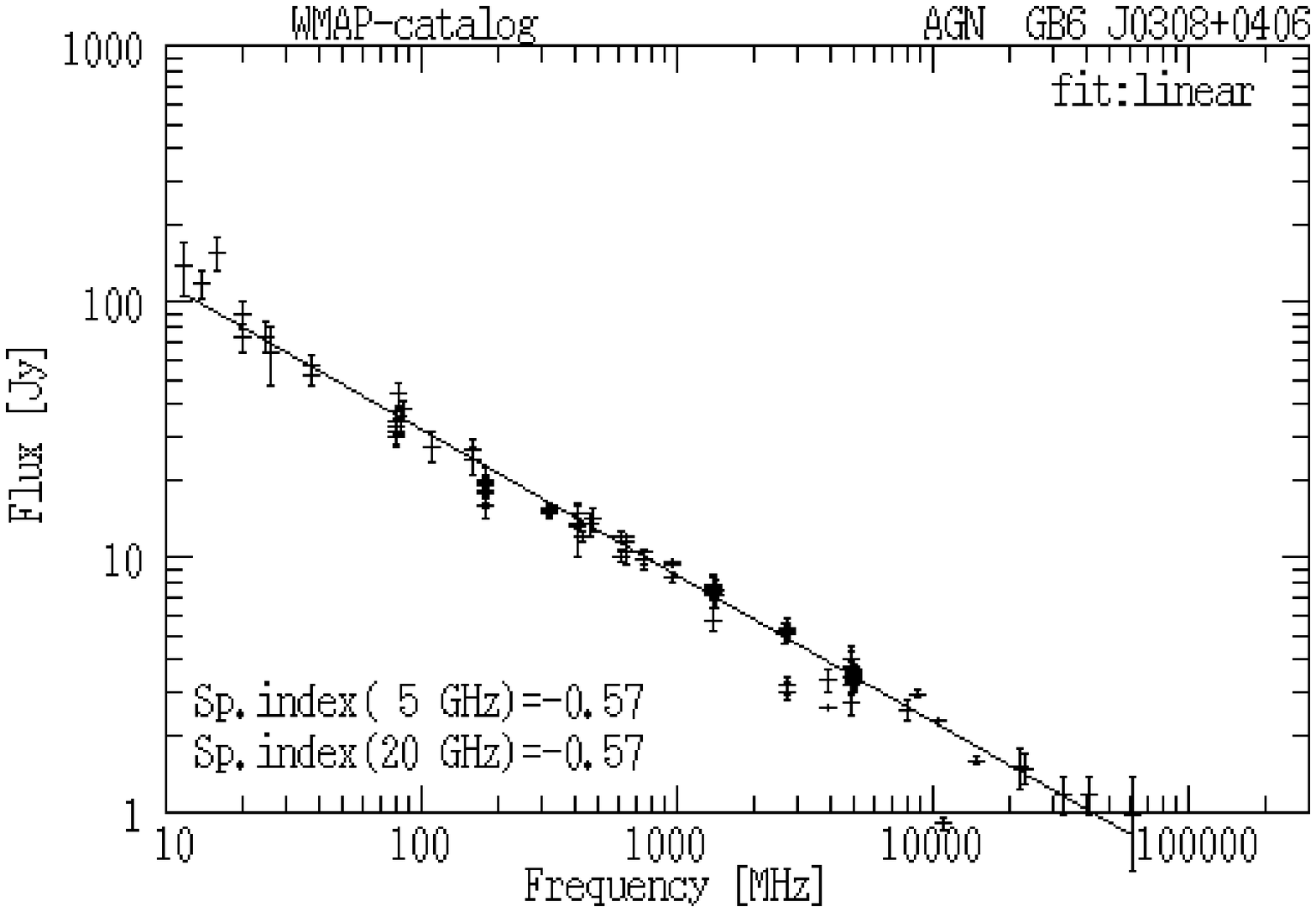,width=8.2cm,angle=0}}}\end{figure}
\begin{figure}\centerline{\vbox{\psfig{figure=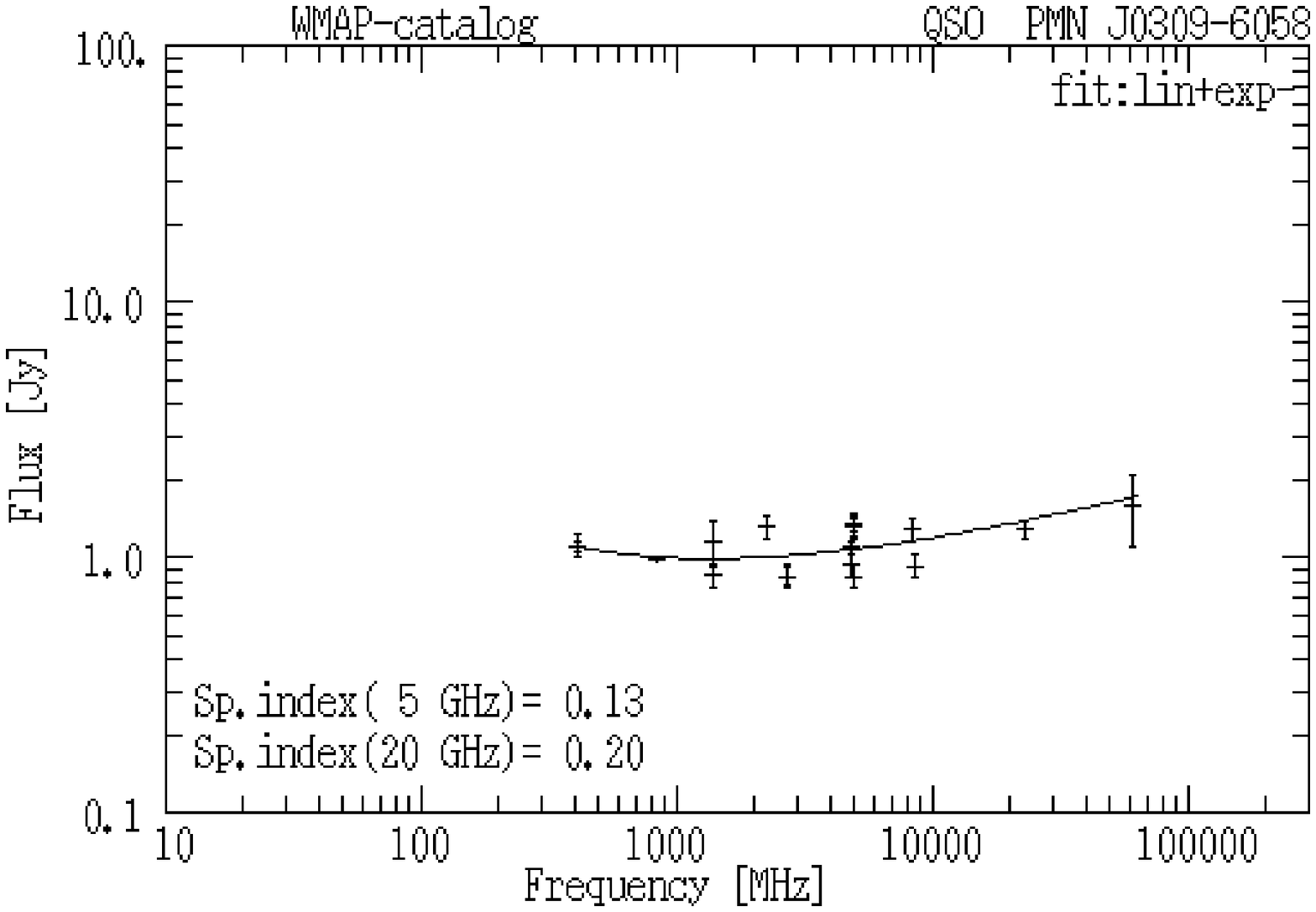,width=8.2cm,angle=0}}}\end{figure}
\begin{figure}\centerline{\vbox{\psfig{figure=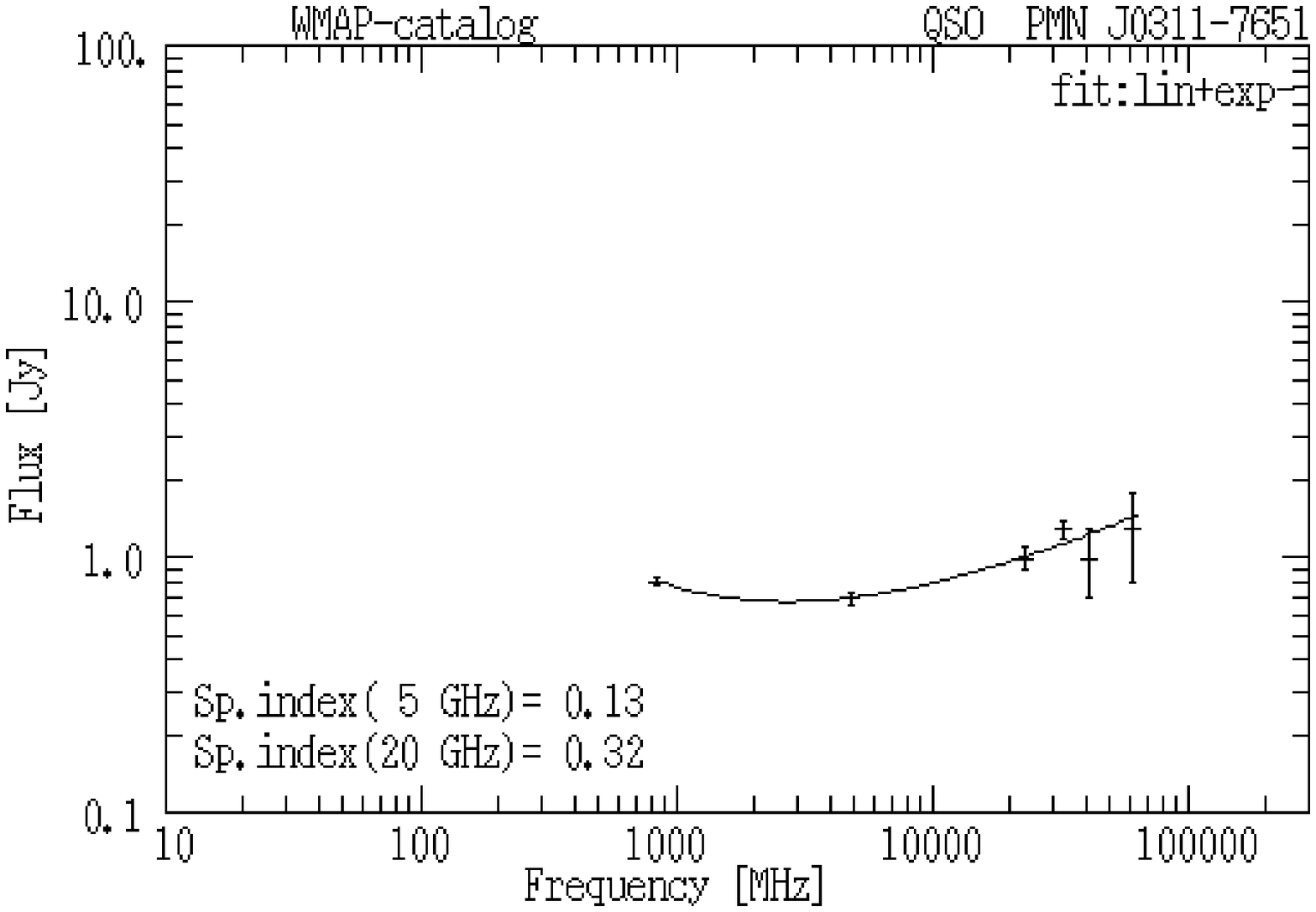,width=8.2cm,angle=0}}}\end{figure}\clearpage
\begin{figure}\centerline{\vbox{\psfig{figure=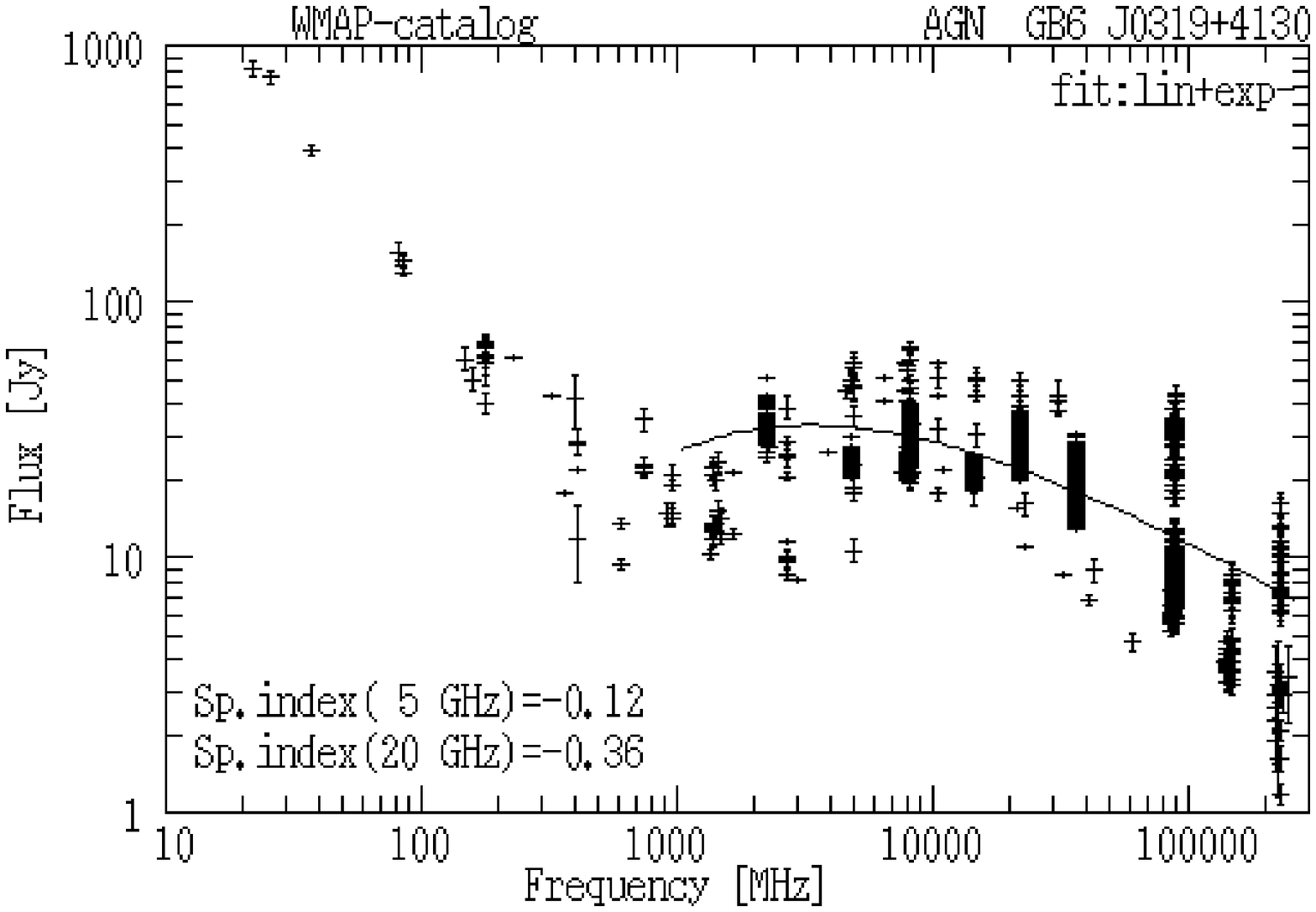,width=8.2cm,angle=0}}}\end{figure}
\begin{figure}\centerline{\vbox{\psfig{figure=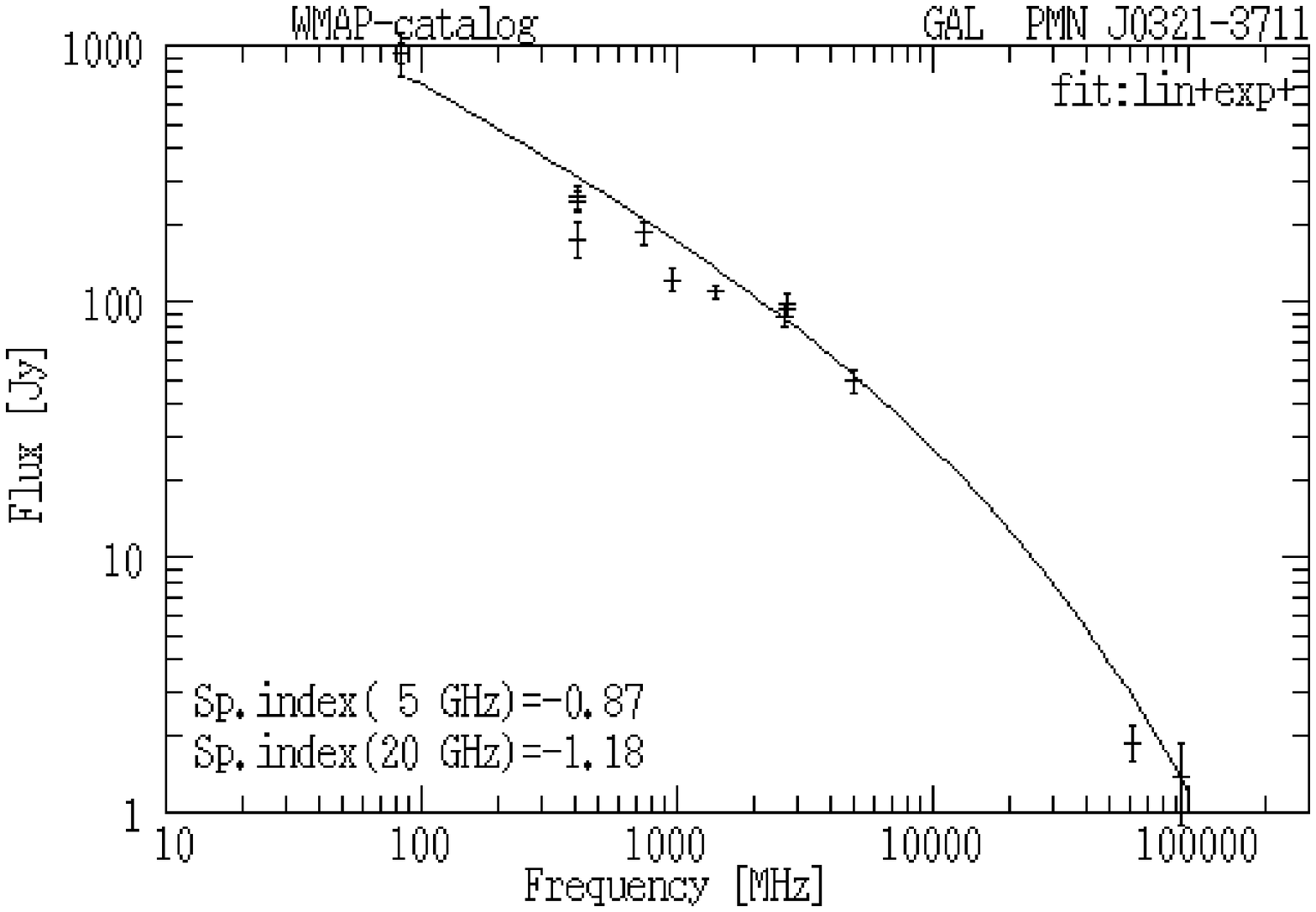,width=8.2cm,angle=0}}}\end{figure}
\begin{figure}\centerline{\vbox{\psfig{figure=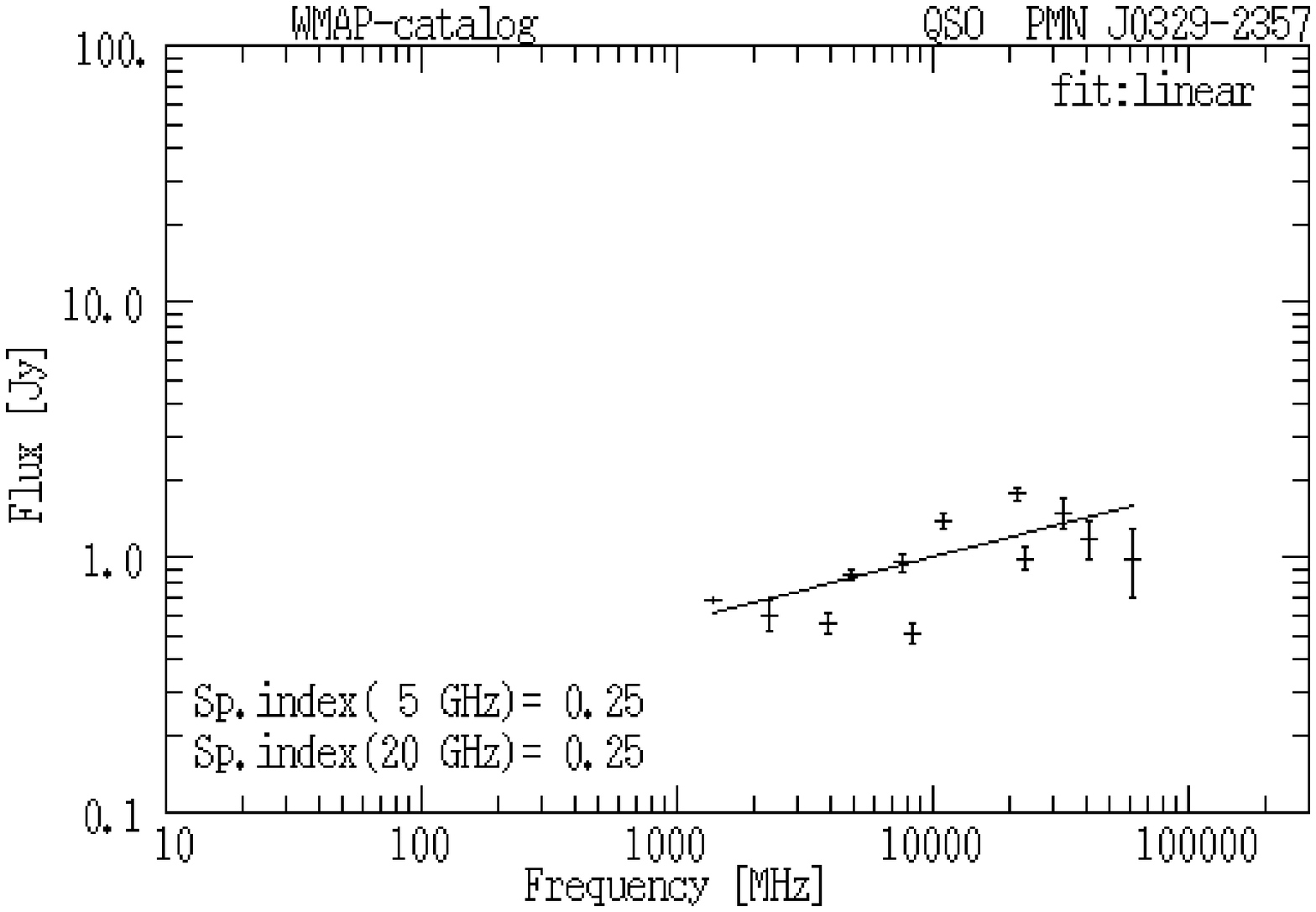,width=8.2cm,angle=0}}}\end{figure}
\begin{figure}\centerline{\vbox{\psfig{figure=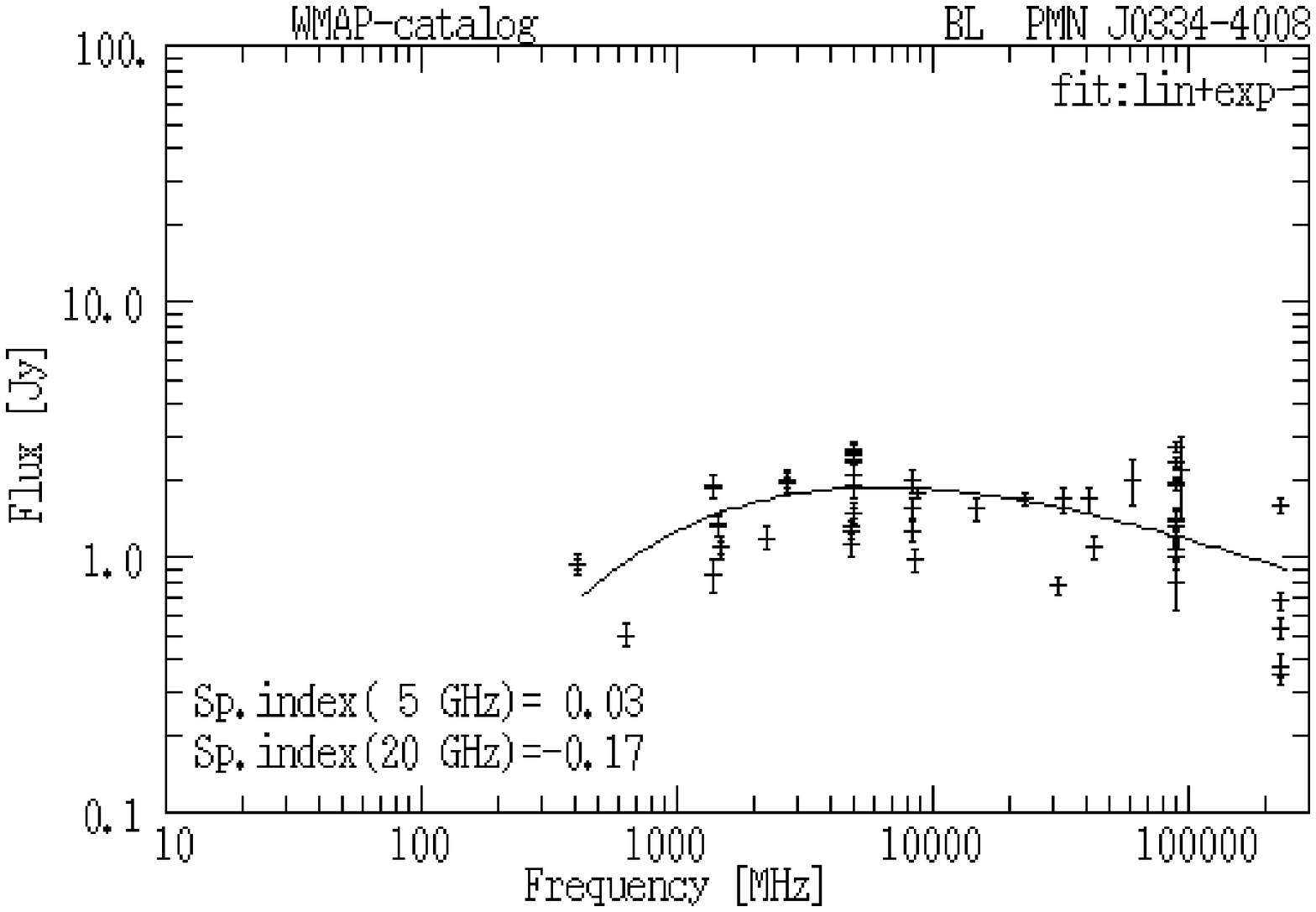,width=8.2cm,angle=0}}}\end{figure}
\begin{figure}\centerline{\vbox{\psfig{figure=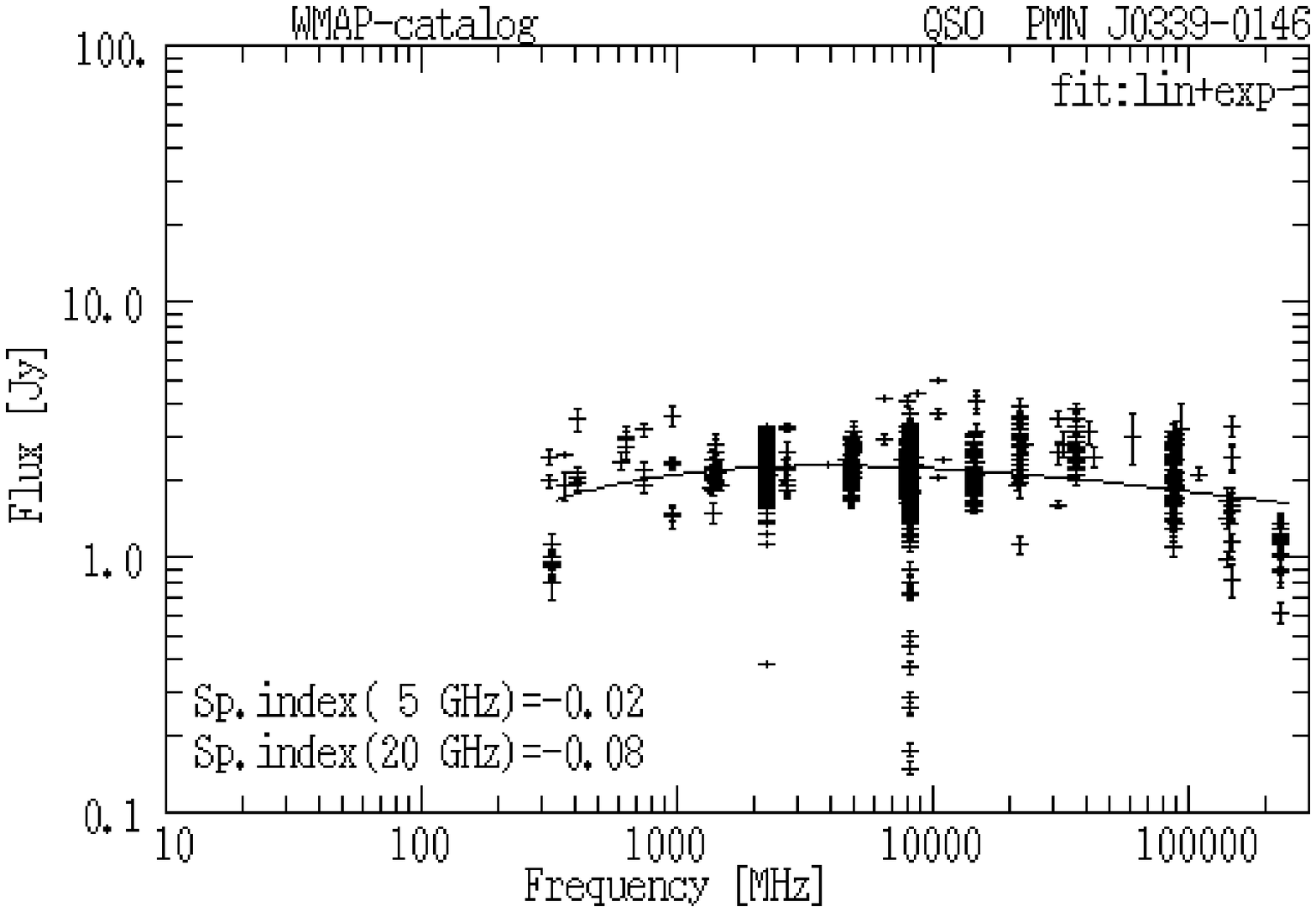,width=8.2cm,angle=0}}}\end{figure}
\begin{figure}\centerline{\vbox{\psfig{figure=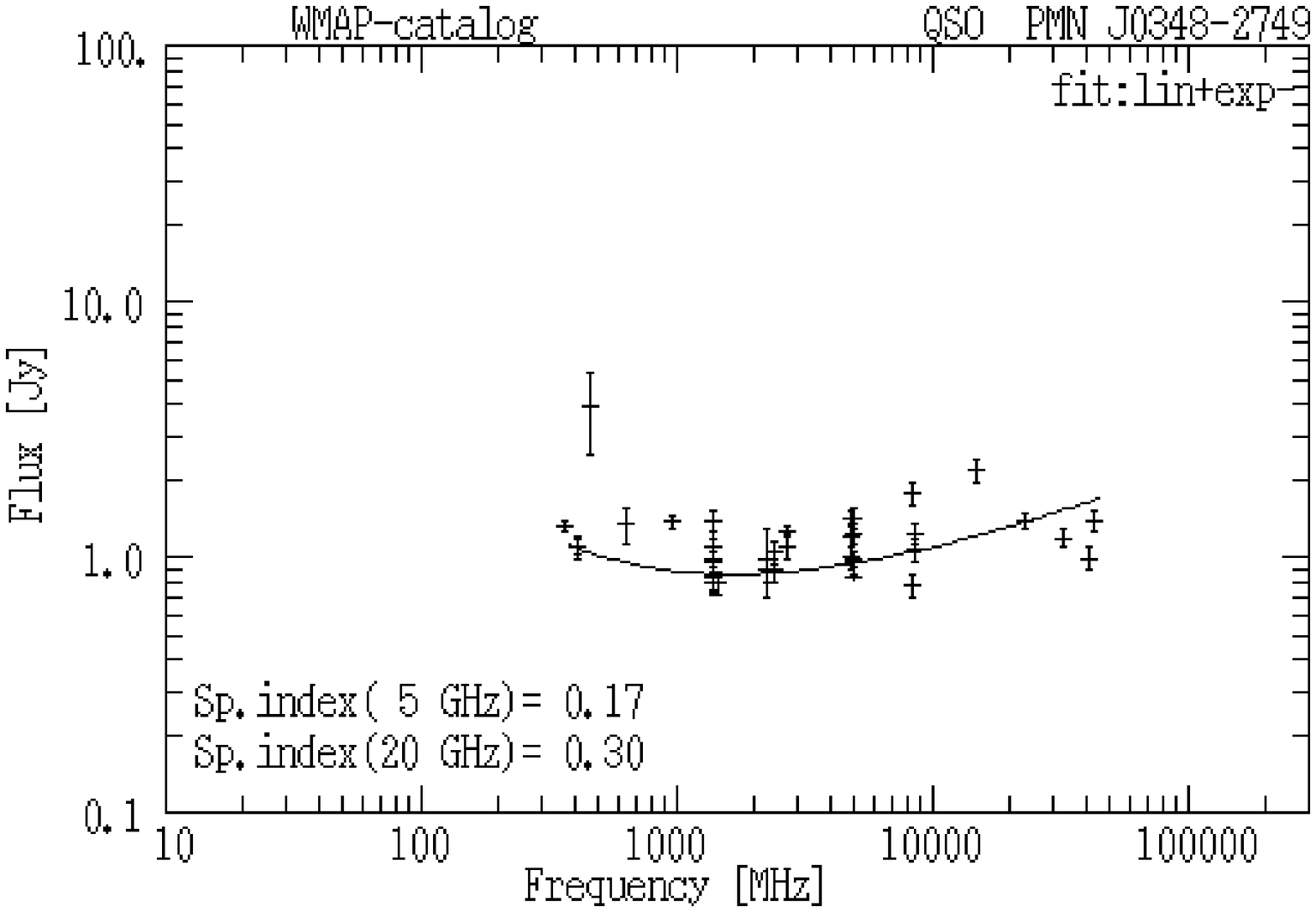,width=8.2cm,angle=0}}}\end{figure}
\begin{figure}\centerline{\vbox{\psfig{figure=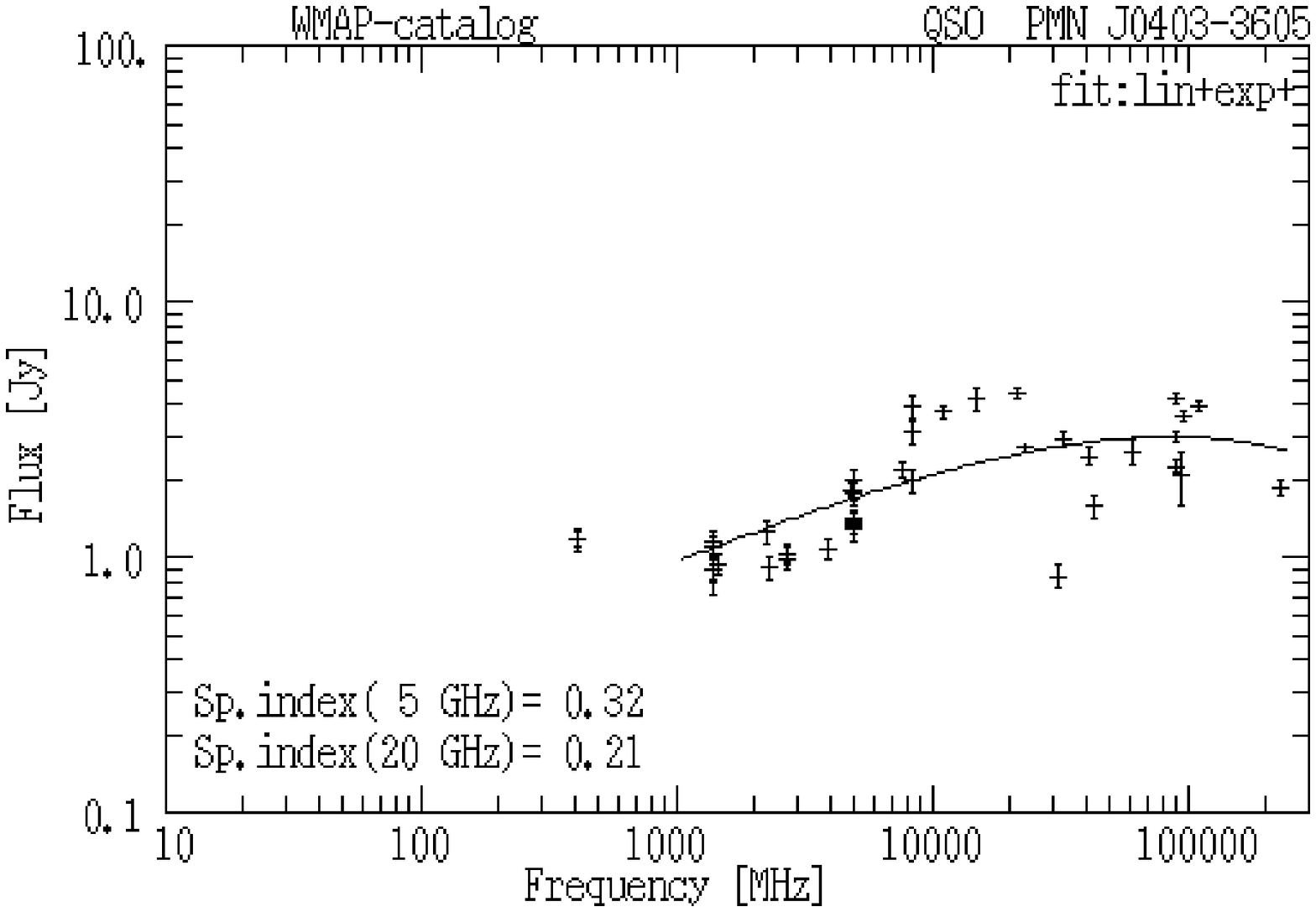,width=8.2cm,angle=0}}}\end{figure}
\begin{figure}\centerline{\vbox{\psfig{figure=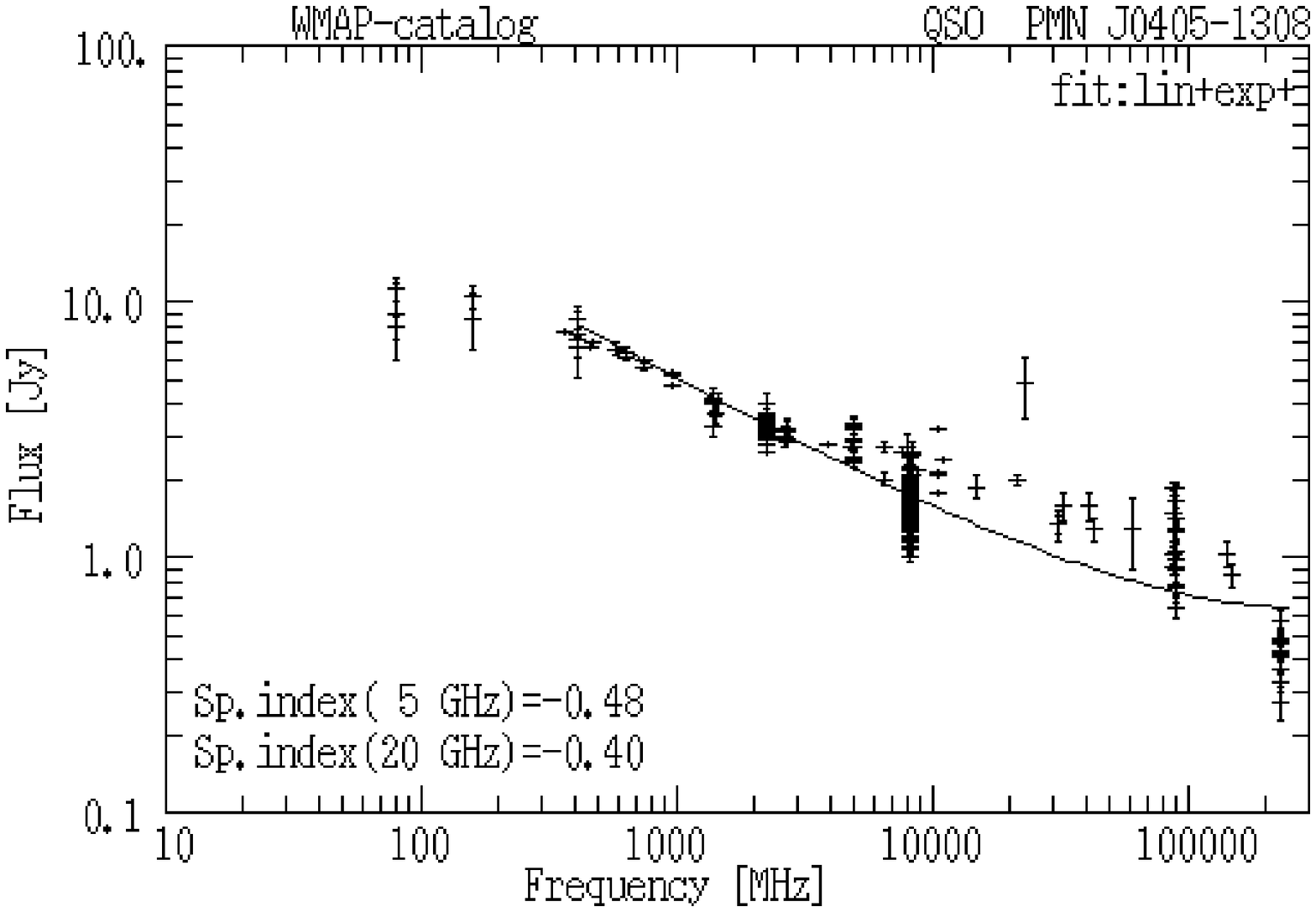,width=8.2cm,angle=0}}}\end{figure}\clearpage
\begin{figure}\centerline{\vbox{\psfig{figure=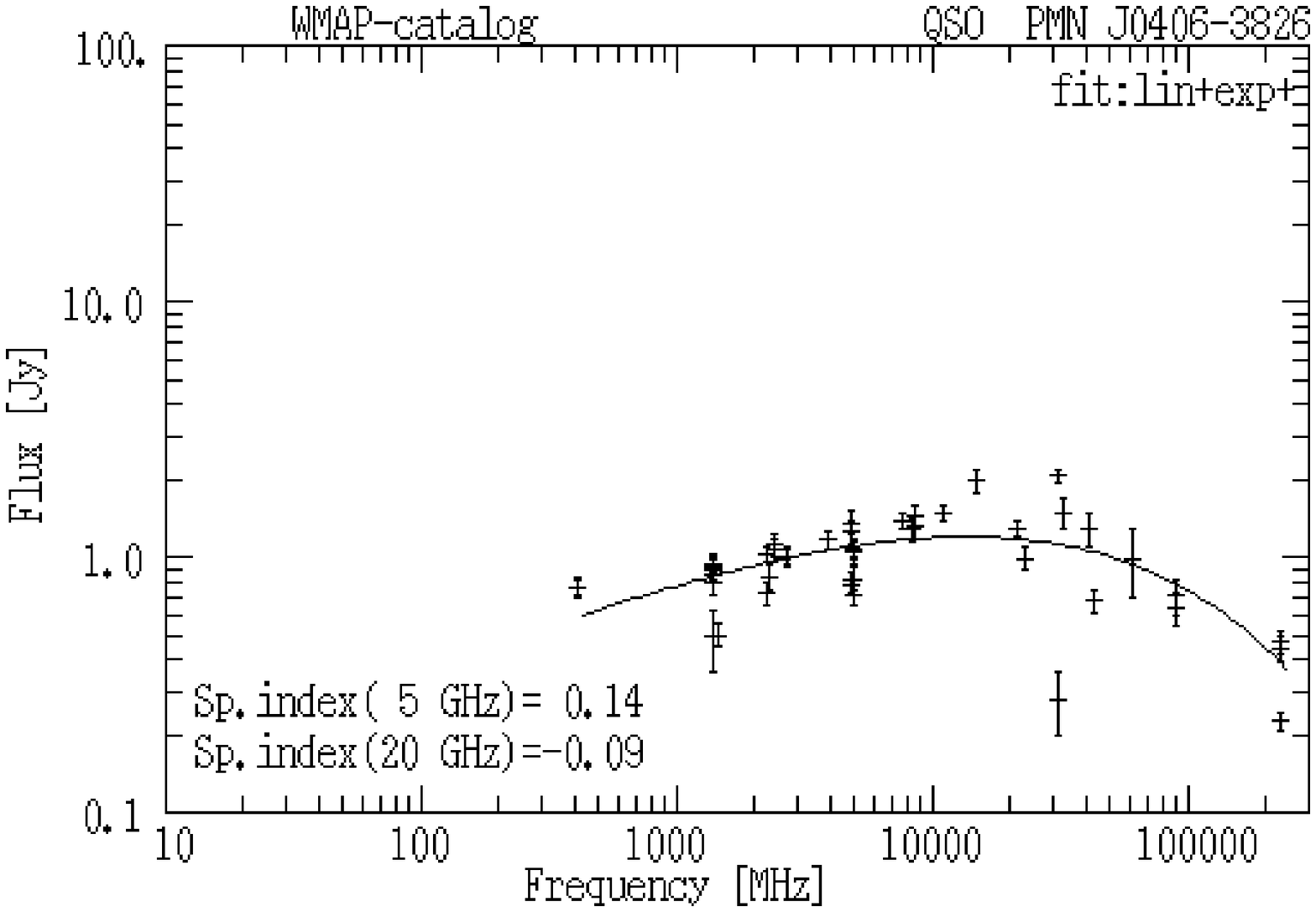,width=8.2cm,angle=0}}}\end{figure}
\begin{figure}\centerline{\vbox{\psfig{figure=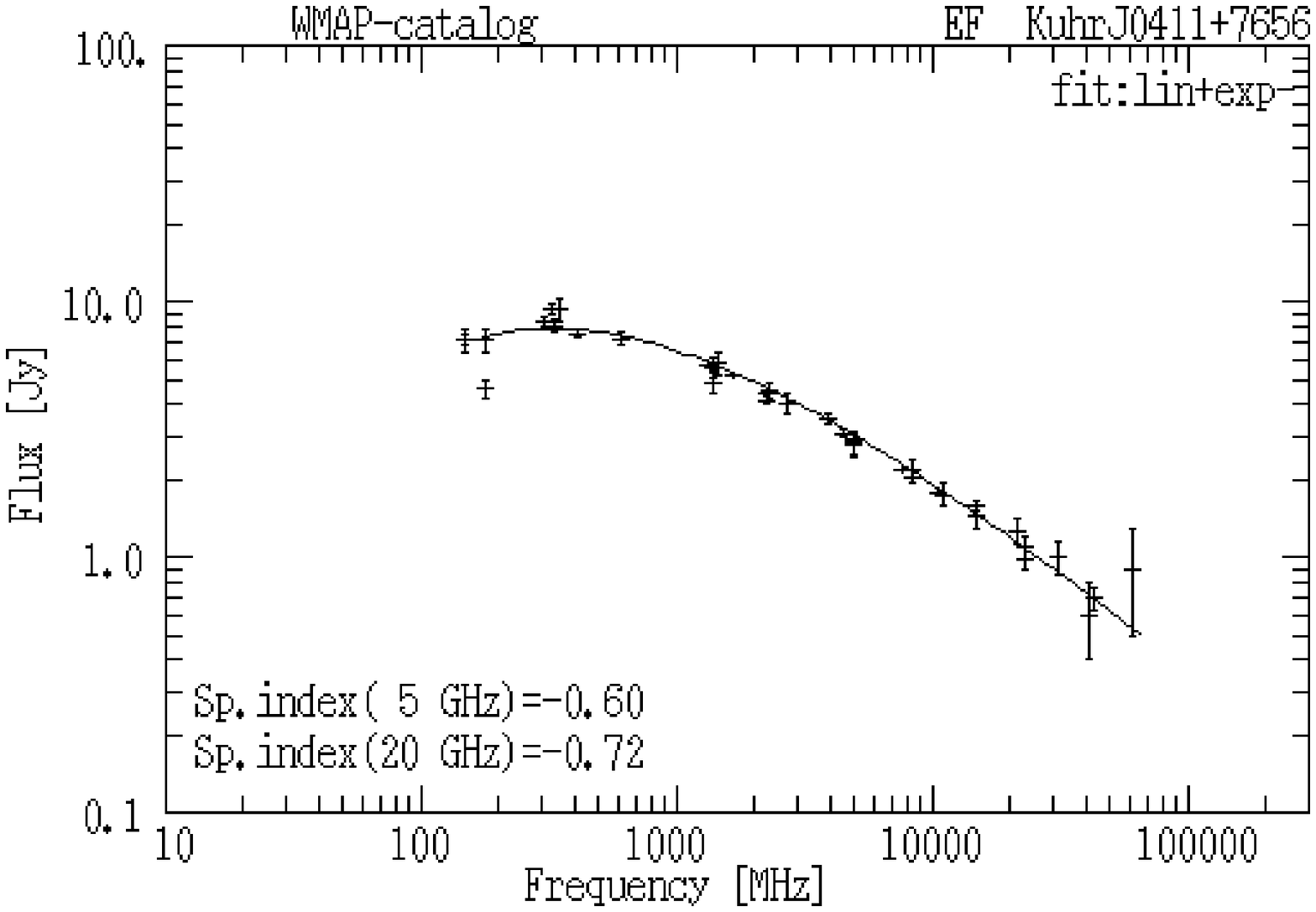,width=8.2cm,angle=0}}}\end{figure}
\begin{figure}\centerline{\vbox{\psfig{figure=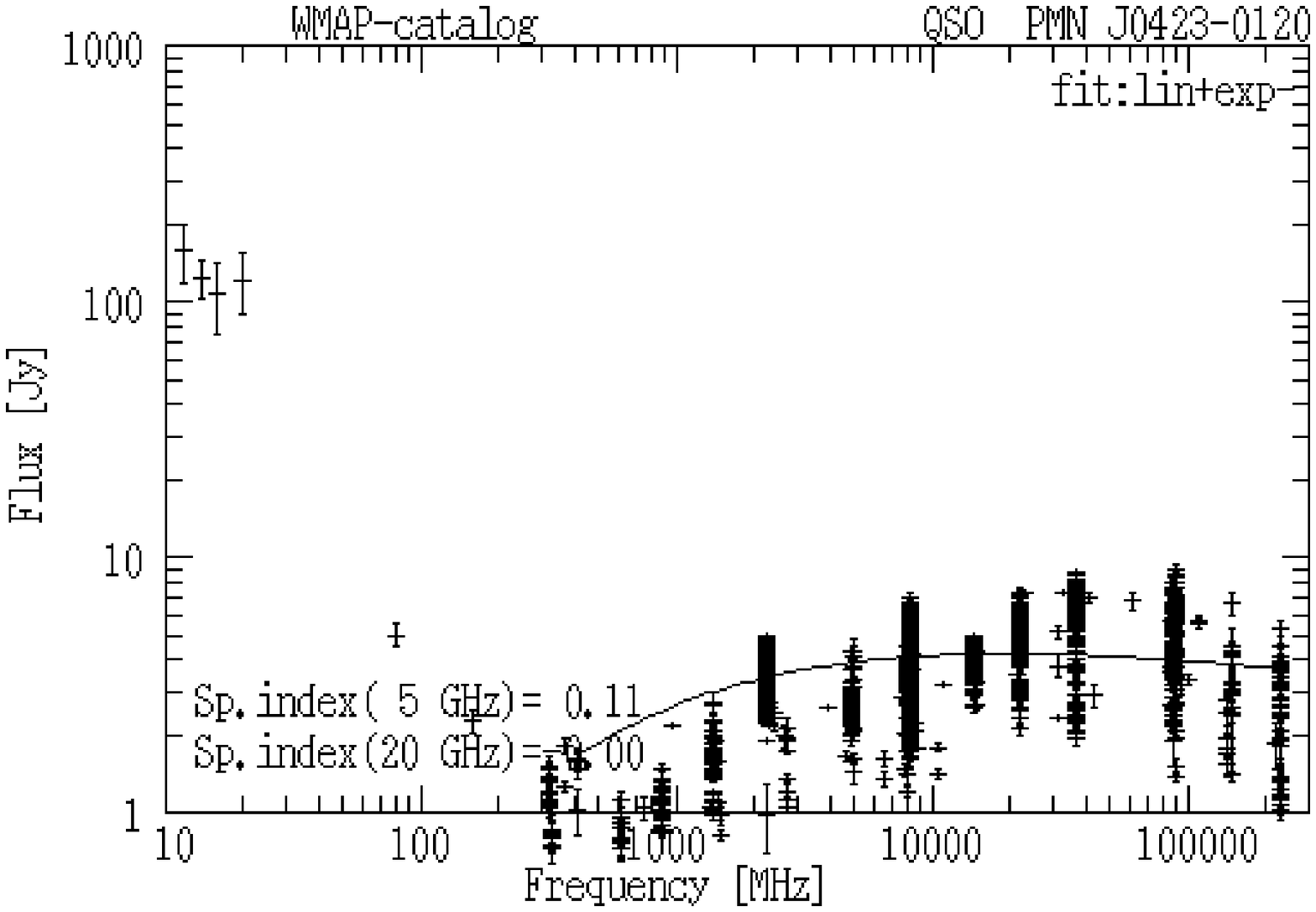,width=8.2cm,angle=0}}}\end{figure}
\begin{figure}\centerline{\vbox{\psfig{figure=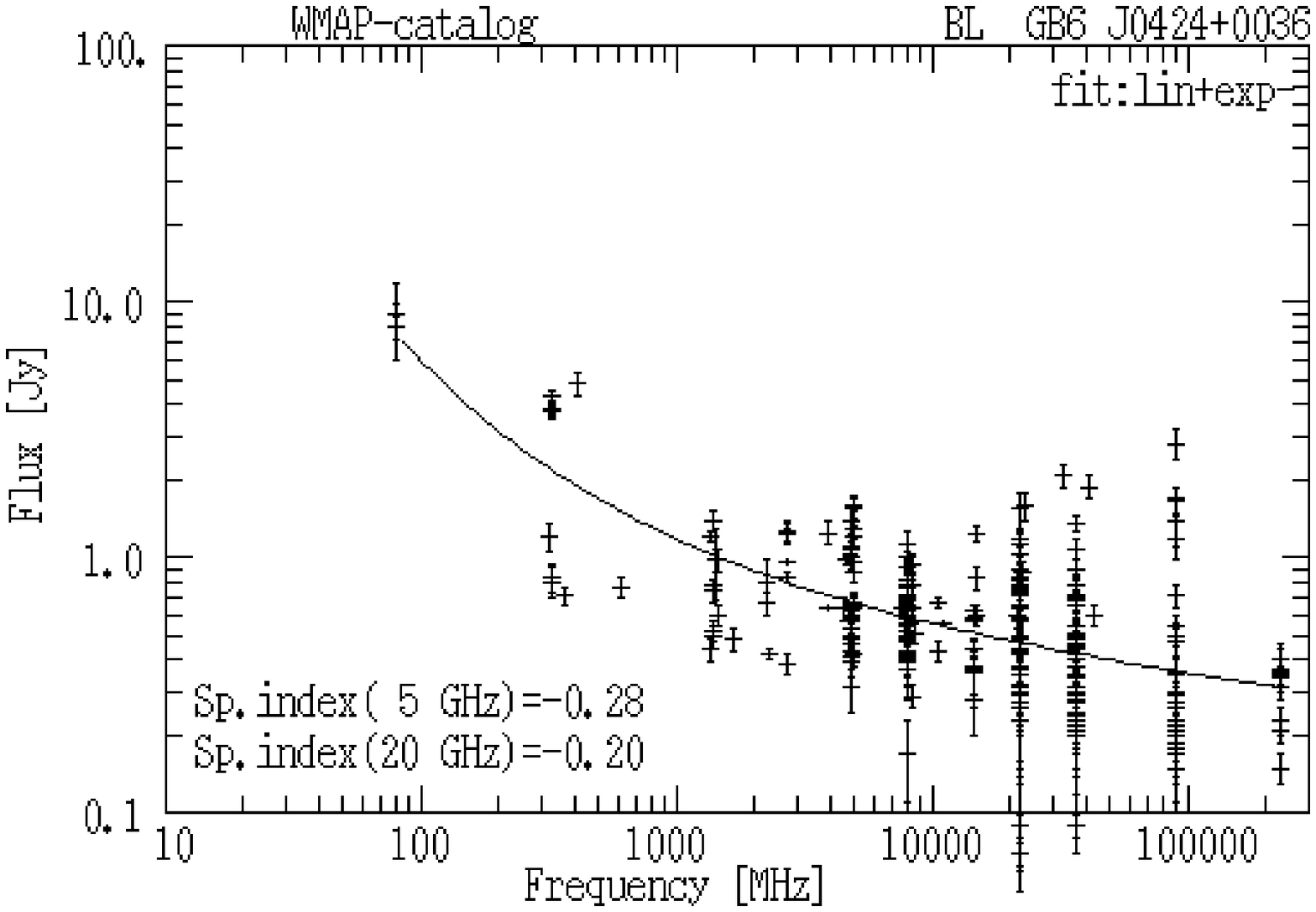,width=8.2cm,angle=0}}}\end{figure}
\begin{figure}\centerline{\vbox{\psfig{figure=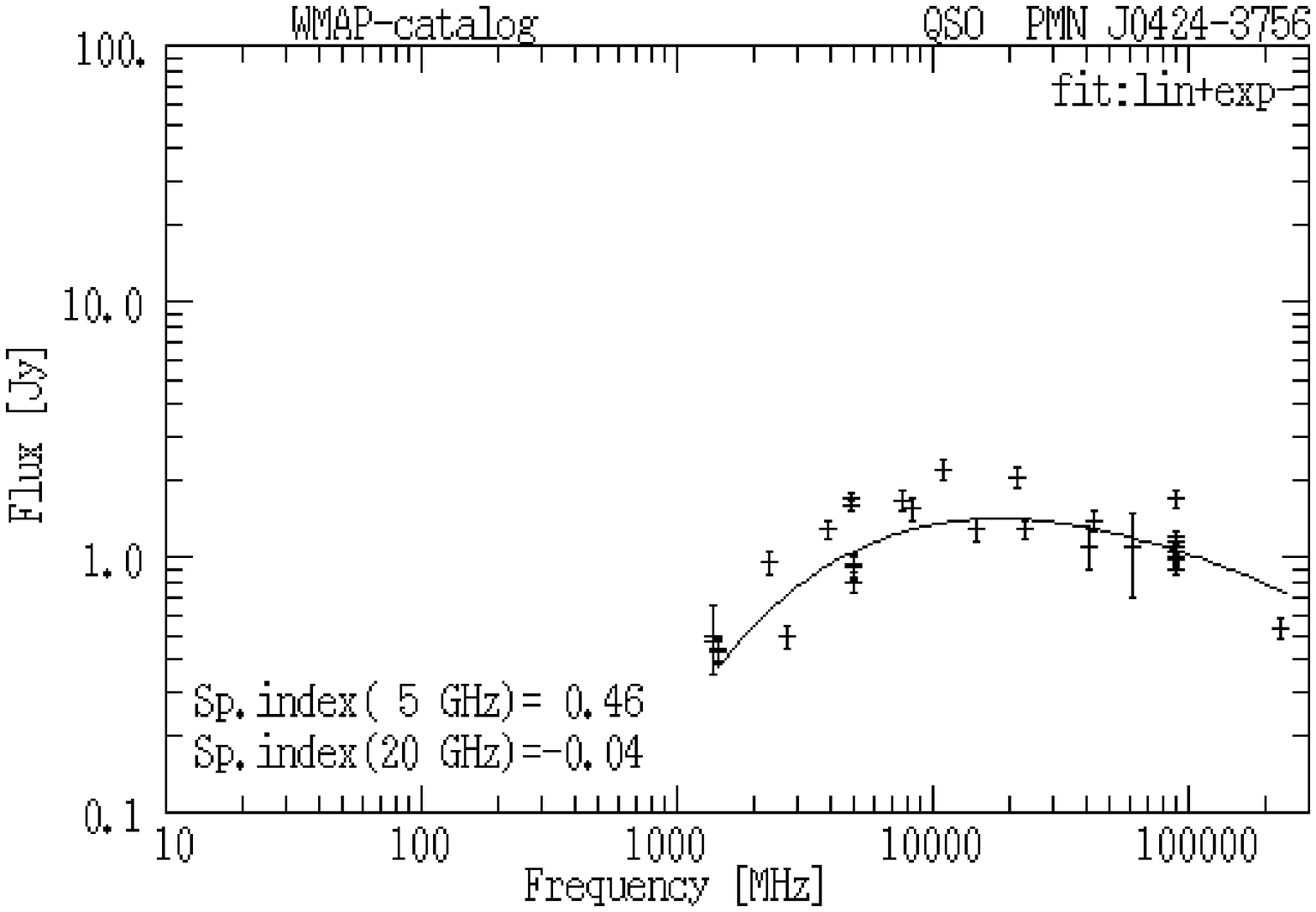,width=8.2cm,angle=0}}}\end{figure}
\begin{figure}\centerline{\vbox{\psfig{figure=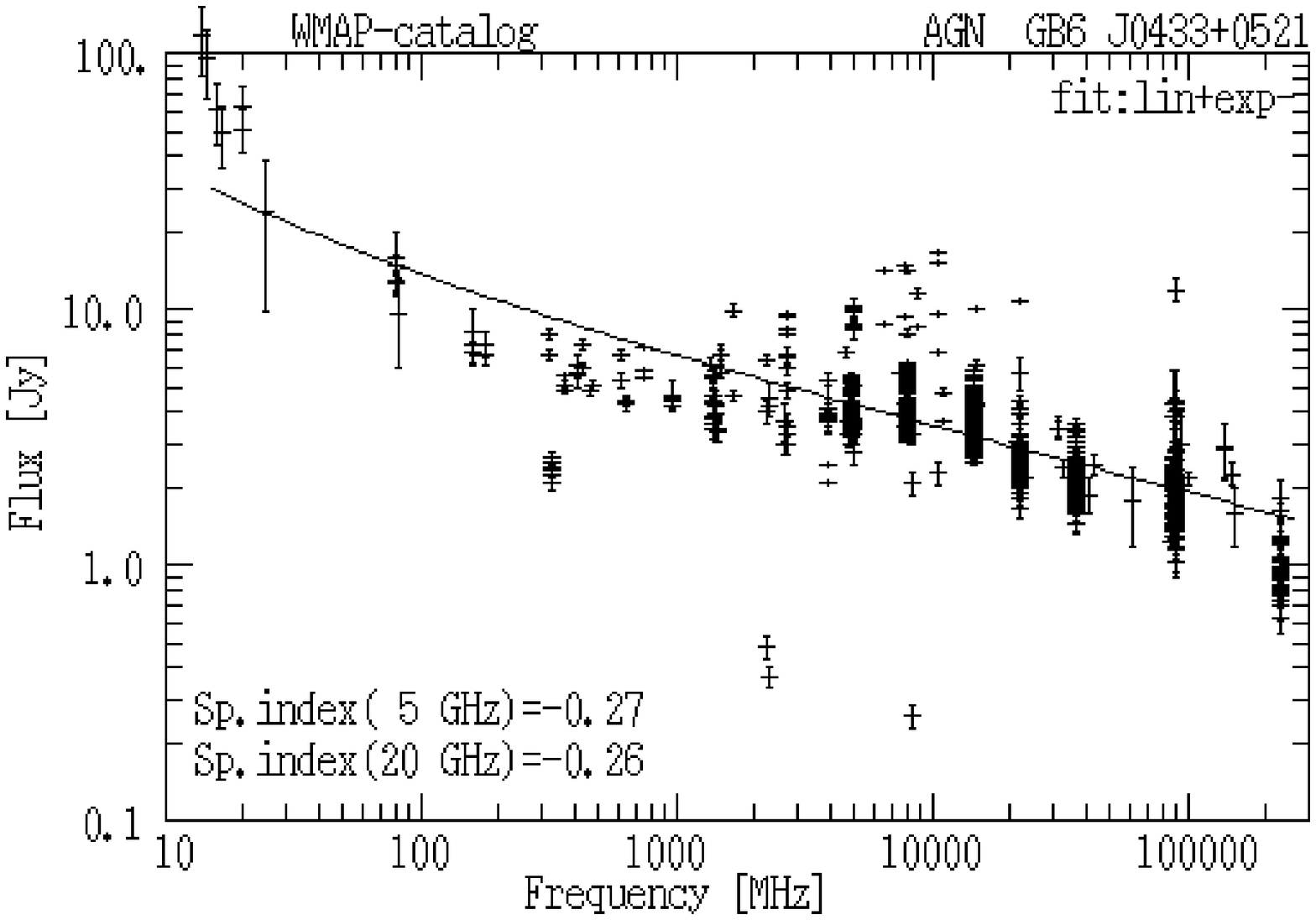,width=8.2cm,angle=0}}}\end{figure}
\begin{figure}\centerline{\vbox{\psfig{figure=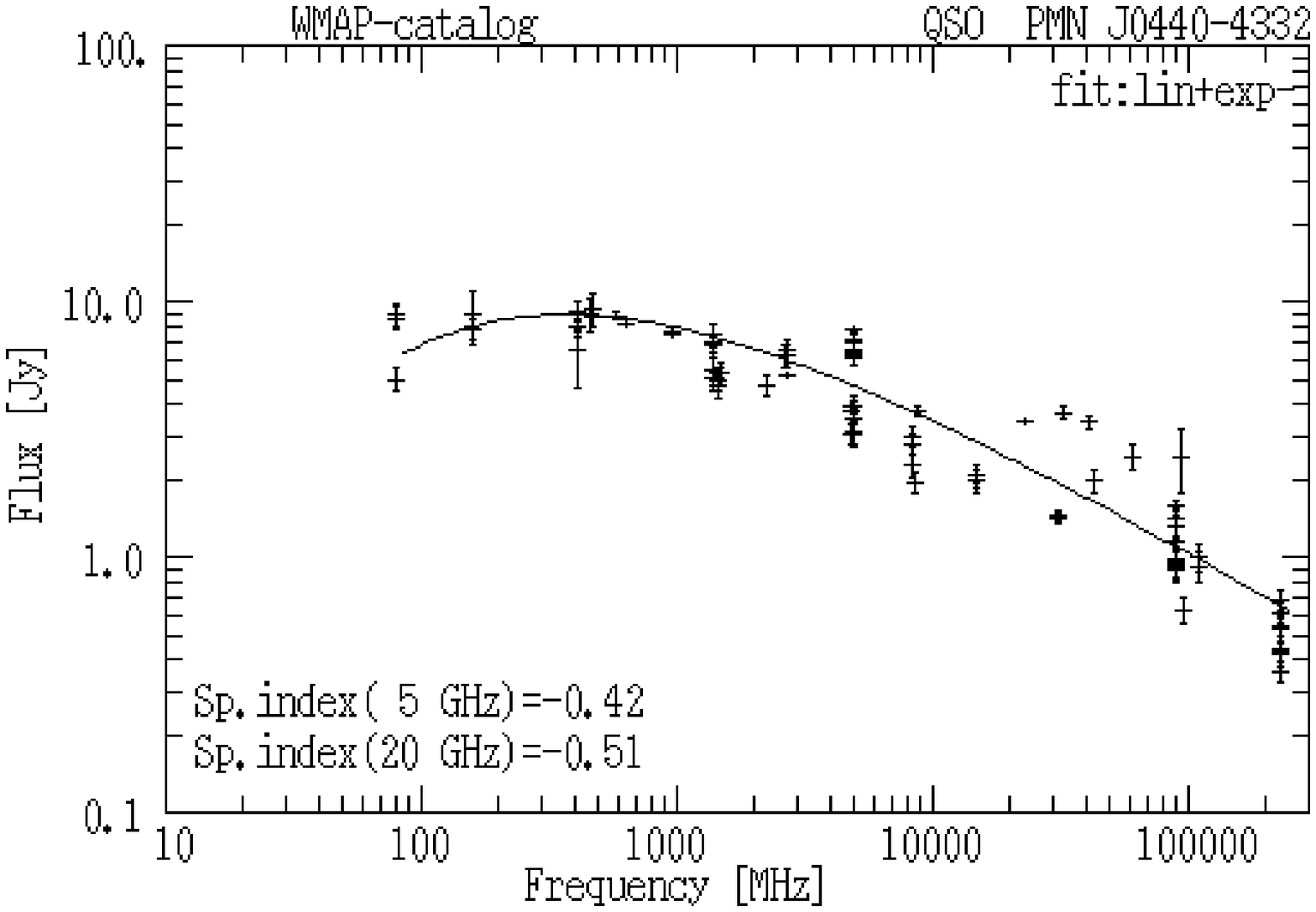,width=8.2cm,angle=0}}}\end{figure}
\begin{figure}\centerline{\vbox{\psfig{figure=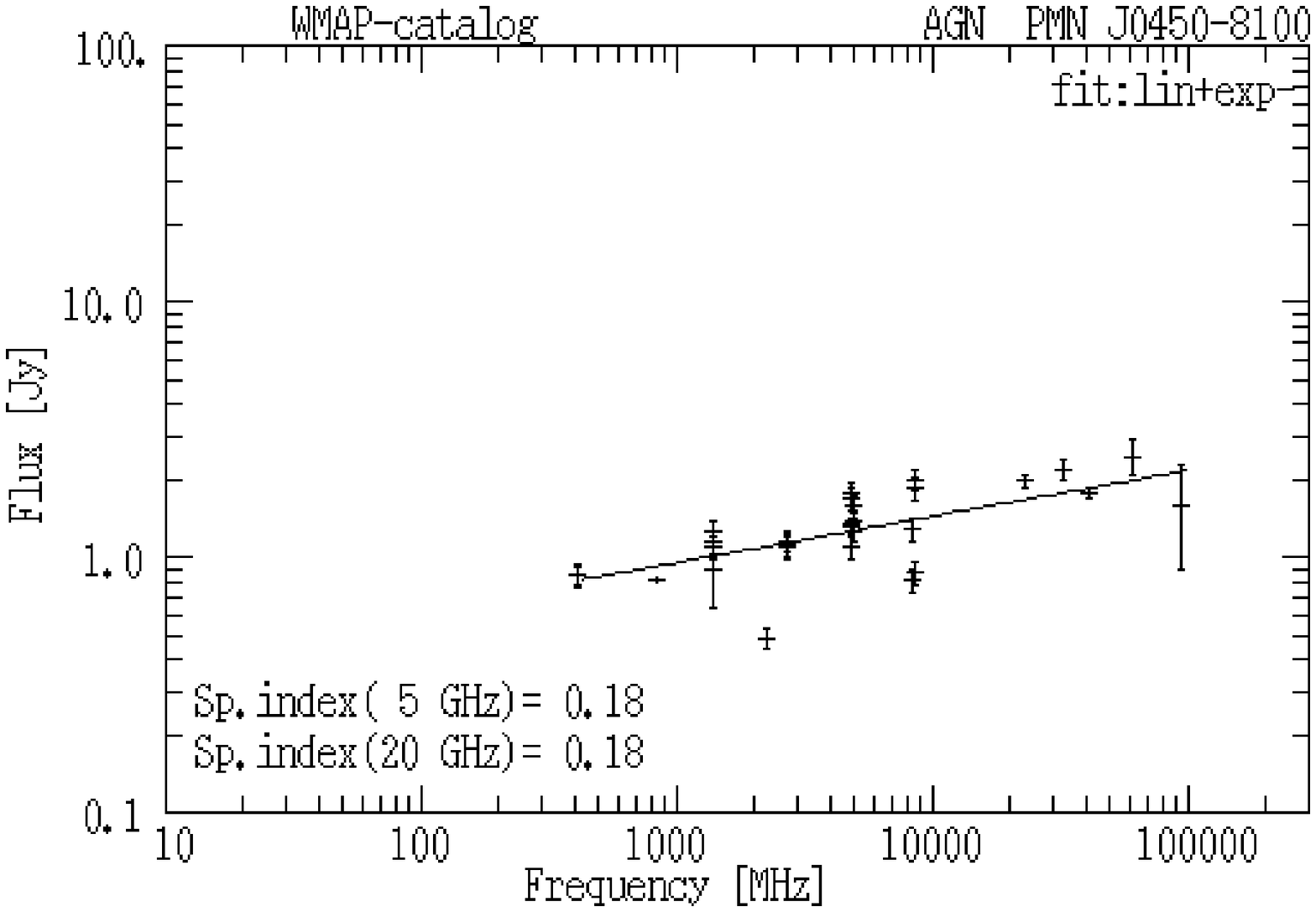,width=8.2cm,angle=0}}}\end{figure}
\end{document}